\documentclass[preprint,12pt]{elsarticle} 

\usepackage{lineno}
\modulolinenumbers[5]
\usepackage{color}
\usepackage[usenames,dvipsnames]{xcolor}
\usepackage{fullpage}
\usepackage{amsmath}
\usepackage{amssymb}
\usepackage{mathrsfs}
\usepackage{newfloat}
\usepackage{graphicx}
\usepackage{stmaryrd}
\usepackage{subfig}
\usepackage{epstopdf}
\usepackage{multirow}
\usepackage{stackengine}
\usepackage{caption}
\usepackage{tikz}
\usepackage{multicol}
\usetikzlibrary{arrows,positioning,calc}
\usepackage{pstool}
\usepackage{textcomp}
\usepackage[hidelinks]{hyperref}
\hypersetup{colorlinks,bookmarksopen,linkcolor=blue,citecolor=blue,urlcolor=blue}
\usepackage{float}
\usepackage{makecell}
\usepackage[export]{adjustbox}
\usepackage{pdfpages}
\usepackage{booktabs} 

\usepackage[normalem]{ulem}
\definecolor{gray}{RGB}{111,111,111}

\definecolor{myblue}{RGB}{0,115,189}
\definecolor{mygreen}{rgb}{0.19,0.61,0.21}

\newcommand{\bs}[1]{{\boldsymbol{#1}}}

\newcommand{\sgn}[1]{\mathrm{sgn}\left(#1\right)}
\newcommand{\mtrx}[1]{{\boldsymbol{\mathsf{#1}}}}
\newcommand{\tensor}[1]{{\boldsymbol{#1}}}
\newcommand{\tensorfour}[1]{{\mathbb{#1}}}
\newcommand{\column}[1]{{\underline{#1}}}

\DeclareFloatingEnvironment{algorithm}
\graphicspath{{figs/}}

\newcommand{\ignore}[1]{}

\newcommand{\full}{{Full}}
\newcommand{\fepn}{{FE-PN}}
\newcommand{\qcfix}{{QC fix}}
\newcommand{\qcadaptive}{{QC ZZ}}
\newcommand{\fecz}{{FE-CZ}}
\newcommand{\qccoarse}{{QC coarse}}
\newcommand{\qcfine}{{QC fine}}

\newcommand{\processZone}{{distributed damage zone}}

\journal{Comput. Methods Appl. Mech. Engrg.}

\begin{document}


\begin{frontmatter}
\title{Comparative study of multiscale computational strategies for materials with discrete microstructures\tnoteref{titlefoot}}

\author[CTU]{K. Mike\v{s}\corref{correspondingauthor}}
\ead{Karel.Mikes@cvut.cz}

\author[TUe]{F. Bormann}
\ead{f.bormann@m2i.nl}

\author[CTU,TUe]{O.~Roko\v{s}}
\ead{O.Rokos@tue.nl}

\author[TUe]{R.H.J.~Peerlings}
\ead{R.H.J.Peerlings@tue.nl}

\address[CTU]{Department of Mechanics, Faculty of Civil Engineering, Czech Technical University in Prague, Th\'{a}kurova~7, 166~29 Prague~6, Czech Republic}

\address[TUe]{Mechanics of Materials, Department of Mechanical Engineering, Eindhoven University of Technology, P.O.~Box~513, 5600~MB~Eindhoven, The~Netherlands}
\cortext[correspondingauthor]{Corresponding author.}

\tnotetext[titlefoot]{The post-print version of this article is published in \emph{Comput. Methods Appl. Mech. Engrg.}, \href{https://doi.org/10.1016/j.cma.2021.113883}{10.1016/j.cma.2021.113883}. This manuscript version is made available under the \href{https://creativecommons.org/licenses/by-nc-nd/4.0/}{CC-BY-NC-ND~4.0} license.}

\begin{abstract}

The evolution of local defects such as dislocations and cracks often determines the performance of engineering materials. For a proper description and understanding of these phenomena, one typically needs to descend to a very small scale, at which the discreteness of the material emerges.
Fully-resolved discrete numerical models, although highly accurate, often suffer from excessive computing expenses when used for application-scale considerations.
More efficient multiscale simulation procedures are thus called for, capable of capturing the most significant microscopic phenomena while being computationally tractable for macroscopic problems. Two broad classes of methods are available in the literature, which conceptually differ significantly. The first class considers the fully-resolved discrete system, which is subsequently reduced through suitable mathematical tools such as projection and reduced integration. The second class of methods first homogenizes the discrete system into an equivalent continuum formulation, into which the main phenomena are added through specific enrichments.
This paper provides a thorough comparison of the two different modeling philosophies in terms of their theory, accuracy, and performance. To this goal, two typical representatives are adopted: the Quasicontinuum method for the first class, and an effective continuum with an embedded cohesive zone model for the second class. 
Two examples are employed to demonstrate capabilities and limitations of both approaches. In particular, dislocation propagation and pile-up against a coherent phase boundary is considered at the nanoscale level, whereas a three-point bending test of a concrete specimen with crack propagation is considered at the macroscale level. In both cases, the accuracy of the two methods is compared against the fully-resolved discrete reference model.
It is shown that whereas continuum models with embedded cohesive zones offer good performance to accuracy ratios, they might fail to capture unexpected more complex mechanical behavior such as dislocation reflection or crack branching. The Quasicontinuum method, on the other hand, offers more flexibility and higher accuracy at a slightly higher computational cost.

\end{abstract}


\end{frontmatter}

%
%
\section{Introduction} 
%
Numerical modeling of material behavior is an important ingredient in engineering. At sufficiently large scales, most materials can be considered as a continuum, whereas at smaller scales, materials such as foams, textiles, concrete, or paper, reveal discrete microstructures. At the smallest scale, i.e.,~nano-scale, all materials become discrete and their behavior is dictated by the underlying atomic lattice. When modeling of localized mechanical phenomena such as for instance the nucleation and propagation of dislocations, cracks or other defects, is of interest, the inherently non-local behavior of the underlying discrete microstructure comes into play. To capture those non-localities, accurate numerical models may be required.

The most accurate option is to use discrete models that fully resolve the microstructure and hence inherently incorporate the underlying non-locality at the level of the individual interactions. Typical examples at different scales are molecular statics~\cite{shastry1996molecular,budarapu2014adaptive,parthasarathy1993molecular} or dynamics~\cite{decelis1983molecular,swadener2002molecular,yamakov2002dislocation} for atomic systems, or lattice or beam networks for mesoscopically discrete materials.
Sophisticated discrete material models for various materials such as 
paper~\cite{liu20102,bosco2015predicting}, 
textile~\cite{beex2013experimental}, 
fiber-reinforced composites~\cite{bavzant1990random,zhou1995failure}, 
 other fibrous materials~\cite{wilbrink2013discrete,ridruejo2010damage}, or concrete~\cite{schlangen1997fracture,cusatis2011lattice,lilliu20033d} can be found in the literature. 
For typical engineering applications, however, full-scale models suffer from prohibitive computational costs requiring simplifications, especially in problems with a large separation of scales between the application length scale and the length scale of the underlying lattice, or in problems in which multiple model evaluations are required such as optimization, parameter identification, or stochastic modeling.

\begin{figure}
	\centering
	\subfloat[two-way simplification]{\includegraphics[scale=0.9]{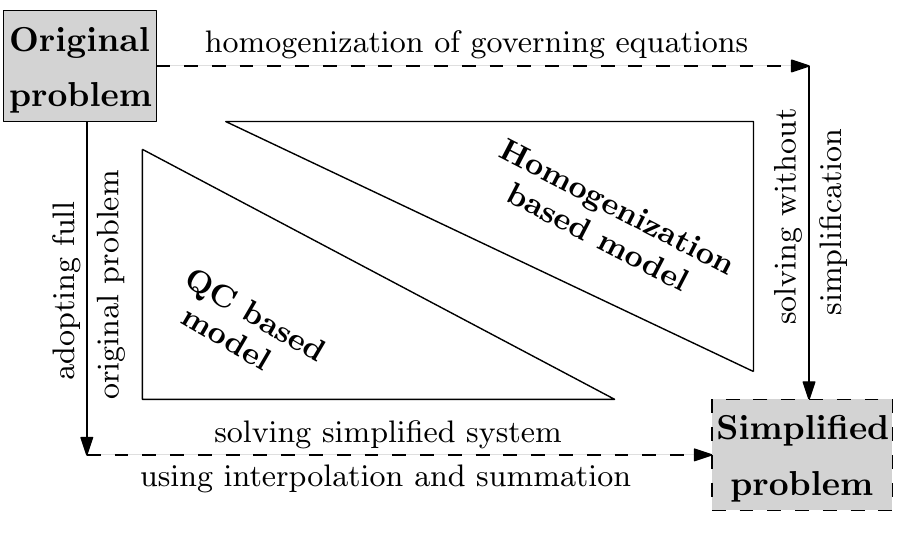}}
	\hspace{2em}	
	\subfloat[graphical representations]{\includegraphics[scale=0.9]{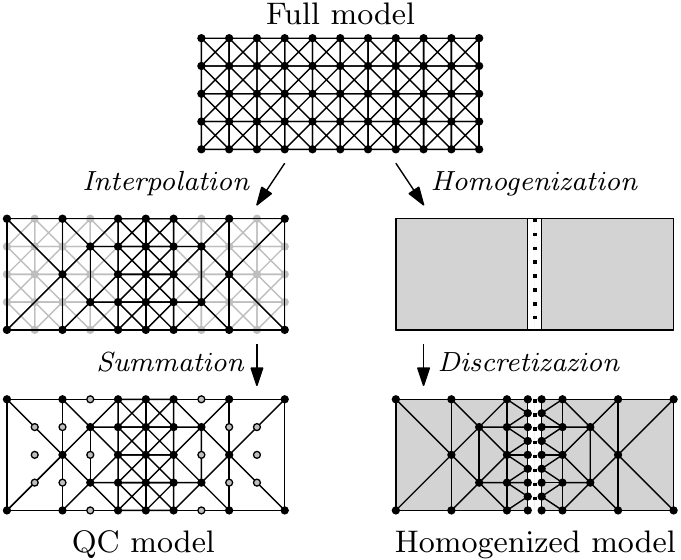}}		
	\caption{(a)~A schematic illustration showing two ways of simplification of the original fully-resolved discrete model. In the first approach, the original system is reduced through suitable numerical and mathematical techniques, solving subsequently a simplified system (the left-bottom path). In the second approach, the original system is homogenized first to provide an equivalent continuum model with embedded key features, which is subsequently solved using standard techniques for continuum problems (the top-right path). (b)~A graphical representation of the individual steps required by the two simplification approaches.}
	\label{fig:int-simplification}
\end{figure}

Two broad classes of conceptually distinct methodologies are available to provide computationally efficient numerical tools for such cases. The first class relies on considering directly the underlying model, which is subsequently reduced by suitable numerical and mathematical techniques such as projection on a reduced kinematical basis and reduced integration. The second class of methods first derives equivalent homogeneous governing equations from the underlying discrete microstructure. Afterwards, suitable enrichments are added to model the key mechanical phenomena. The resulting system is then solved using standard numerical techniques for continuum problems such as the Finite Element~(FE) method. For the purposes of this paper, the QuasiContinuum~(QC) method, originally introduced by Tadmor et al.~\cite{tadmor1996quasicontinuum}, is adopted as a typical representative for the first class of approaches, whereas a continuum formulation with an embedded cohesive zone model is considered as a representative for the second class of approaches. The reduction of both classes is implemented at different levels of the problem formulation, as depicted in Fig.~\ref{fig:int-simplification}.
%
%
\subsection{Quasicontinuum based approach}
In the QC approximation, the original complexity of the full underlying problem is reduced to simplified model in two steps:
\begin{enumerate}[(i)]\setlength{\itemsep}{0pt}\setlength{\parskip}{0pt}\setlength{\parsep}{0pt}  
	\item \emph{Interpolation} is used to define the positions of all atoms or lattice sites of the original system from a set of representative atoms, the so-called repatoms;
	\item \emph{Summation} efficiently estimates the stored energy and internal forces of the system in which the interpolation step has been applied, based on a set of the so-called sampling atoms.
\end{enumerate}

The interpolation step considers only a small set of repatoms to represent the kinematic behavior of the entire system, typically chosen as vertices of an overlaid triangular Finite Element~(FE) mesh. To accurately capture any localized phenomena, the considered domain is split into two parts. The first part is the area of high interest where all atoms of the underlying lattice are considered as repatoms and thus also as nodes of the triangular mesh, i.e., the overlaid triangulation is fully refined down to the level of the underlying lattice. Elsewhere, interpolation is adopted to coarse-grain the underlying lattice; in this region individual triangular elements contain multiple atoms or lattice sites. The fully-resolved region may be adaptively adjusted during the simulation to capture the relevant evolution of the microstructure. The main advantage of the QC approach is that since the response of the reduced model is governed by the underlying `exact' discrete model, one expects a high fidelity, which is limited only by the coarsening away from the region of interest. The price to pay is (possibly) a more expensive simulation.

The QC method was originally introduced in 1996~\cite{tadmor1996quasicontinuum, tadmor1996mixed} for atomic systems with long-range conservative interaction potentials and has been successfully used for the simulation of localized phenomena, such as crack nucleation and propagation~\cite{klein1998crack,miller1998quasicontinuum}, motion and interaction of dislocations~\cite{phillips1999hierarchical}, or processes related to nanoindentation~\cite{knap2001analysis,tadmor1999nanoindentation}.
Later, the QC method was extended to regular discrete lattice networks with short-range nearest-neighbor interactions with both conservative~\cite{beex2011quasicontinuum} and non-conservative~\cite{beex2014multiscale_dissipative,beex2014multiscale_sliding} interaction potentials including dissipation and fibre sliding.
A further extension was provided for planar beam lattices~\cite{beex2015higher,beex2014quasicontinuum} and periodic beam lattices in a co-rotational framework~\cite{phlipot2019quasicontinuum}. Additional research focused on goal-oriented adaptivity~\cite{arndt2007goal,memarnahavandi2015goal}, a meshless QC method~\cite{kochmann2014meshless}, or an energy-based variational formulation for regular lattices with plasticity~\cite{rokovs2016variational} and localized damage~\cite{rokovs2017adaptive,rokovs2017extended}.
Recently, the QC method was also extended to irregular lattices~\cite{mikevs2017quasicontinuum} or polymer networks~\cite{ghareeb2020adaptive}. A generalization to metallic lattice materials was introduced in~\cite{chen2020generalized}, where different finite element shape functions are used for different types of lattice nodes.
%
%
\subsection{Homogenization based approach}
In the second approach, homogenization of the underlying discrete system is first carried out, from which continuum governing equations follow. Usually a Cauchy continuum is adopted, which neglects any non-local interactions. A specific enrichment thus needs to be added, which is usually implemented through an appropriate interface model, e.g.,~a cohesive zone model for cracks. These interfaces allow for mutual displacements in tangential and normal directions, governed by appropriate effective constitutive laws.

The positioning and orientation of such interfaces is assumed to be known a priori and fixed hereafter, although more general models can be found in the literature, involving cohesive zones in between all elements~\cite{xu1994numerical}, or an extended FE method with cohesive zone formulation~\cite{benvenuti2008regularized, asferg2007consistent, wang2015numerical}. 
Often, the interfaces positions can be set a priori on the basis of available knowledge of the underlying physics and to limit the computational expense. A crack path can in certain cases be estimated with sufficiently high accuracy to be fixed in all simulations, for instance. 
Thus, by design, this approach is capable of capturing only those phenomena that are considered in the homogenized model. 
In situations in which the positions of interfaces cannot be estimated in advance, homogenization based approaches are usually no longer advantageous. Even though generalized homogenized models (which, e.g., consider cohesive zones in between all elements) can be used even without any prior information on the position of the expected localized phenomena, it is usually computationally less expensive directly employ adaptive QC method in such situations, since induced computing costs might be comparable or exceed those of the QC method or even of a full solution.
Although many physical phenomena are excluded from the homogenized modeling, one generally expects this approach to be more efficient. The material parameters of the homogenized system as well as properties of the cohesive interfaces are fully determined by the underlying lattice physics. In most cases, a linear elastic isotropic material model is sufficient to provide an accurate description of the lattice behavior far from any localized phenomena, whereas in the close vicinity of a crack or a dislocation large deformations and strains may occur; these are, however, neglected in this contribution for simplicity. Interface constitutive laws range from simple closed-form uncoupled to numerical and coupled laws derived directly from numerical homogenization. Typical representatives of the homogenization based approaches, both considered in this paper, are the Peierls--Nabarro model introduced in~\cite{peierls1940size, nabarro1947dislocations, schoeck1994generalized, hirth1983theory,bormann2020peierls} and cohesive zone model discussed in~\cite{barenblatt1962mathematical,dugdale1960yielding,de2003numerical,elices2002cohesive,jin2006comparison,SUN2012227}. Yet another representative is the Discrete Dislocation Dynamics~(DDD) approach, e.g., \cite{van1995discrete, cleveringa2000discrete}, which is not considered here further.
%
%
\subsection{Objectives}
The goal of this paper is a thorough comparison of the above two modeling philosophies in terms of their theoretical framework, accuracy, and numerical efficiency. To this end, two examples are considered with microstructures at the nano- and meso-scale level. At the nanoscale, dislocation propagation and pile-up against a coherent phase boundary in a two-dimensional atomic system with a hexagonal stacking of atoms is adopted. At the mesoscale, a three-point bending test of a concrete specimen with crack propagation through a two-dimensional regular X-braced lattice structure is examined. In both cases, a Quasicontinuum model and a FE approach equipped with cohesive zone interfaces, using either the Peierls--Nabarro model or a traction--separation law, are compared against a fully-resolved discrete model. The accuracy and efficiency of both homogenization techniques are discussed and evaluated.

\subsection{Outline and notation}
The remainder of this paper is organized as follows. Section~\ref{sec:Microscale} considers the molecular statics problem of dislocation propagation, starting with the theory of the fully-resolved discrete atomic problem, followed by a Quasicontinuum formulation, and closing with the FE Peierls--Nabarro homogenization based approach. In Section~\ref{sec:Mesoscale}, an analogous comparison is made for the case of a three-point bending test considered at the mesoscale, mimicking quasi-brittle damage in a concrete specimen. The paper closes with a summary and conclusions in Section~\ref{sec:Conclusion}.

Throughout the paper, scalars are denoted~$ a $, vectors~$ \vec{a} $, position vectors in a two-dimensional continuum~$ \vec{x} = x\vec{e}_x + x\vec{e}_y $, and atom position vectors~$ \vec{r} = r_x\vec{e}_x + r_y\vec{e}_y $. Second-order tensors are denoted~$ \tensor{\sigma} $, matrices~$ \mtrx{A} $, column matrices~$ \column{a} $, scalar products of two vectors~$ \vec{a} \cdot \vec{b} = a_i b_j $, single contractions of two second-order tensors~$ \tensor{\sigma} \cdot \tensor{\varepsilon} = \sigma_{ik}\varepsilon_{kj}\vec{e}_i\vec{e}_j $, double contractions of two second-order tensors~$ \tensor{\sigma}:\tensor{\varepsilon} = \sigma_{ij}\varepsilon_{ji}$, whereas a transpose is denoted as~$ \tensor{\sigma}^\mathsf{T}$, where $ \sigma_{ij}^\mathsf{T} = \sigma_{ji} $. The gradient and divergence operators are denoted as~$ \displaystyle \vec{\nabla} \vec{a}(\vec{x}) = \frac{\partial a_j(\vec{x})}{\partial x_i} \vec{e}_i \vec{e}_j $ and~$ \displaystyle \vec{\nabla} \cdot \vec{a}(\vec{x}) = \frac{\partial a_i(\vec{x})}{\partial x_i} $, respectively. Throughout this contribution, Einstein's summation convention is adopted on repeated indices~$i$ and~$j$, and~$\vec{e}_x$ and~$\vec{e}_y$ denote the basis vectors of a two-dimensional Cartesian coordinate frame.
%
%
\section{Molecular statics: dislocation transmission across a phase boundary}
\label{sec:Microscale}
%
%
\subsection{Full atomic model}
\label{ssec:Micro-Full}
\begin{figure}
	\centering
	\subfloat[fully-resolved discrete atomic system]{\includegraphics[trim=0cm 0.cm .0cm 0.0cm,clip=true,scale=1]{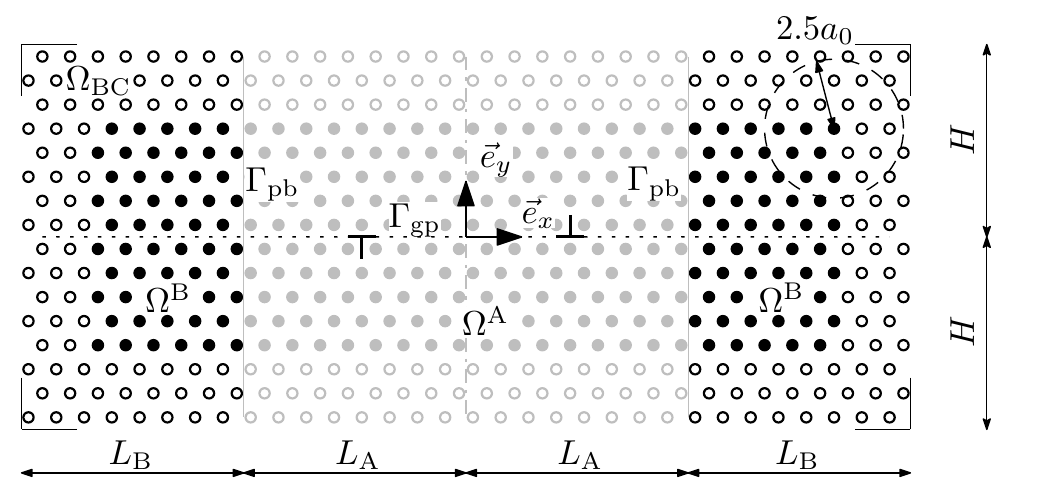}\label{fig:geometry-atm}}
	\hspace{2em}
	\subfloat[applied shear deformation]{\includegraphics[trim=0cm -3cm .0cm 0.5cm, clip=true, width=0.28\linewidth]{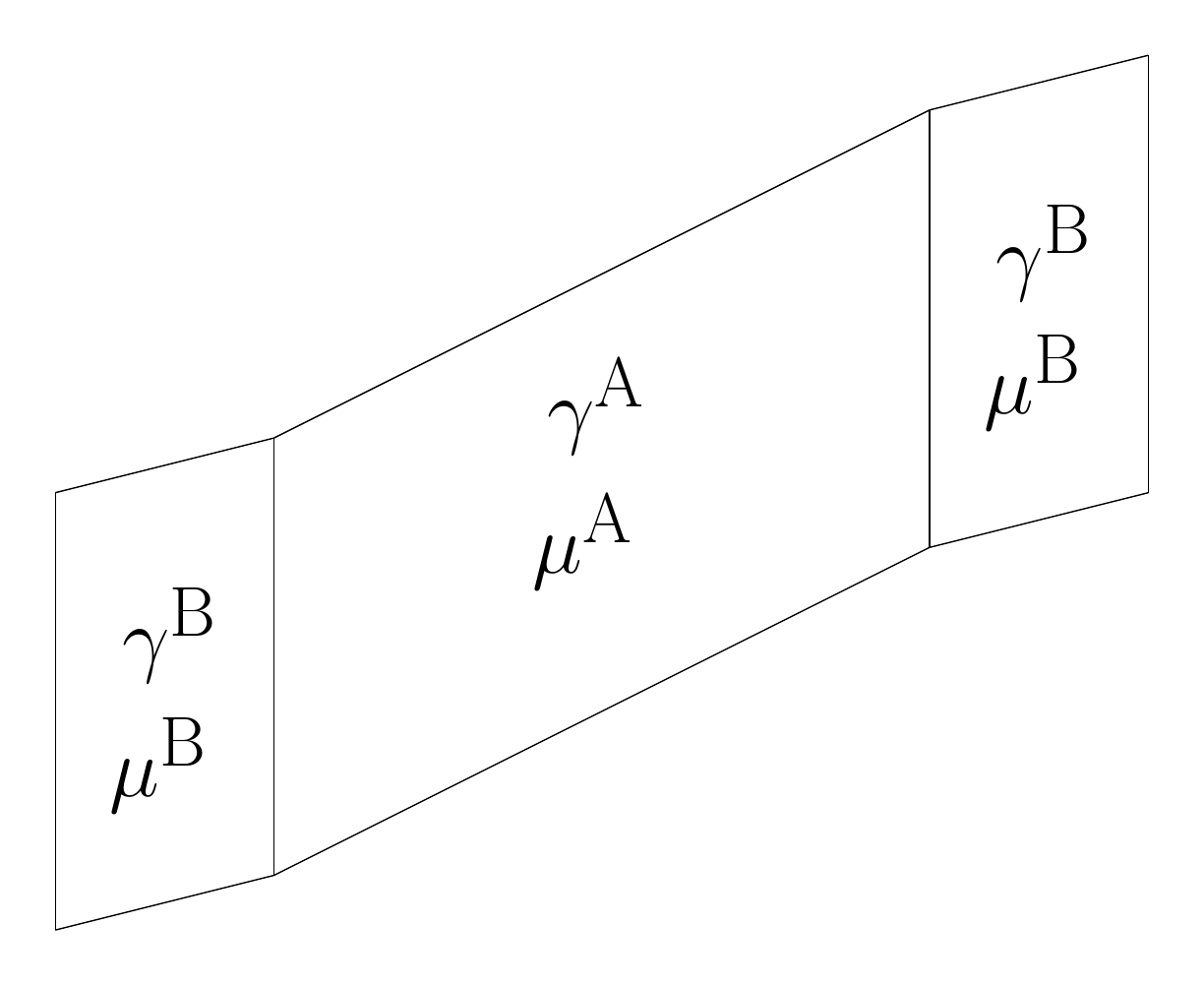}\label{fig:geometry-shear}}
	\caption{(a)~A sketch of the atomic representation of a small illustrative~$32 \times 16$ atom bi-crystal, analogous to the $1,024 \times 512$ atom bi-crystal used for the simulation of edge dislocation dipoles interacting with a coherent phase boundary. The entire domain~$\Omega$ consists of two phases, A (gray dots) and~B (black dots). Boundary conditions are applied by fixing the positions of all boundary atoms (positioned within the~$r_\mathrm{cut}$ distance from the specimen's boundary~$\partial\Omega$), denoted by the hollow circles. A horizontal glide plane is marked by the dotted line, the material interface between the two types of crystals by the solid gray line, whereas the employed cut-off radius is indicated by the dashed circle. (b)~A sketch of the applied shear deformation which---assuming an ideal, defect-free, and linear-elastic specimen---initially induces approximately constant shear stress throughout the entire domain.}
	\label{fig:geometry}
\end{figure}

%
To analyze a dislocation pile up against a material interface in two dimensions, a shear test with two different material phases is considered. Dislocation dipoles are nucleated at the center of the specimen, and are pushed towards boundaries by externally applied shear stress. As illustrated in Fig.~\ref{fig:geometry-atm}, the adopted bi-crystal, occupying a domain~$\Omega$, consists of a soft Phase~A that is flanked by a stiffer Phase~B. The individual subdomains $\Omega^\mathrm{A}$ and $\Omega^\mathrm{B}$ are separated by a phase boundary $\Gamma_{\mathrm{pb}}$. The considered dimensions are summarized in Tab.~\ref{Tab:Size}. 
Inside~$\Omega$, hexagonal lattice structures of equal spacings~$a_0$ are considered in both phases, comprising a total of~$n_{\mathrm{ato}}$ atoms~$\alpha$, collected in an index set~$N = N^\mathrm{A} \cup N^\mathrm{B}$ where $N^\mathrm{A}$ and $N^\mathrm{B}$ are sets of atoms associated with the individual phases. The crystal orientation is chosen such that one set of glide planes is oriented perpendicular to the two vertical coherent phase boundaries, whereas the other sets of glide planes are inclined by an angle of~$ \pm 60^\circ $.

\begin{figure}
	\centering
	\subfloat[hexagonal atomic lattice]{\includegraphics[trim=0cm -1.0cm .0cm 0.0cm,clip=true, scale=0.6] {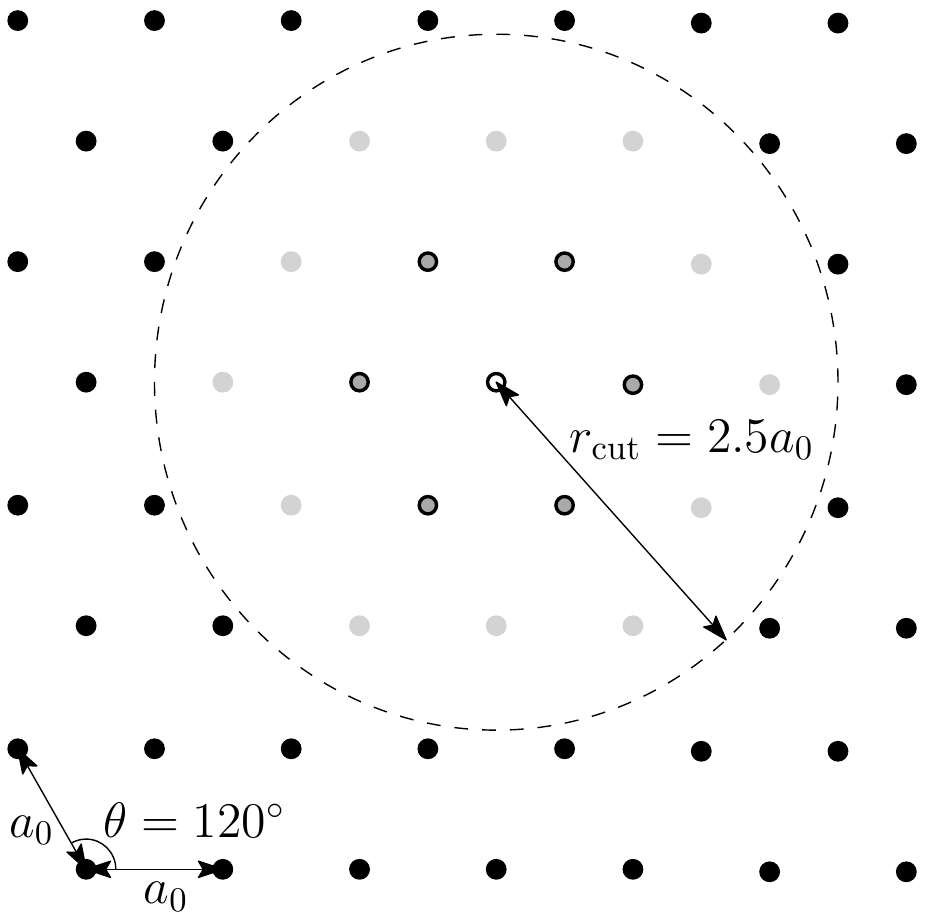} \label{fig:LJ-a}}
	\hspace{1em}
	\subfloat[interatomic pair potential]{\includegraphics[trim=0cm .0cm 0.7cm 0.0cm, clip=true,scale=0.5]{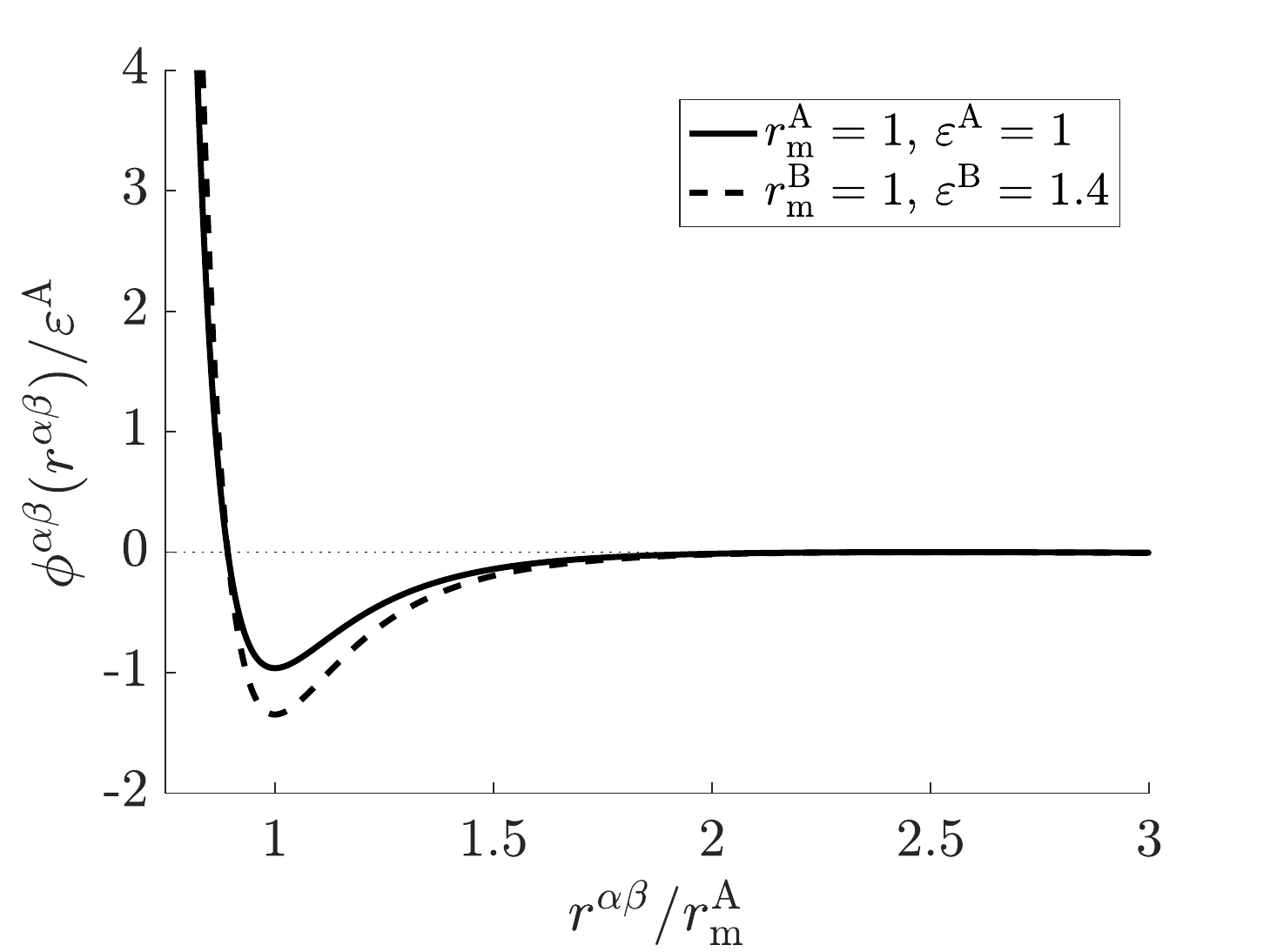} \label{fig:LJ-b}}
	\caption{(a)~The hexagonal lattice used for atomic simulations in two dimensions; current atom (white), nearest-neighbor atoms (bounded gray), and next-to-nearest atoms (gray). (b) Two versions of the Lennard--Jones potential corresponding to material contrast ratio~$\rho = 1.4$.}
	\label{fig:LJ}
\end{figure}

The interatomic pair potentials employed within both phases, denoted~$\phi^\mathrm{A}$ and~$\phi^\mathrm{B}$, are of the Lennard--Jones~(LJ) type. To account for nearest and next-to-nearest interatomic interactions only, a cut-off radius~$r_{\mathrm{cut}}=2.5\,r_\mathrm{m}$ is adopted, as indicated in Figs.~\ref{fig:geometry-atm} and~\ref{fig:LJ-a} by the dashed circles. To ensure smoothness of the potential at the cut-off radius, cf. Fig.~\ref{fig:LJ-b}, the LJ potential is modified according to
\begin{equation}
\phi(r) = \phi_{lj}(r) - \phi_{lj}(r_{\mathrm{cut}}) - (r-r_{\mathrm{cut}})\phi_{lj}'(r_{\mathrm{cut}}), 
\quad \text{with} \quad 
\phi'(r) = \frac{\mathrm{d}\phi(r)}{\mathrm{d}r},
\label{eq:ljsf}
\end{equation}
where~$r$ is a scalar measuring the distance between a pair of atoms, and
\begin{equation}
\phi_{lj}(r) = \varepsilon\left[\left(\frac{r_{\mathrm{m}}}{r}\right)^{12} -2\left(\frac{r_{\mathrm{m}}}{r}\right)^{6}\right]
\label{eq:lj}
\end{equation}
is the standard LJ interatomic pair potential. In Eq.~\eqref{eq:lj}, $\varepsilon$ denotes the depth of the energy well and~$r_\mathrm{m}$ the distance for which the interaction energy reaches its minimum, as indicated in Fig.~\ref{fig:LJ-b}. For further details see, e.g.,~\cite{TadmorModel}. Mixing of the interatomic potentials across the phase boundary~$\Gamma_\mathrm{pb}$ is considered through averaging, i.e.,
\begin{equation}
\phi^{\alpha\beta}(r^{\alpha\beta}) = \frac{1}{2}\left[ \phi^\alpha(r^{\alpha\beta}) + \phi^\beta(r^{\alpha\beta}) \right],
\label{eq:averaging}
\end{equation}
where~$\vec{r}^\alpha = r_x^\alpha\vec{e}_x + r_y^\alpha\vec{e}_y$ denotes a current position of an atom~$\alpha$, $r^{\alpha\beta} = \| \vec{r}^{\alpha\beta} \|_2$ denotes the Euclidean distance between a pair of atoms~$\alpha$ and~$\beta$, $\vec{r}^{\alpha\beta} = \vec{r}^\beta - \vec{r}^\alpha$ is a vector of their relative positioning, and
\begin{equation}
\phi^\alpha = \left\{
\begin{array}{@{}l@{}l}
\phi^\mathrm{A}, & \quad \mbox{if} \quad \alpha \in N^\mathrm{A}, \\
\phi^\mathrm{B}, & \quad \mbox{if} \quad \alpha \in N^\mathrm{B},
\end{array}
\right. \quad \alpha = 1, \dots, n_{\mathrm{ato}}.
\label{eq:ljphase}
\end{equation}
Material parameters are specified in Tab.~\ref{tab:mat}, which are all normalized with respect to~$\varepsilon^\mathrm{A}$ and~$r_\mathrm{m}^\mathrm{A}$.
\begin{table}
	\caption{Geometric parameters used in the atomic simulations. A rectangular specimen of a size~$2(L_\mathrm{A}+L_\mathrm{B}) \times 2 H$ is considered with a hexagonal stacking of atoms and a lattice spacing of~$a_0$.}
	\renewcommand*{\arraystretch}{1.3}
	\centering
	\begin{tabular}{c|cccc}
		parameters  & $L_\mathrm{A}$  & $L_\mathrm{B}$ & $H$	&	$h_0$  \\ \hline
		Size $1{,}024\times512$ &      $256 \, a_0$      &      $256 \, a_0$       &      $256 \, h_0$	&	$\sqrt{3} \, a_0/2$      \\
	\end{tabular}
	\label{Tab:Size}
\end{table}
\begin{table}
	\caption{Parameters of pair potentials corresponding to the two phases, A and~B, as a function of a material contrast ratio~$\rho \in \{ 1.4, 4.0 \}$.}
	\renewcommand*{\arraystretch}{1.3}	
	\centering
	\begin{tabular}{c|rrrr}
		parameters & \multicolumn{1}{c}{$\varepsilon/\varepsilon^{\mathrm{A}}$} & \multicolumn{1}{c}{$r_{\mathrm{m}}/r_{\mathrm{m}}^{\mathrm{A}}$} & \multicolumn{1}{c}{$a_0/r_{\mathrm{m}}^{\mathrm{A}}$} & \multicolumn{1}{c}{$r_{\mathrm{cut}}/a_0$} \\\hline
		Phase~A & $1$ & $1$ & $0.99296702$ & $2.5$ \\
		Phase~B & $\rho$ & $1$ & $0.99296702$ & $2.5$ 		
	\end{tabular}
	\label{tab:mat}
\end{table}

If only the nearest-neighbor interactions are considered and there are no surface effects within a homogeneous crystal, i.e.,~an infinite homogeneous crystal is assumed, the lattice spacing~$a_0 = r_\mathrm{m}$ yields a stress-free configuration. For next-to-nearest interactions, however, such a lattice spacing invokes a pre-stress within the lattice, as calculated by a non-vanishing Virial stress tensor~$\tensor{\sigma}$ defined as 
\begin{equation}
\tensor{\sigma} = \frac{1}{2V} \sum_{\substack{\alpha,\beta=1\\\alpha\neq\beta,\, r^{\alpha\beta}<r_\mathrm{cut}}}^{n_\mathrm{ato}}[\phi^{\alpha\beta}(r^{\alpha\beta})]'\ \frac{\vec{r}^{\alpha\beta} \otimes \vec{r}^{\alpha\beta}}{r^{\alpha\beta}},
\label{eq:virialstress}
\end{equation}
where~$V$ denotes a volume considered in the deformed configuration over which the stress is computed, see, e.g.,~\cite[Section~11.5.2]{TadmorModel} for more details. To attain a stress-free state, the system reduces its initial interatomic spacing~$a_0 = r_\mathrm{m}$ to~$a_0 = 0.99296702 \, r_\mathrm{m}$. Although such a change is geometrically negligible, it has significant influence on the macroscopic mechanical behavior of the resulting system.

For the application of  the boundary conditions the behavior of an ideal, defect-free, and linear-elastic specimen is assumed under a state of constant shear stress~$\tau$, i.e.,
\begin{equation}
\tau(t) = \mu^\mathrm{A}\gamma^\mathrm{A}(t) = \mu^\mathrm{B}\gamma^\mathrm{B}(t).
\label{eq:constant-shear}
\end{equation}
As illustrated in Fig.~\ref{fig:geometry-shear}, this is achieved by prescribing shear strains $\gamma$ inversely proportional to the effective shear moduli $\mu$ associated with the individual phases. In particular, the shear angles correspond to~$t\overline{\gamma}$ in Phase~A, and~$t\overline{\gamma}/\rho$ in Phase~B, where~$\rho$ is a material contrast ratio, and~$\overline{\gamma}$ is a target shear angle in Phase~A, and~$t\in[0,T]$ is a parametrization pseudo-time. The requirement of constant shear stress translates to prescribing atom positions of the boundary atoms as
\begin{equation}
\begin{aligned}
\vec{r}^\alpha &= \vec{r}_0^\alpha + \left[t\overline{\gamma}\,r_{0,x}^\alpha\right]\vec{e}_y, && \text{for} \quad \alpha \in N^\mathrm{A} \cap N^{\mathrm{BC}},\\
\vec{r}^\beta &= \vec{r}_0^\beta + \frac{t\overline{\gamma}}{\rho}\left[r_{0,x}^\beta+(\rho-1)L_\mathrm{A}\,\sgn{r_{0,x}^\beta}\right]\vec{e}_y, && \text{for} \quad \beta \in N^\mathrm{B} \cap N^{\mathrm{BC}},\\
\end{aligned}
\quad t \in [0,1],
\label{eq:bcs}
\end{equation}
where~$N^{\mathrm{BC}} \subset N$ are the atoms to which boundary conditions are applied, i.e.,~atoms positioned within the~$r_{\mathrm{cut}}$ distance from the specimen's boundary~$\partial\Omega$, indicated by the hollow circles in Fig.~\ref{fig:geometry-atm}.

The mechanical configuration of the system at each time step~$t_k$ for a time discretization of the time horizon
\begin{equation}
t \in [0,T], \quad 0 = t_0 < t_1 < \dots < t_{n_T} = T,	
\label{eq:timediscretization}
\end{equation}
is governed by the minimization of the total potential energy of the system, i.e.,
\begin{equation}
\column{r}_k \in \underset{{\scriptsize\column{v}}}{\text{arg min}}\ \mathcal{V}(\column{v}).
\label{eq:ms_min}
\end{equation}
Here, $\column{r}_k = \column{r}(t_k)$ is a relaxed configuration at a time instant~$t_k$, $\column{r}$ is a column matrix storing the components of the position vectors in the deformed configuration~$\vec{r}^{\alpha}$ corresponding to all atoms~$\alpha = 1, \dots, n_{\mathrm{ato}}$, $\column{v}$ denotes an admissible configuration of the system reflecting the boundary conditions of Eq.~\eqref{eq:bcs} applied at a time instant~$t_k$, and
\begin{equation}
\mathcal{V}(\column{r}) = \frac{1}{2}
\sum_{\substack{\alpha,\beta = 1\\\alpha\neq\beta,\,r^{\alpha\beta}<r_\mathrm{cut}}}^{n_\mathrm{ato}}
\phi^{\alpha\beta}(r^{\alpha\beta})
\label{eq:energy}
\end{equation}
is the internal free energy\footnote{For pair potentials, the sum over all atoms in Eq.~\eqref{eq:energy} may be replaced by a sum over all interatomic interactions, reducing thus the associated computing cost by a factor of two. An additional speed-up can be achieved by, e.g., parallelization.}. No externally applied forces are considered, as only the prescribed displacements are used to load the atomic system, recall Eq.~\eqref{eq:bcs}. The minimization problem of Eq.~\eqref{eq:ms_min} is solved using a Trust-region algorithm, see, e.g.,~\cite{conn2000trust}, which has been implemented within an in-house code.

To avoid potential occurrence of multiple glide planes, the numerical solver is initiated at the beginning of each time step~$t_k$ towards the preferred glide plane~$\Gamma_\mathrm{gp} = \{\vec{x} \in \mathbb{R}^2 : y = 0 \}$ by means of the analytical Volterra solution; see Fig.~\ref{fig:geometry-atm}, where the preferred glide plane is depicted by the dotted line. A more detailed description of the same initialization procedure can be found in~\cite[Section~2.2]{bormann2020peierls}.
%
%
\subsection{Quasicontinuum model}
\label{ssec:Micro-QC}
\begin{figure}
	\centering
	\includegraphics[trim=2cm 1cm 1.4cm 0.5cm, clip=true, width=0.7\linewidth]{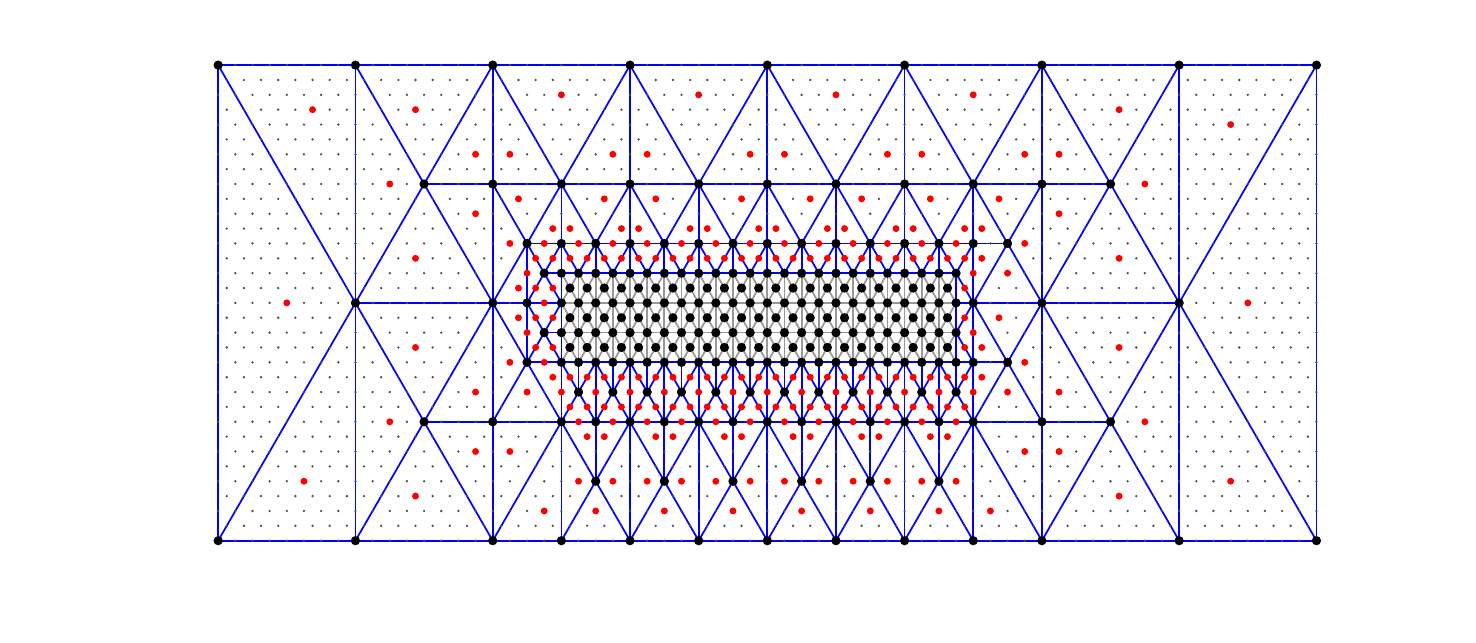}
	\caption{A typical triangulation of a small atomic domain with a predefined fully-resolved region. Repatoms are shown as black dots, sampling atoms as red dots, and all remaining atoms as gray dots. In the fully-resolved region the triangulation matches the underlying atomic structure (gray triangles), whereas elsewhere interpolation is used (blue triangles).}
	\label{fig:atm-geo-qc}
\end{figure}

The Quasicontinuum~(QC) methodology is a concurrent multiscale technique, originally introduced in~\cite{tadmor1996quasicontinuum}. The main idea consists in combining an accurate but expensive full atomic description inside regions of high interest with a cheaper continuum-like approximation elsewhere. This is achieved through a finite element mesh overlaid on top of the underlying lattice, as shown in Fig.~\ref{fig:atm-geo-qc}, where all mesh nodes (vertices of linear triangular elements) coincide with atoms. This set of atoms, the so-called repatoms, determines positions of all the remaining atoms through interpolation
\begin{equation}
\column{r} = \mtrx{\Phi} \column{r}_\mathrm{rep}.
\label{eq:qc-interp}
\end{equation}
In Eq.~\eqref{eq:qc-interp}, $\column{r}_\mathrm{rep}$ is a column storing positions of all repatoms, whereas~$\mtrx{\Phi} = [\column{\varphi}_1, \dots, \column{\varphi}_{2n_\mathrm{rep}}]$ stores the interpolation basis functions~$\column{\varphi}_i$ associated with the triangular mesh, see~\cite{TadmorModel} for more details. By gradually refining the triangulation, a seamless transition between the fully-resolved discrete description in the region of high interest and an efficient continuum description elsewhere is achieved. A common practice is to choose the mesh such that the number of repatoms~$n_\mathrm{rep}$ is much smaller than the total number of atoms~$n_\mathrm{ato}$, reducing substantially the number of Degrees Of Freedom (DOFs). This in turn yields computational savings, since the minimization problem of Eq.~\eqref{eq:ms_min} is carried out with respect to~$\column{r}_\mathrm{rep}$ instead of~$\column{r}$.

In order to construct the potential energy~$\mathcal{V}$ in Eq.~\eqref{eq:energy} all lattice sites need to be visited, which is expensive and unnecessary. The second Quasicontinuum step is thus introduced to approximate the energy by
\begin{equation}
\mathcal{V}(\column{r}) \approx \sum_{\alpha \in N_\mathrm{S}}
w^{\alpha}v^{\alpha}(\column{r}),
\label{eq:qc-summ}
\end{equation}
where the energy associated with a lattice site~$\alpha$ is specified as
\begin{equation}
v^{\alpha}(\column{r}) = \frac{1}{2}
\sum_{\substack{\beta=1\\\alpha\neq\beta,\,r^{\alpha\beta}<r_{\mathrm cut}}}^{n_\mathrm{ato}}
\phi^{\alpha\beta}(r^{\alpha\beta}),
\label{eq:qc-summ1}
\end{equation}
and where~$w^\alpha$ is a summation weight associated with each sampling atom~$\alpha \in N_\mathrm{S}$. A set of all sampling atoms, $N_\mathrm{S}$ (indicated in Fig.~\ref{fig:atm-geo-qc} as red dots), is selected to accurately represent the energy of the entire system. The summation rule is tailored to a particular interpolation scheme used, for which multiple options exist, as reported, e.g., in~\cite{amelang2015summation, beex2015higher, luskin2009analysis}. In what follows, the central summation rule of Beex \emph{et al.}~\cite{beex2014central} is used. A QC method thus minimizes the approximate energy of Eq.~\eqref{eq:qc-summ} with respect to the reduced kinematic variable~$\column{r}_\mathrm{rep}$ related to the configuration of the entire atomic system through interpolation of Eq.~\eqref{eq:qc-interp}.

The area of high interest, i.e.,~the fully-resolved region and hence the associated triangulation, need to adaptively evolve in time to properly accommodate dislocation motion while retaining QC efficiency. To this end, various adaptive criteria for atomic lattices can be found in the literature, see, e.g., \cite{miller2002quasicontinuum, phlipot2019quasicontinuum, memarnahavandi2015goal, tembhekar2017automatic, kwon2009efficient, arndt2008error}.
In this work, the Zienkiewicz--Zhu (ZZ) error estimator \cite{zienkiewicz1987simple, zienkiewicz1992superconvergent,Shenoy:1999} is used as a refinement criterion, which is applied to the deformation gradient tensor computed within each element, see also~\cite[][Section~12.6.5]{TadmorModel}. That is, the local error inside each element is estimated and compared against a pre-selected threshold value~$\mathrm{ZZ}_\mathrm{tr}$, to determine elements which need refinement. The fully-resolved region is updated sequentially as follows:
\begin{enumerate}[(i)]\setlength{\itemsep}{0pt}\setlength{\parskip}{0pt}\setlength{\parsep}{0pt}  
\item At each time increment, the QC system is equilibrated for a fixed fully-resolved region;
\item The triangulation is checked by the mesh refinement criterion (using the ZZ error indicator) and refined if necessary;
\item All required atoms are added as repatoms, the set of sampling atoms is amended, the interpolation matrix is updated, and the equilibrium is restored again.
\end{enumerate}
This procedure is repeated until the mesh refinement criterion is satisfied for all mesh elements, proceeding subsequently to a new load increment in~(i). Since in the here considered problem the same dislocation path is followed by trailing dislocations, no coarsening is considered for the atomic problem.
%
%
\subsection{Peierls--Nabarro formulation}
\label{ssec:Micro-PN}
\begin{figure}
	\centering
	\includegraphics[width=0.9\linewidth]{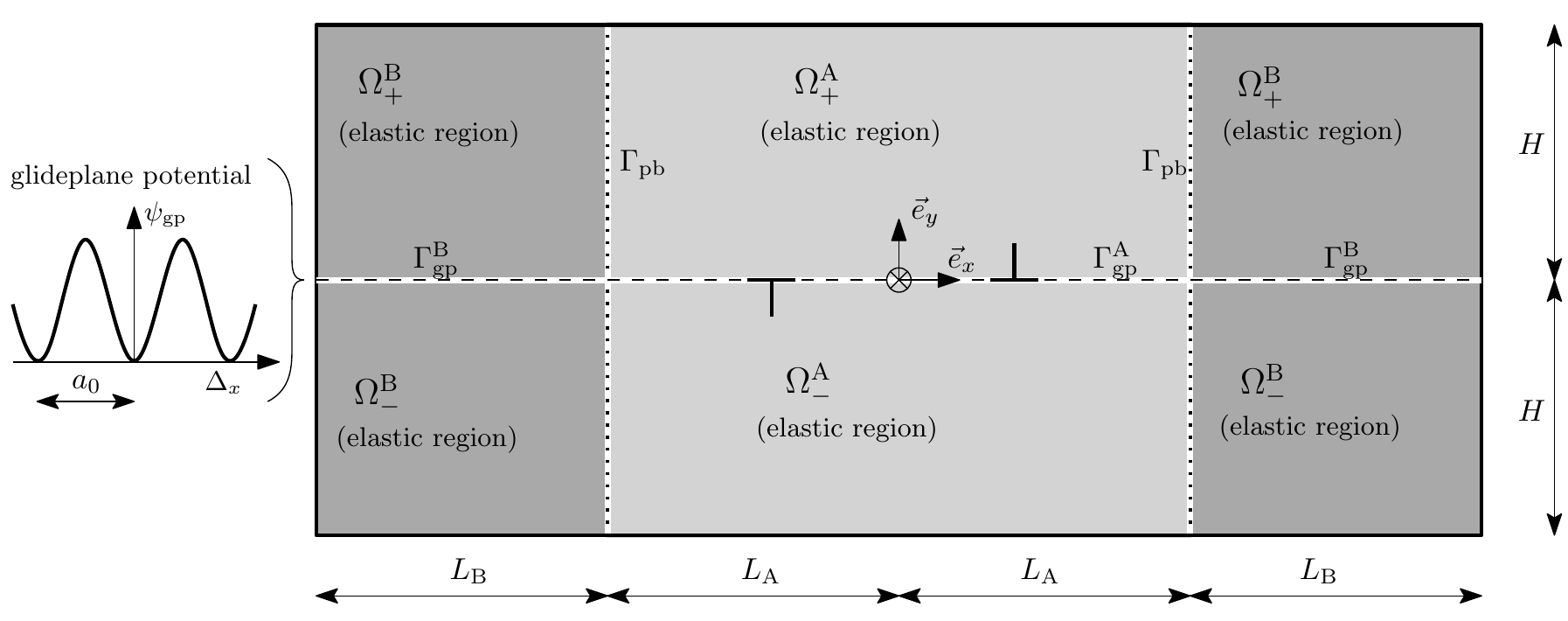}
	\caption{The \fepn{} model for edge dislocation dipoles interacting with the phase boundary in a two-phase microstructure. A continuous representation of Phase~A (bright gray solid) and Phase~B (dark gray solid), glide plane (dashed line), and phase boundaries (dotted lines). Dislocation dipoles $\bot$ are generated by the Frank--Read source denoted with $\otimes$.}
	\label{fig:atm-geo-PN}
\end{figure}

The homogenization based simplification of the full atomic problem of Section~\ref{ssec:Micro-Full}, shown in Fig.~\ref{fig:geometry-atm}, employs the FE Peierls--Nabarro model~(abbreviated \fepn{}; for more details see~\cite{bormann2019computational}). The problem domain~$\Omega$ is split into two regions, $\Omega^{\mathrm{A}}$ and~$\Omega^{\mathrm{B}}$. 
These regions are separated by perfectly and fully coherent phase boundaries of zero thickness~$\Gamma_{\mathrm{pb}}$, which are normal to~$\vec{e}_x$. Displacement and traction continuity conditions are enforced on $\Gamma_{\mathrm{pb}}$ as
\begin{equation}
\begin{aligned}
\vec{u}^{\mathrm{A}} &= \vec{u}^{\mathrm{B}}, && \mathrm{on}\quad\Gamma_{\mathrm{pb}}, \\
\boldsymbol{\sigma}^{\mathrm{A}}\cdot\vec{e}_x &= \boldsymbol{\sigma}^{\mathrm{B}}\cdot\vec{e}_x, && \mathrm{on}\quad\Gamma_{\mathrm{pb}},
\end{aligned}
\end{equation}
where~$\tensor{\sigma}$ is the Cauchy stress tensor, and~$\vec{u}$ is the displacement field. Both phases~$\Omega^i$, $i\in\left\{\mathrm{A,B}\right\}$, are modeled as an elastic continuum enriched with a Peierls--Nabarro glide plane~$\Gamma_\mathrm{gp}$. Any dislocation motion, and in fact all non-linearity in the system, is confined to this plane, inducing additional energy through a corresponding misfit potential. The glide plane is considered perpendicular to the phase boundary~$\Gamma_\mathrm{pb}$, i.e., oriented along~$\vec{e}_x$. $\Gamma_{\mathrm{gp}}$ is assumed as continuous throughout the entire domain~$\Omega$, horizontally splitting each phase~$\Omega^i$ into two subdomains~$\Omega^i_\pm$, $i\in\left\{\mathrm{A,B}\right\}$, as shown in Fig.~\ref{fig:atm-geo-PN}. A shear deformation is prescribed to the considered domain as a function of time~$ t \in [0, T] $ on the external boundary~$ \partial\Omega $, inducing initially constant shear stress throughout the specimen according to Eq.~\eqref{eq:constant-shear}.

In analogy to the full atomic system and the QC formulation, the evolution of the \fepn{} model is governed by minimization of the total potential energy (specified per unit thickness because plane strain conditions are assumed), i.e.,
\begin{equation}
\vec{u}_k \in \underset{{\scriptsize\vec{v}}}{\text{arg min}}\ \Psi(\vec{v}).
\label{eq:minpn}
\end{equation}
In Eq.~\eqref{eq:minpn}, $\vec{u}_k = \vec{u}(t_k)$ is the relaxed displacement field at a time instant~$t_k$ and $\vec{v}$ is a kinematically admissible displacement field respecting the kinematic constraints prescribed on~$\partial\Omega$. The total potential energy is expressed as
\begin{equation}
\Psi(\vec{u})=\int_{\Omega\setminus\Gamma_{\mathrm{gp}}}\psi_{\mathrm{e}}(\vec{u})\,\mathrm{d}\Omega+\int_{ \Gamma_{\mathrm{gp}} }\psi_{\mathrm{gp}}(\vec{u})\,\mathrm{d}\Gamma,
\label{eq:internal-energy_atm}
\end{equation}
where~$\psi_{\mathrm{e}}$ is the elastic strain energy density (considered inside~$\Omega_\pm^i$), $\psi_{\mathrm{gp}}$ is the glide plane potential (localized along the glide plane~$\Gamma_{\mathrm{gp}}^i$); any external forces have been neglected.

Within all subdomains~$\Omega_\pm^i$, small strain linear elasticity is assumed, i.e.,
\begin{equation}
\psi_{\mathrm{e}}=\frac{1}{2}\,\tensor{\varepsilon}:\tensorfour{D}:\tensor{\varepsilon},
\label{eq:linearElastic}
\end{equation}	
where
\begin{equation}
\tensor{\varepsilon}=\frac{1}{2}\big(\vec{\nabla}\vec{u} + (\vec{\nabla}\vec{u})^\mathsf{T}\big)
\end{equation}
is the small strain tensor, and~$\vec{\nabla}$ denotes the gradient operator in the reference configuration. The energy density of Eq.~\eqref{eq:linearElastic} yields the linear elastic constitutive law
\begin{equation}
\tensor{\sigma}=\tensorfour{D}:\tensor{\varepsilon},
\label{eq:stress}
\end{equation}
in which~$\tensorfour{D}$ is a phase-specific isotropic fourth-order elasticity tensor.
%
%
\subsection{Calibration}
The elasticity tensor $\tensorfour{D}$ used in the FE-PN model in Eqs.~\eqref{eq:linearElastic} and~\eqref{eq:stress}, which is fully isotropic due to the considered hexagonal lattice, is obtained through homogenization of the atomic lattice as follows (see also~\cite{TadmorModel}). First, a numerically homogenized stiffness tensor~$\tensorfour{D}^\mathrm{atm}$, with components
\begin{equation}
D_{ijkl}^\mathrm{atm} = \frac{1}{2V}
\sum^{n_\mathrm{ato}}_{\substack{\alpha,\beta = 1\\\alpha\neq\beta,\,r^{\alpha\beta}<r_\mathrm{cut}}}
\left[\phi''(r^{\alpha\beta})-\frac{\phi'(r^{\alpha\beta})}{r^{\alpha\beta}}\right]
\frac{r^{\alpha\beta}_{i} r^{\alpha\beta}_{j} r^{\alpha\beta}_{k} r^{\alpha\beta}_{l}}{(r^{\alpha\beta})^2},
\ \text{where} \
\phi''(r) = \frac{\mathrm{d}^2\phi^{\alpha\beta}(r)}{\mathrm{d}r^2},
\label{eq:stiffness}
\end{equation}
is computed on the basis of the periodic unit cell of the atomic lattice. 
In Eq.~\eqref{eq:stiffness}, $V$ denotes the volume of the simulation cell in the deformed configuration, whereas~$r_i^{\alpha\beta}$ ($i = 1,2$) are the components of the relative position vectors~$\vec{r}^{\alpha\beta}$, cf. Eq.~\eqref{eq:virialstress} and the discussion thereof. For nearest and next-to-nearest interactions the stiffness tensor~$\tensorfour{D}^\mathrm{atm}$ has been derived analytically for a unit hexagonal lattice by assembling the contributions of all relevant atoms (within the dashed circle in Fig.~\ref{fig:LJ-a}). The computed parameters (tensor components) for Phase~A are listed in Tab.~\ref{tab:homogenized} (the corresponding atomic properties are specified in Tab.~\ref{tab:mat}); the parameters of Phase~B are obtained simply by scaling those of Phase~A with the material contrast ratio~$\rho \in \{ 1.4, 4.0 \}$. Next, an elastic isotropic plane strain stiffness tensor~$\tensorfour{D}(E_\mathrm{iso},\nu_\mathrm{iso})$ is considered with the Young's modulus~$E_\mathrm{iso}$ and Poisson's ratio~$\nu_\mathrm{iso}$, in which effective parameters
\begin{eqnarray}
E_{\mathrm{iso}} = \frac{5}{6}D_{1111}^\mathrm{atm} \quad \text{and} \quad \nu_\mathrm{iso} = \frac{1}{4}
\end{eqnarray}
are obtained to match~$\tensorfour{D}^\mathrm{atm} = \tensorfour{D}(E_\mathrm{iso},\nu_\mathrm{iso})$. 
\begin{table}
	\caption{Homogenized constitutive parameters corresponding to Phase~A. The constitutive parameters for Phase~B are obtained by scaling with the material contrast ratio~$\rho$.}
	\renewcommand*{\arraystretch}{1.3}		
	\centering
	\begin{tabular}{c|cc}
		parameter & $D_{1111}^\mathrm{atm} = D_{2222}^\mathrm{atm}$ & $D_{1122}^\mathrm{atm} = D_{1212}^\mathrm{atm}$ \\\hline
		Phase~A & $102.520 \, \varepsilon/r_m^2$ & $~~34.173 \, \varepsilon/r_m^2$ ($=D_{1111}^\mathrm{atm}/3$) \\
	\end{tabular}
	\label{tab:homogenized}
\end{table}

\begin{figure}
	\centering
	\subfloat[GSFE, sketch of considered simulation box]{\includegraphics[trim=0cm 0cm 0cm 0cm, clip=true, width=0.35\linewidth]{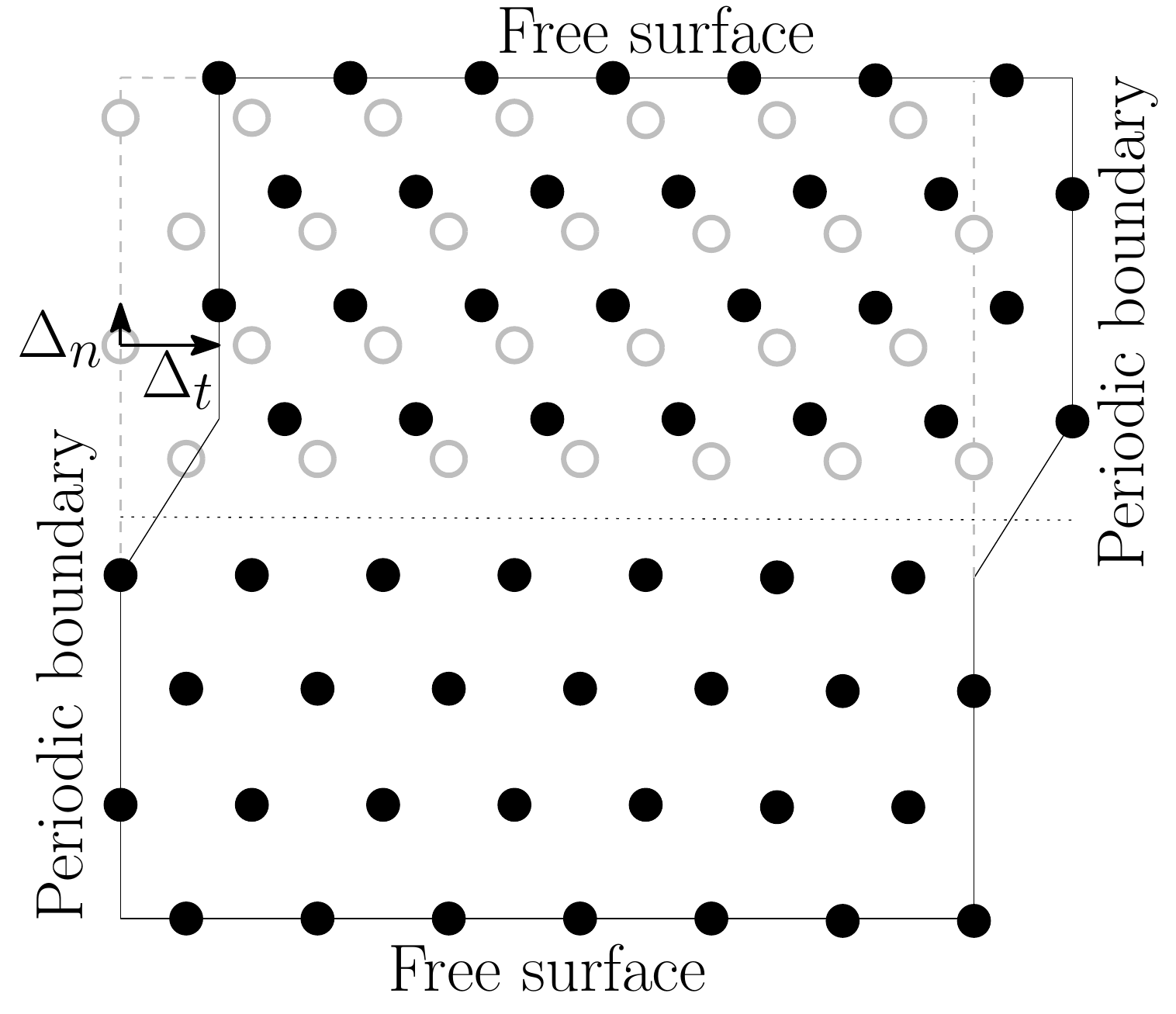}\label{fig:atm-gp-en-a}}
	\hspace{1.75em}
	\subfloat[normalized GSFE for Phase~A]{\includegraphics[trim=0cm 0cm 0cm 0cm, clip=true, width=0.45\linewidth]{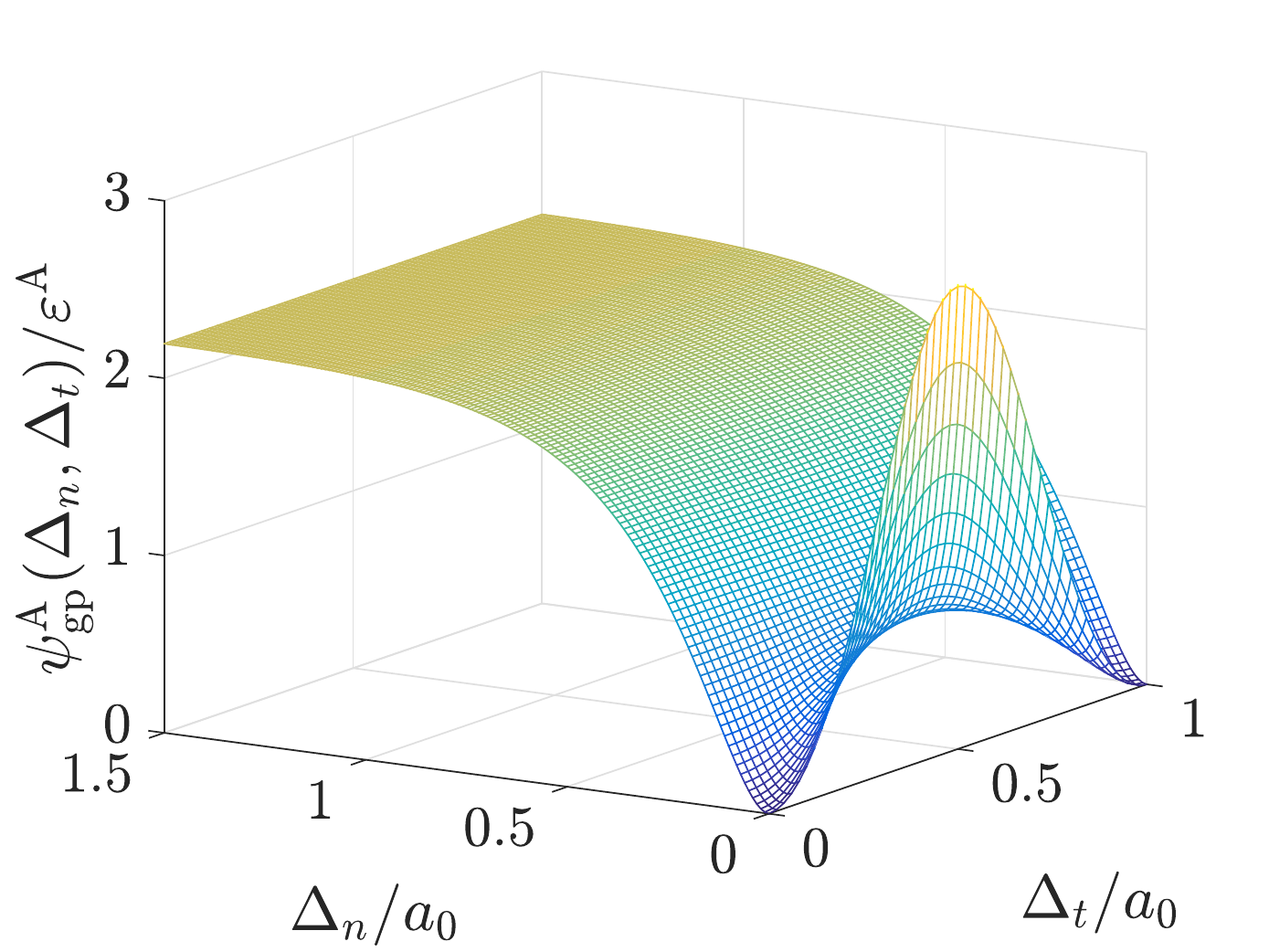}\label{fig:atm-gp-en-b}}
	\caption{(a) Geometry of a rectangular simulation box with applied displacements $\Delta_n$ and $\Delta_t$ for the computation of the glide plane potential. Original positions of atoms are shown as gray circles, whereas atoms in the deformed configuration are shown as black dots. (b) Glide plane potential corresponding to Phase~A, denoted~$\psi_{\mathrm{gp}}^{\mathrm{A}}(\Delta_n,\Delta_t)$, normalized with respect to~$\varepsilon^\mathrm{A}$.}
	\label{fig:atm-gp-en}
\end{figure}

The glide plane potential~$\psi_{\mathrm{gp}}$ is a function of the disregistry profile~$\vec{\Delta}$, i.e.,~of the displacement jump across the glide plane~$\Gamma_{\mathrm{gp}}$,
\begin{equation}
\vec{\Delta} = \llbracket \vec{u} \rrbracket=\vec{u}_+-\vec{u}_-  \quad\mathrm{on}\quad\Gamma_{\mathrm{pb}},
\label{eq:Disregistry_atm}
\end{equation} 
which splits into a tangential~$\Delta_t$ and a normal~$\Delta_n$ part, i.e.,~$\vec{\Delta} = \Delta_t \vec{e}_t + \Delta_n \vec{e}_n$. To capture the effect of periodicity of the underlying atomic lattice, $\psi_{\mathrm{gp}}$ is a non-convex periodic function of~$\Delta_t$ with a period~$a_0$. Different expressions have been introduced for $\psi_{\mathrm{gp}}$ in the literature, as discussed and compared, e.g.,~in~\cite{bormann2020peierls}. Here it is based on the Generalized Stacking Fault Energy~(GSFE) of the underlying atomic lattice, see, e.g.,~\cite{Vitek:1968} where the GSFE was obtained from atomic calculations, or~\cite{Duesbery:1989} where density functional theory was employed. In this paper, the GSFE is constructed through atomic calculations using a rectangular simulation box of size~$20 a_0 \times 12 a_0 \sqrt{3}$ with periodicity conditions between the vertical boundaries and free horizontal surfaces, as shown in Fig.~\ref{fig:atm-gp-en-a}. A lattice with the stress-free spacing~$a_0$ is considered, and the upper part is rigidly displaced as~$\vec{\Delta} = \Delta_t\vec{e}_t + \Delta_n\vec{e}_n$. The GSFE computed for Phase~A is shown in Fig.~\ref{fig:atm-gp-en-b}, whereas the GSFE for Phase~B is obtained again by multiplying with the corresponding material contrast ratio~$\rho$. Within the specimen domain~$\Omega$, the glide plane potential is considered as a piecewise constant function. This is possible because the lattice considered in both phases is perfectly aligned, having the same lattice spacing~$a_0$. If the lattices within the different phases were not perfectly aligned with the same lattice spacing, or if they were oriented differently with respect to each other, additional simulations of the phase interface might be required.

Standard finite elements are used to discretize the elastic regions $\Omega_{\pm}^{i}$. The glide plane $\Gamma_{\mathrm{gp}}$ is discretized by interface elements which are inserted between the bulk elasticity elements above and below it. Mechanical equilibrium is established by minimizing the total potential energy~$\Psi$ of Eq.~\eqref{eq:internal-energy_atm} with respect to the DOFs of the considered FE triangulation. The non-convexity of~$\psi_{\mathrm{gp}}$ is addressed by a truncated Newton optimization algorithm elaborated in~\cite{bormann2018application}. Individual dislocations are initialized in analogy to the approach used for the fully-resolved atomic system, as described at the end of Section~\ref{ssec:Micro-Full}; for more details see~\cite[][Section~3.2]{bormann2019computational}.
%
%
\subsection{Results and comparison}
\label{ssec:Micro-Results}
\begin{figure}
	\centering
	\subfloat[\fepn{}]{\includegraphics[trim=1.8cm 1.3cm 1.8cm 1.3cm, clip=true, width=0.45\linewidth]{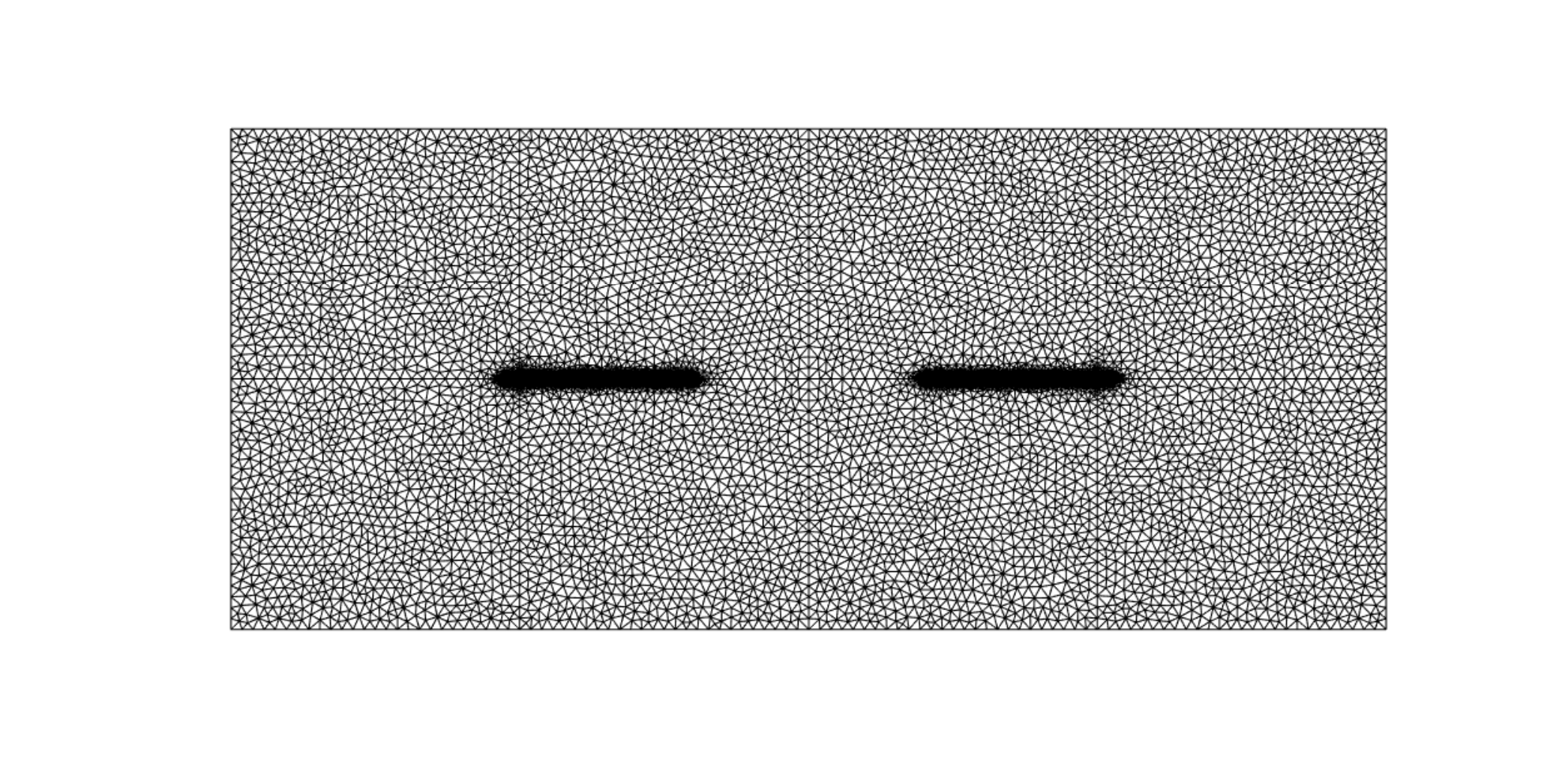}\label{fig:atm-mesh-PN}}
	\hspace{0.03\linewidth}
	\subfloat[\qcfix{}]{\includegraphics[trim=0cm -0.9cm 0cm 0cm, clip=true, width=0.4\linewidth]{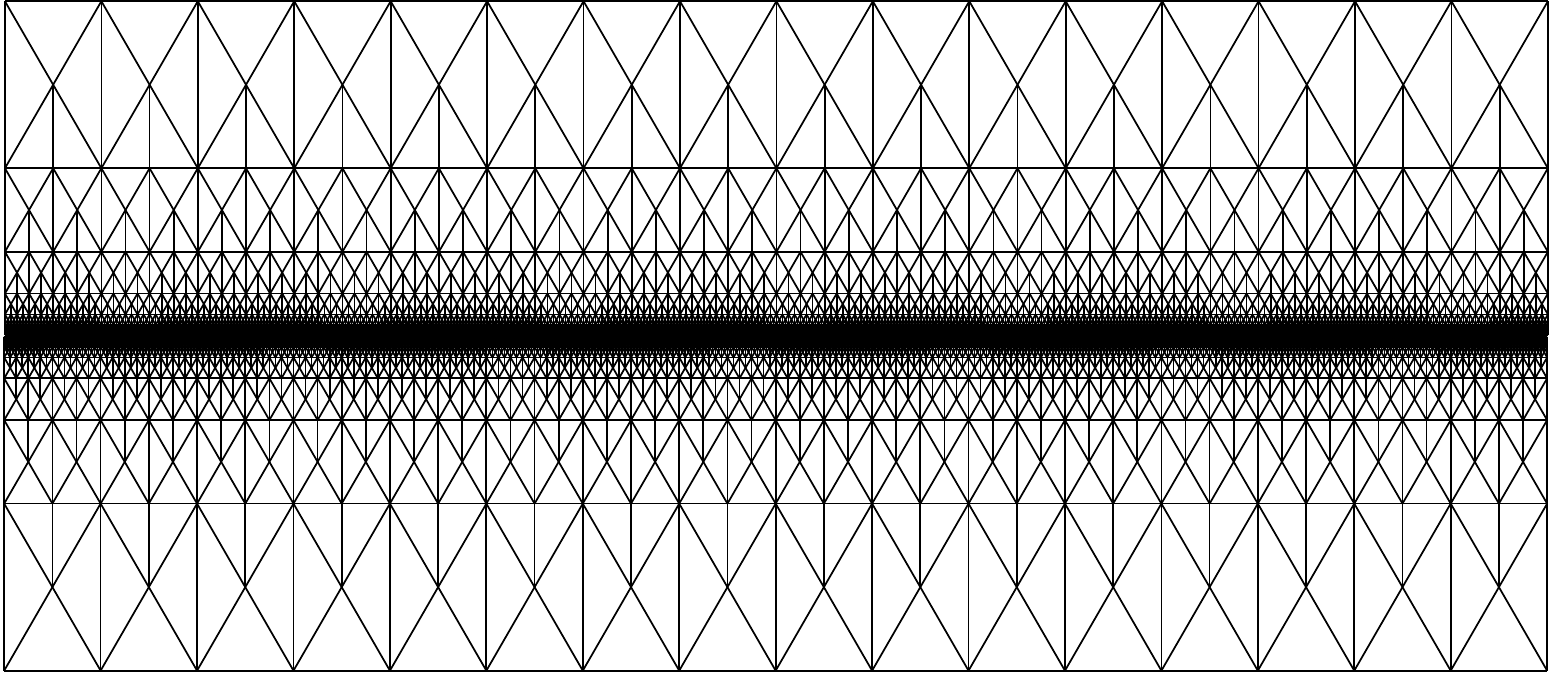}\label{fig:atm-mesh-QCf}}
	\\
	\hspace{0.02\linewidth}
	\subfloat[\qcadaptive{} initial]{\includegraphics[trim=0cm 0cm 0cm 0cm, clip=true, width=0.4\linewidth]{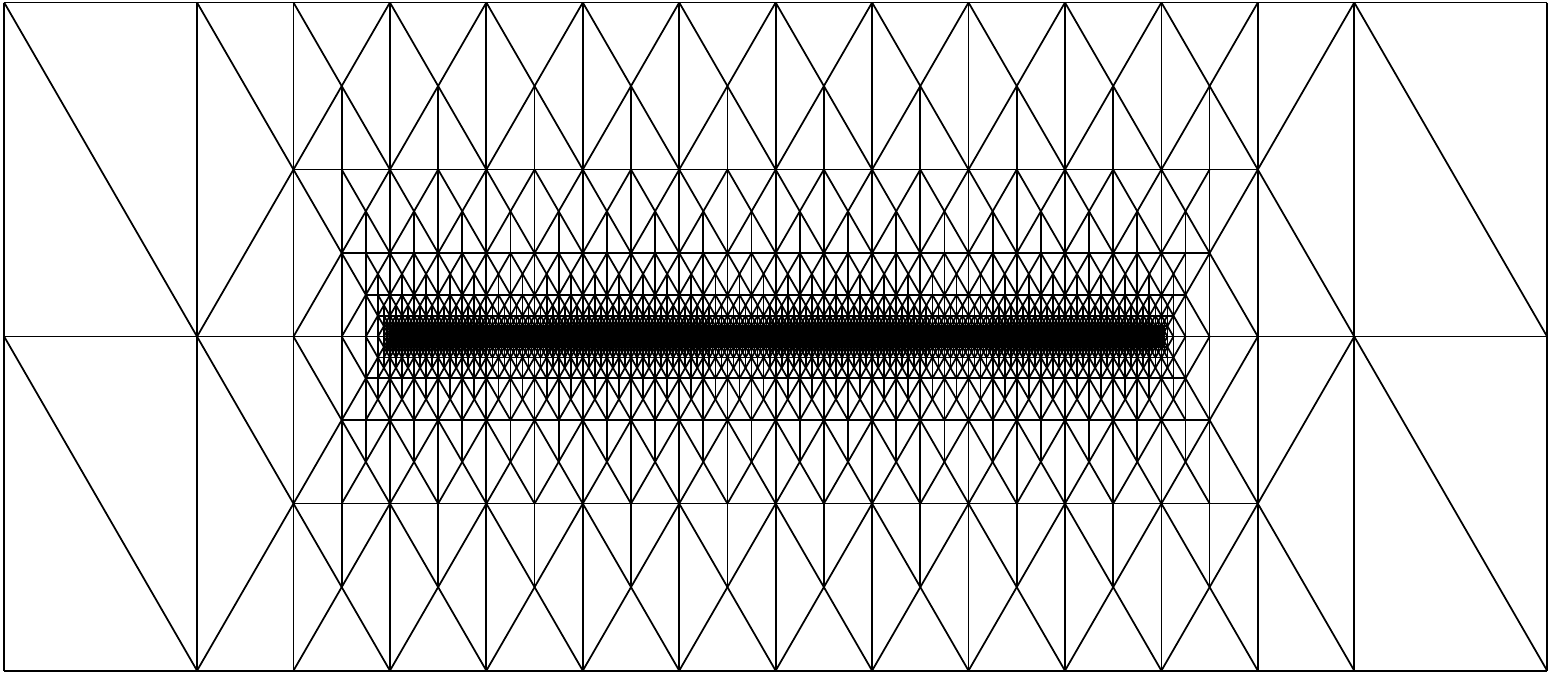}\label{fig:atm-mesh-QCzz}}
	\hspace{0.05\linewidth}
	\subfloat[\qcadaptive{} final]{\includegraphics[trim=0cm 0cm 0cm 0cm, clip=true, width=0.4\linewidth]{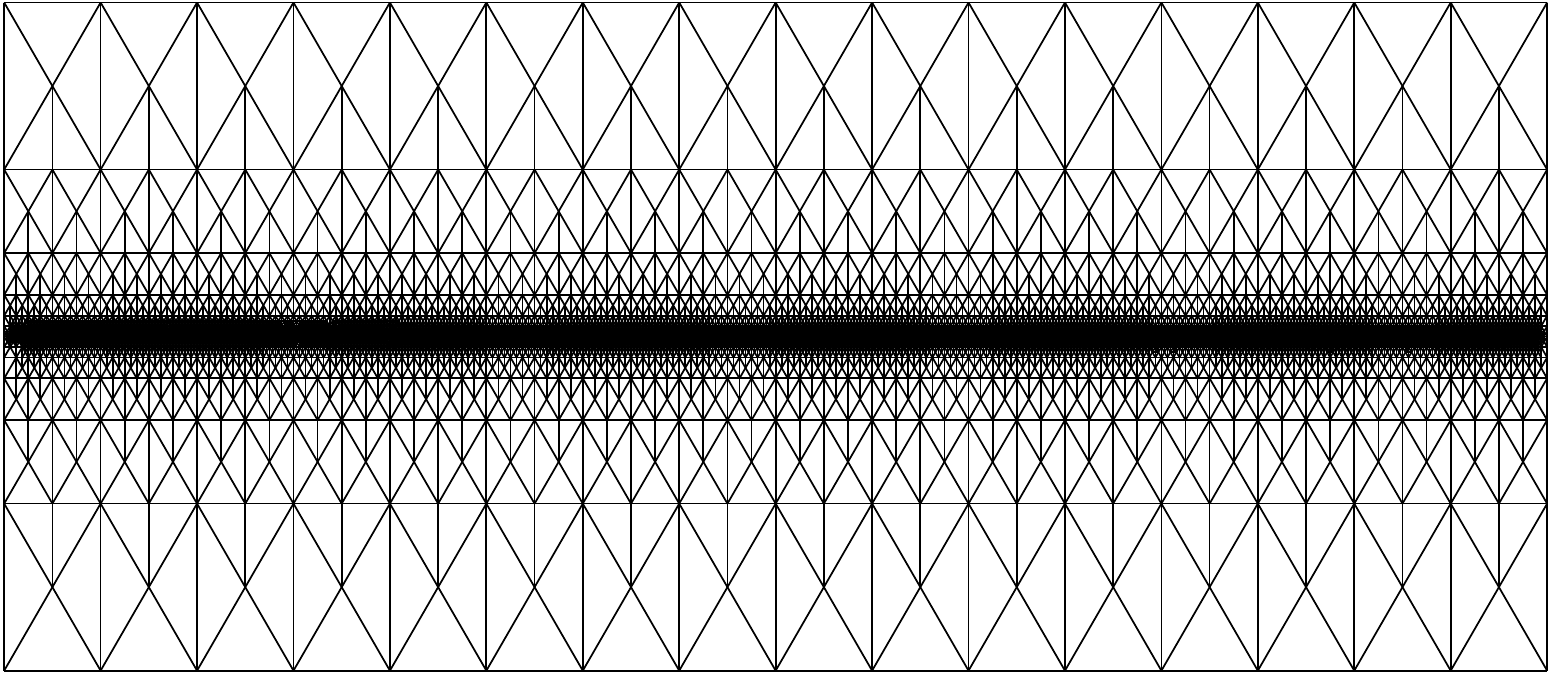}\label{fig:atm-mesh-QCzz2}}
	\caption{Employed initial triangulations for dislocation transmission across phase boundaries, corresponding to~(a) \fecz{} model, (b)~\qcfix{} model, (c)~\qcadaptive{} model. (d)~Triangulation for the \qcadaptive{} model at the end of the simulation.}
	\label{fig:atm-mesh}
\end{figure}

The behavior of the fully-resolved atomic system of Section~\ref{ssec:Micro-Full} (referred to as \full{}) is described first, considering the material contrast ratio~$ \rho = 1.4$. Two QC systems of Section~\ref{ssec:Micro-QC} are considered: (i)~QC with a fixed mesh that is fully refined along the entire glide plane (referred to as \qcfix{}), and~(ii) QC with a mesh that has only a small fully-resolved region situated around the Frank--Read source equipped with adaptivity and a Zienkiewicz--Zhu error indicator of critical threshold ZZ\textsubscript{tr}$ = 0.001$, referred to as \qcadaptive{}. The \fepn{} model of Section~\ref{ssec:Micro-PN} is considered for only one discretization with a fixed mesh refined along the glide plane down to an element size~$h=a_0/16$. Although such an excessively small element size is not strictly necessary, see~\cite{bormann2019computational, bormann2020peierls} where $4$--$8$ times coarser mesh was employed to obtain adequate macroscopic results such as transmission stress, it is used here to test the best accuracy of the \fepn{} method, in particular to accurately capture the shape of the dislocation core. For coarser meshes a substantially better speed-up can be expected compared to the results presented here. The initial triangulations associated with the individual systems are shown in Fig.~\ref{fig:atm-mesh}, whereas the corresponding (initial/final) number of DOFs are listed in Tab.~\ref{tab:atm-res-num}. All reduced methods are compared against the full atomic system in terms of the total potential energy evolution, dislocation positions, critical transmission stresses, disregistry profiles, and computational effort.

\begin{figure}
	\flushleft
	\subfloat[dislocation pile-up, $\rho = 1.4$]{\includegraphics[width=0.42\linewidth]{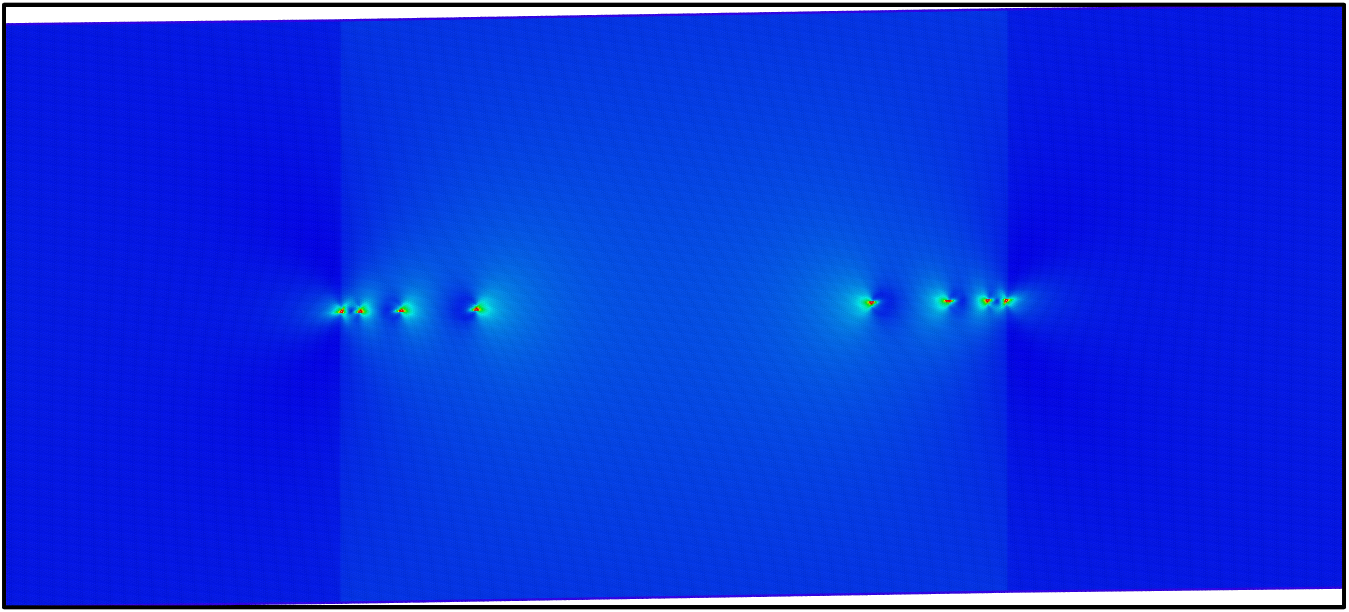}\label{fig:atm-res-position-14-1}}
	\hspace{1em}
	\subfloat[dislocation pile-up, $\rho = 4.0$]{\includegraphics[width=0.42\linewidth]{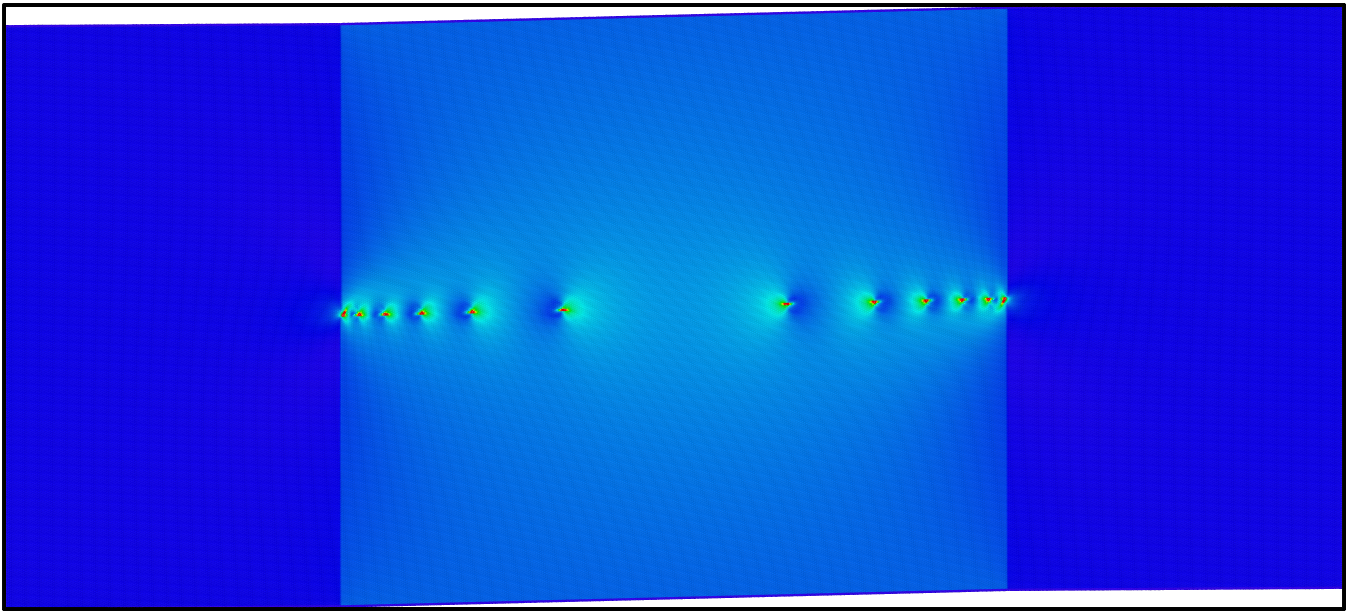}\label{fig:atm-res-position-4-1}}\\
	\vspace{-6em}
	\subfloat[dislocation transmission, $\rho = 1.4$]{\includegraphics[width=0.42\linewidth]{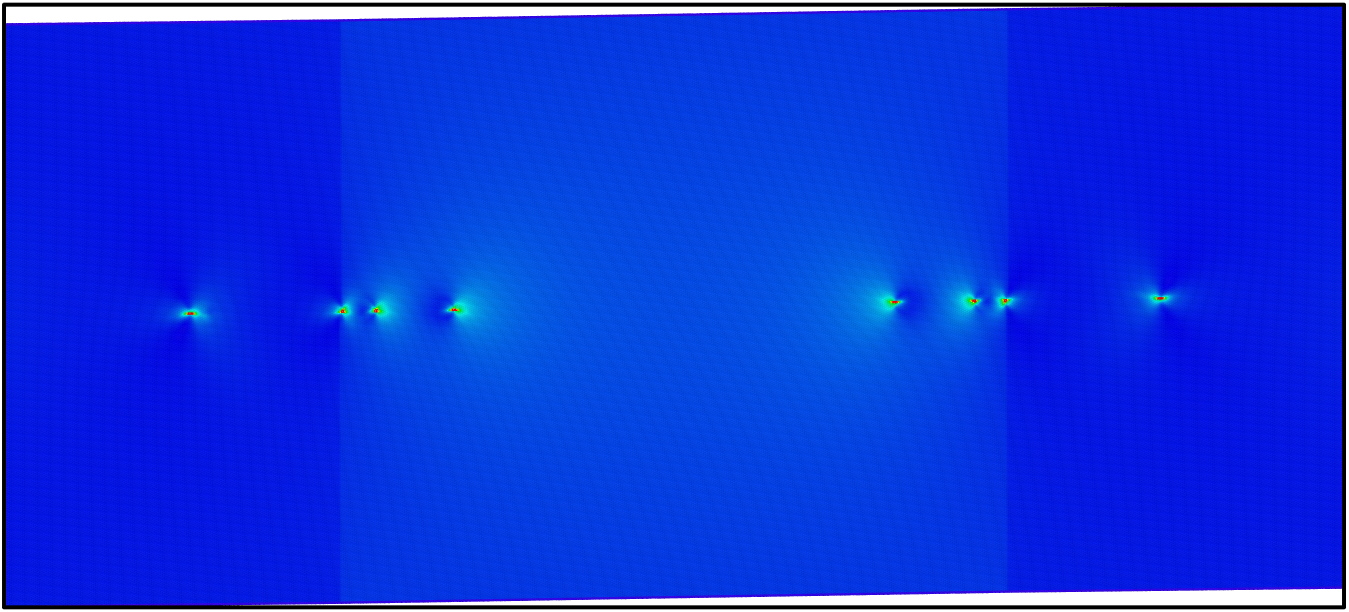}\label{fig:atm-res-position-14-2}}
	\hspace{1em}
	\subfloat[dislocation reflection, $\rho = 4.0$]{\includegraphics[width=0.42\linewidth]{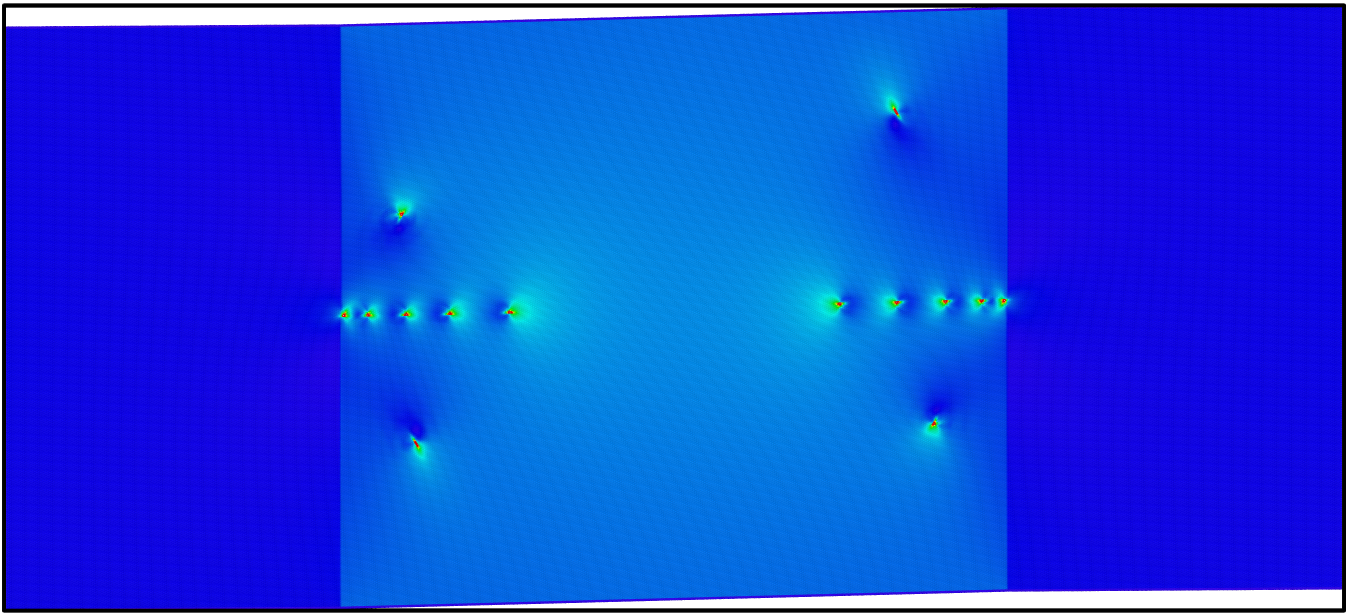}\label{fig:atm-res-position-40-2}}
	\put(20,30){\stackunder{\includegraphics[width=0.065\linewidth]{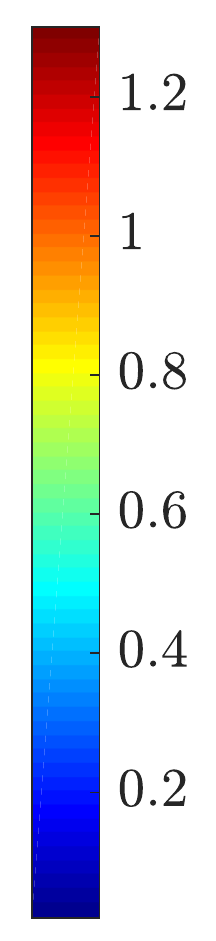}}{\hspace{-0.5em}\scriptsize $LLD/a_0$}}
	\caption{Local lattice disregistry of Eq.~\eqref{eq:lld} for the full atomic model: (a)~one step before, and~(c) after transmission of the leading dislocation the for stiffness ratio $\rho =1.4$; one step before~(b), and after~(d) reflection of the leading dislocation for the stiffness ratio $\rho =4.0$.}
	\label{fig:atm-res-14-40}
\end{figure}

Upon loading, for the fully-resolved atomic system at a certain time instance a critical resolved shear stress is reached (the activation stress of a Frank--Read source), resulting from the shear deformation applied on the remote boundary. At that moment, a new dislocation dipole is nucleated, which moves symmetrically under the increasing load and local Peach--Koehler force towards the phase boundary, where it is obstructed as a result of the phase contrast. By increasing the applied shear further, new dislocations are emitted and a dislocation pile-up is established in front of the interface. Further increasing the applied shear induces dislocation transmission into the neighboring Phase~B. For visualization purposes, the quantity called Local Lattice Disregistry associated with an atom $\alpha$~($LLD^\alpha$), represents how much the current configuration of the hexagonal lattice around this atom differs from its initial configuration, and is defined as a sum of differences between the reference (undeformed) positions of its six nearest-neighbor atoms, $\bs{r}^\gamma_{\rm t}$, and their current deformed positions located closest to their initial positions, $\bs{r}^\beta$, i.e.,
\begin{equation}
LLD^{\alpha} = \sum_{\gamma \in B_\alpha}
\underset{{\scriptsize\substack{\beta;~\beta\neq\alpha }}}{\text{min}}
	||\bs{r}^\gamma_{\rm t}-\bs{r}^{\beta}||_2, \quad \forall \alpha \in N,
\label{eq:lld}
\end{equation}
where $\min||\bs{r}^\gamma_{\rm t}-\bs{r}^{\beta}||_2$ denotes the Euclidean distance of a one particular theoretical position from its closest atom $\beta$ in the actual deformed configuration, and~$B_\alpha$ is the initial set of nearest-neighbors associated with an atom~$\alpha$. The $LLD$ indicator of Eq.~\eqref{eq:lld} is shown in color for all atoms of the full atomic system in Fig.~\ref{fig:atm-res-14-40}, corresponding to the case of a dislocation pile up with subsequent dislocation transmission (i.e.,~$\rho = 1.4$, Figs.~\ref{fig:atm-res-position-14-1} and~\ref{fig:atm-res-position-14-2}).

\begin{figure}
	\centering
	\subfloat[energy evolution paths]{\includegraphics[trim=0cm 0cm 0cm 0cm, clip=true,scale=0.5]{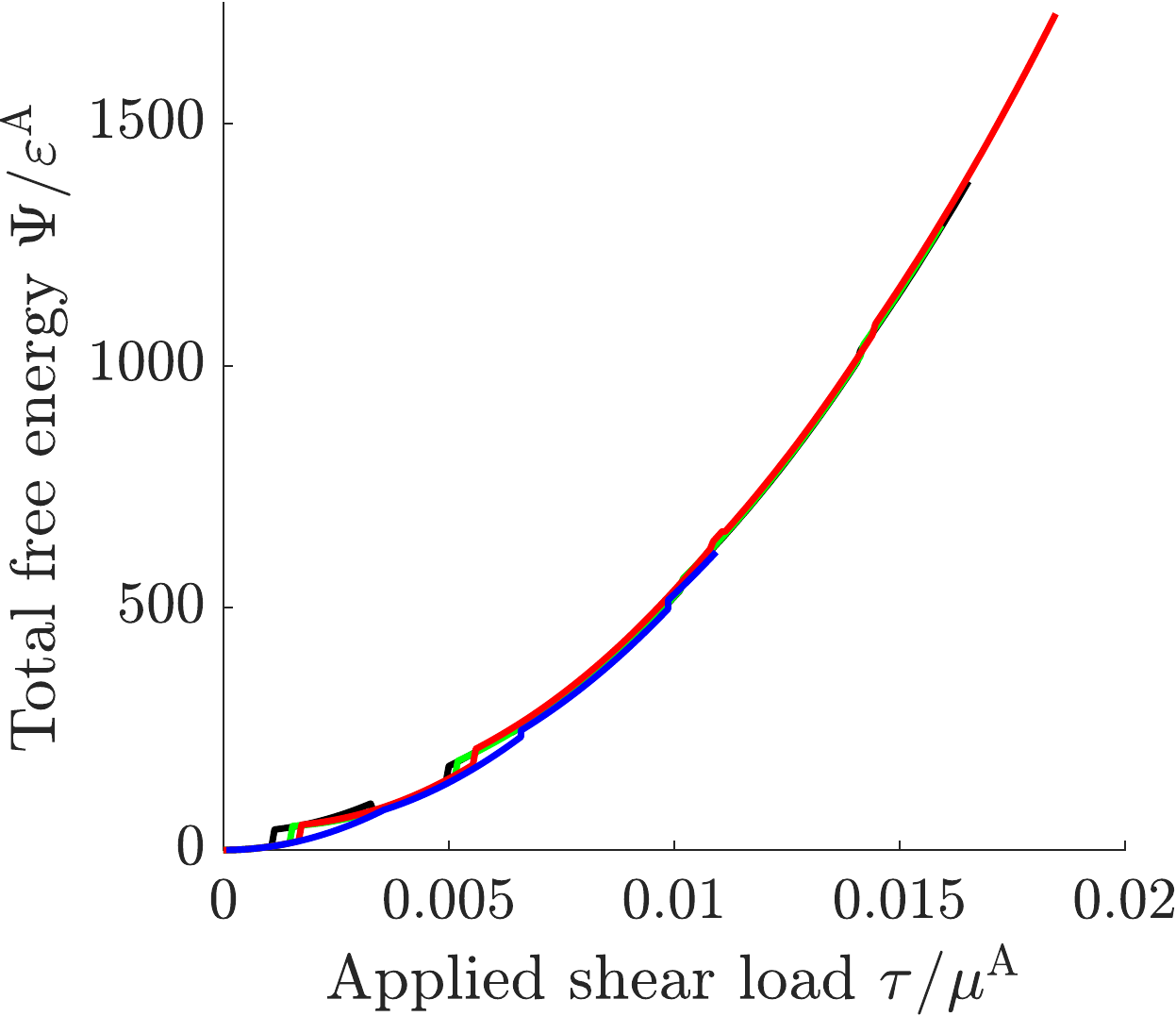}\label{fig:atm-res-energy}}
	\hspace{1em}
	\subfloat[evolution of dislocation positions]{\includegraphics[trim=0cm 0cm 0cm 0cm, clip=true,scale=0.5]{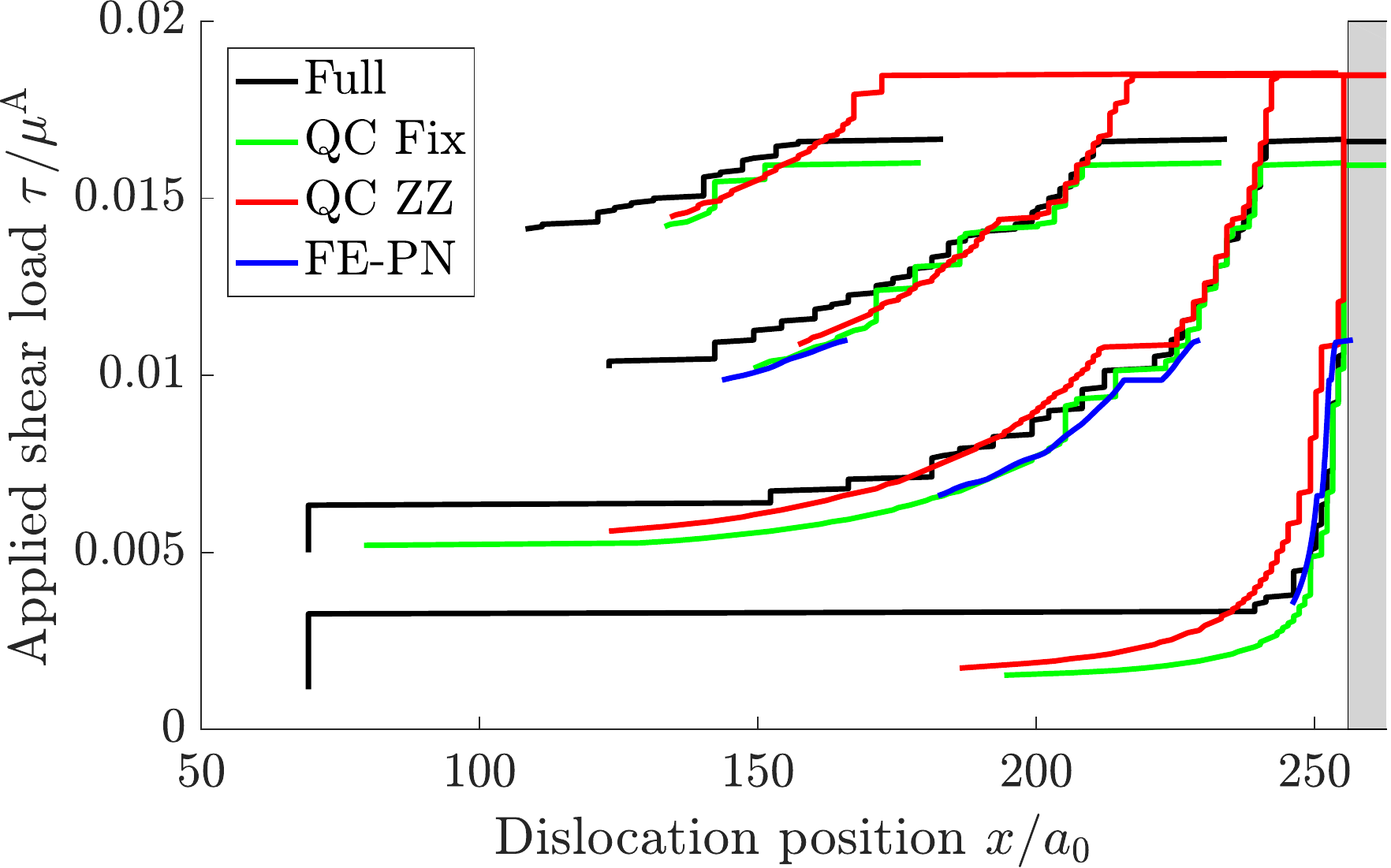}\label{fig:atm-res-position}}
	\caption{The global model response to the formation and evolution of a four-dislocation pile-up under the increasing externally applied shear load~$\tau$ for all considered models. (a)~Evolutions of the total elastic energies, and~(b) evolutions of dislocation positions~$x/a_0$. The gray area indicates the stiffer Phase~B.}
	\label{fig:atm-res-position-energy}
\end{figure}

Global energy evolution profiles for all models are shown in Fig.~\ref{fig:atm-res-energy}, where a good agreement is achieved, with the maximum energy error below~$3\%$ excluding jumps in energy evolutions. These jumps, present mostly in the early stages of the evolutions and observed mainly for the \full{} and both QC models, result from the initiation of new dislocations, which temporarily increases the energy. Upon further loading, the initially immobile but stable dislocations start to propagate towards the phase boundary and the energy drops. Similar behavior is not observed for the FE-PN model due to a vanishingly small Peierls-like barrier, as a consequence of a rather fine discretization (recall that~$h = a_0/16$ has been adopted on the glide plane).

\begin{figure}
	\centering
	\subfloat[full atomic model]{\includegraphics[trim=2.1cm 3.17cm 2.5cm 0.33cm, clip=true, height=0.25\linewidth]{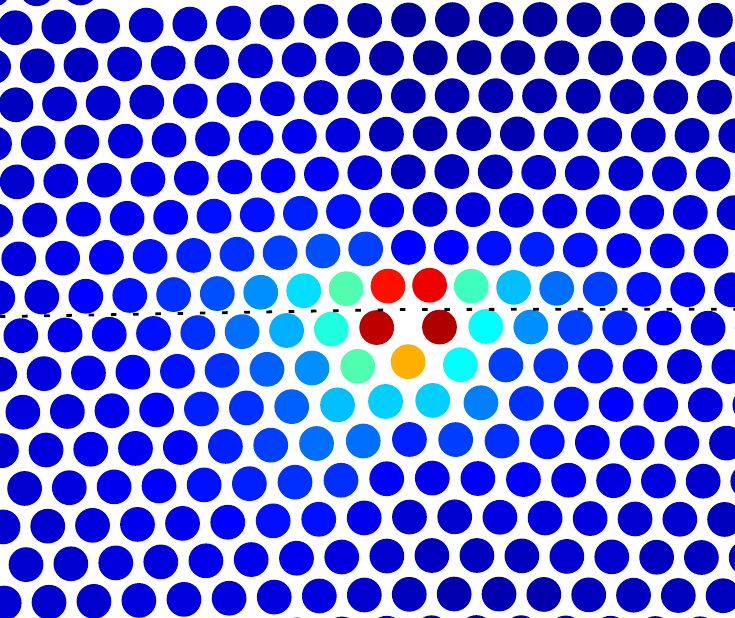}\label{fig:atm-res-core-full}}
	\hspace{1em}
	\subfloat[\fepn{} model ]{\includegraphics[trim=2.1cm 2.8cm 4cm 1.4cm, clip=true, height=0.25\linewidth]{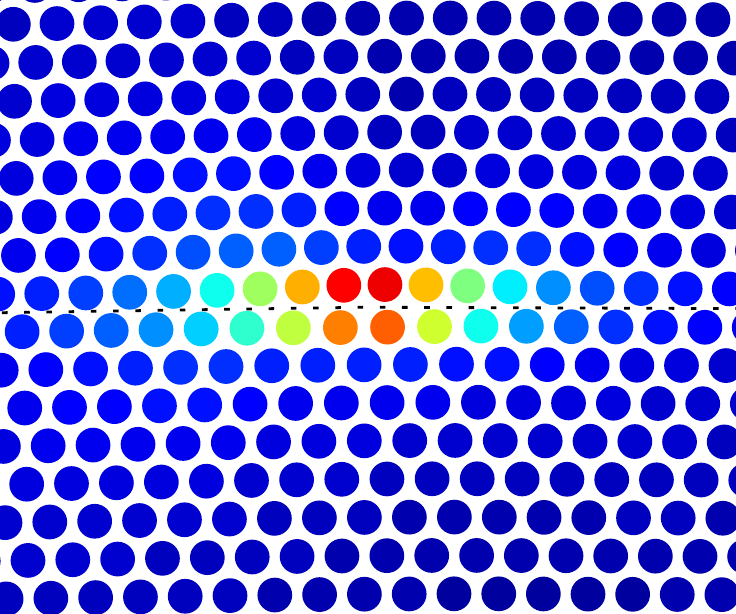}\label{fig:atm-res-core-fepn}}
	\hspace{1em}
	\stackunder{\includegraphics[trim=10.4cm 0.4cm 0cm 0cm, clip=true, height=0.256\linewidth]{atm/atm-LLD_FULLgp.pdf}}{\hspace{-1em}\scriptsize $LLD/a_0$}
	\caption{The detail of the dislocation core. (a)~Deformed configuration obtained for the \full{} atomic model (the color shows the local lattice disregistry of Eq.~\eqref{eq:lld} normalized by the lattice spacing, $LLD/a_{0}$). (b)~Equivalent quantity evaluated from the displacement field of the \fepn{} model projected on the positions of the underlying atomic lattice. The dotted lines denote the glide planes.}
	\label{fig:atm-res-core}
\end{figure}

The corresponding evolutions of dislocation positions along the glide plane~$\Gamma_{\mathrm{gp}}$ are shown in Fig.~\ref{fig:atm-res-position} against the externally applied normalized shear load~$\tau/\mu^\mathrm{A}$. Here we notice that all reduced methods cluster around the results corresponding to the full solution, suggesting good agreement. If there is the same number of dislocations present in the compared models, then the maximum error in the dislocation position does not exceed $10\,a_0$ once the dislocations start propagating towards the interface. However, more significant discrepancy in the dislocation position can be observed at early stages after initiation. All dislocations in the \full{} model and the second dislocation associated with the \qcfix{} and \qcadaptive{} method are initiated significantly earlier and closer to the specimen's center as compared to the \fepn{} method. This can be explained by a vanishingly small Peierls barrier in the \fepn{} method; in all the atomic simulations (\full{} and both QC), despite a negative Peach--Koehler force the dislocation does not annihilate upon initiation due to an existing Peierls barrier. This results in a delayed introduction of new dislocations in the \fepn{} model. Since dislocation initiation is very sensitive to stress distribution along the glide plane, differences in the \full{} and both QC methods are also non-negligible. Numerical settings, such as discretization and interpolation, might have an impact on the evolution of the initial dislocations, affecting their final trajectories. We further notice that the critical transmission stress is underestimated by more than~$30\%$ by the \fepn{} method and slightly overestimated by the \qcadaptive{} method. For clarity, the transmission stresses corresponding to the individual methods are listed in Tab.~\ref{tab:atm-res-num}. The significant underestimation of the critical transmission stress by the \fepn{} method originates from several reasons.
The first reason is an inaccurate representation of the dislocation core structure by the \fepn{} model (because of adopted linear elasticity neglecting large deformations, discreteness, and non-locality of the underlying lattice), which may play a dominant role for dislocation-interface interactions. Dislocation cores obtained for the \full{} and \fepn{} models are compared in Fig.~\ref{fig:atm-res-core}, where we clearly see that the dislocation core provided by the \fepn{} model is almost symmetric with respect to the horizontal glide plane unlike the strongly asymmetric core obtained from the \full{} model.
The second reason is a missing Peierls barrier causing more compressed pile-up, resulting in higher stresses acting on the leading dislocation in the \fepn{} model.
The last but probably the most important reason is that in the atomic models, the leading dislocation dissociates partially into the secondary inclined glide planes (associated with dislocation reflection, discussed later in this section, see also Fig.~\ref{fig:atm-res-position-40-2}).
As a result, the energy of the dislocation is split between the primary and secondary glide planes. On the other hand, because in the \fepn{} model the entire energy is constrained to the primary glide plane associated with the transmission, the dislocation is located closer to the interface and dislocation transmission thus happens earlier.
Additional simulations (not presented here for brevity) confirm that the shear stress needed for dislocation transmission increases significantly once the secondary inclined glide planes are introduced into the \fepn{} model.
\begin{table}
	\caption{Performance of individual computational models considered for atomic problem.}
	\centering
	\renewcommand*{\arraystretch}{1.3}	
	\begin{tabular}{l|rrrr}
		& \shortstack{transmission \\ stress $\tau/\mu^\mathrm{A}$} & \shortstack{dislocation position \\ error, Eq.~\eqref{eq:error_dislocation_positions}} & \shortstack{initial/final \\ DOFs} & \shortstack{computational \\ demand} \\\hline
		\full{} & $(0.01667)$	& $0$	 	& $1{,}047{,}552$		& $1$ \\	
		\qcfix{}		  & $-4.00\%$	& $7.83$	  	& $34{,}576$		& $\approx 1/14$ \\
		\qcadaptive{}  & $+11.16\%$	& $5.83$ 	  	& $15{,}732/19{,}676$	&  $\approx 1/9$ \\
		\fepn{}		  & $-34.01\%$	& $8.92$		& $1{,}181{,}576$	&  $\approx 1/5$ \\
	\end{tabular}
	\label{tab:atm-res-num}
\end{table}

\begin{figure}
	\centering
	\subfloat[$\parallel$ disregistry profile, $\tau = 0.0068\,\mu^\mathrm{A}$]{\includegraphics[trim=0cm 0cm 0cm 0cm, clip=true, scale=0.7]{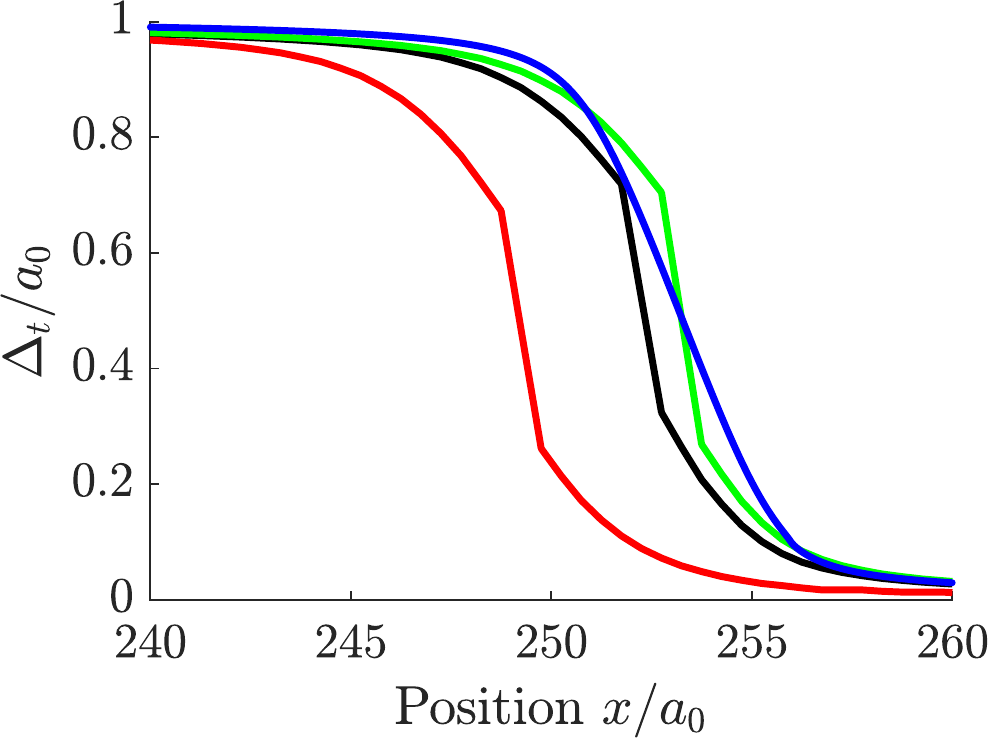}\label{fig:atm-res-disregistry3t}}
	\hspace{1em}
	\subfloat[$\parallel$ disregistry profile, $\tau = 0.01093\,\mu^\mathrm{A}$]{\includegraphics[trim=0cm 0cm 0cm 0cm, clip=true, scale=0.7]{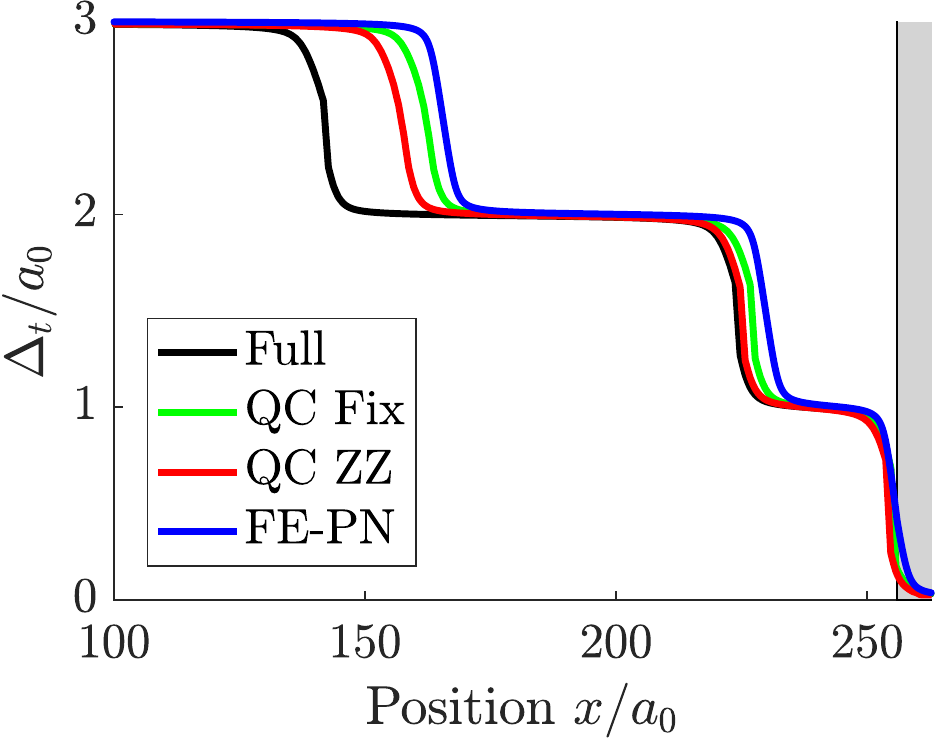}\label{fig:atm-res-disregistry1t}} \\
	\subfloat[$\bot$ disregistry profile, $\tau = 0.0068\,\mu^\mathrm{A}$]{\includegraphics[trim=0cm 0cm 0cm 0cm, clip=true, scale=0.7]{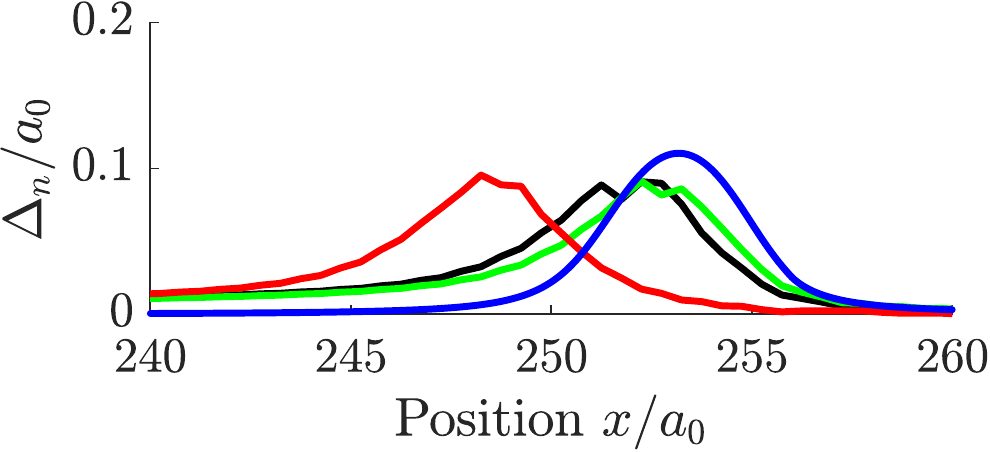}\label{fig:atm-res-disregistry3n}}	
	\hspace{1em}
	\subfloat[$\bot$ disregistry profile, $\tau = 0.01093\,\mu^\mathrm{A}$]{\includegraphics[trim=0cm 0cm 0cm 0cm, clip=true,scale=0.7]	{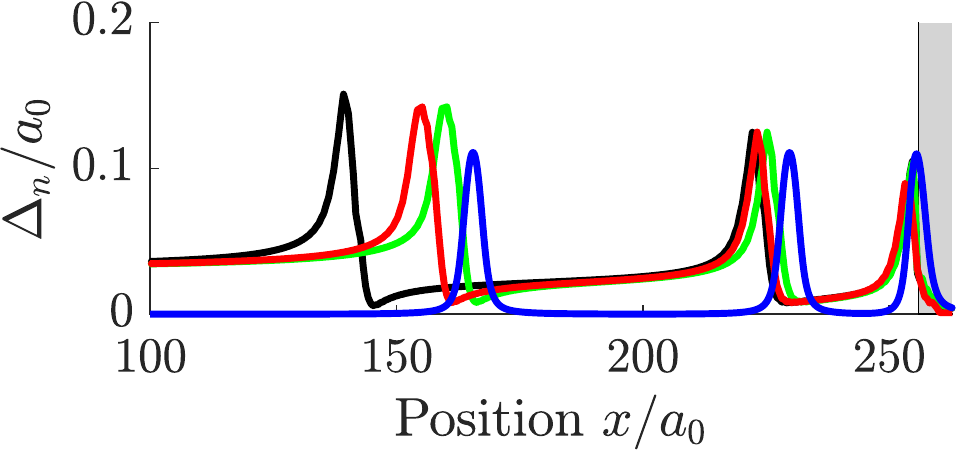}\label{fig:atm-res-disregistry1n}}
	\caption{A comparison of normalized tangential~$\parallel$ and normal~$\bot$ disregistry profiles for different models. Tangential~(a) and normal~(c) disregistry profiles for applied shear stress~$\tau=0.0068\,\mu^{\mathrm{A}}$. Tangential~(b) and normal~(d) disregistry profiles for applied shear stress~$\tau=0.01093\,\mu^{\mathrm{A}}$. The gray area indicates the stiffer Phase~B.}
	\label{fig:atm-res-disregistry}
\end{figure}

Normalized tangential~$\Delta_t/a_0$ and normal~$\Delta_n/a_0$ disregistry profiles, expressed as a function of the normalized coordinate~$x/a_0$ along the glide plane~$\Gamma_{\mathrm{gp}}$, are shown in Fig.~\ref{fig:atm-res-disregistry} for two load levels~$\tau = 0.0068\,\mu^\mathrm{A}$ and~$\tau = 0.01093\,\mu^\mathrm{A}$. In all discrete models (\full{}, \qcfix{}, and \qcadaptive{}), the disregistry profiles are evaluated from the displacements of individual atoms along the glide plane~$\Gamma_{\mathrm{gp}}$, whereas for the \fecz{} model the displacement discontinuity vector~$\vec{\Delta}$ is plotted directly. From the presented results we conclude that, consistently with Fig.~\ref{fig:atm-res-position-energy}, at the lower applied shear level ($\tau = 0.0068\,\mu^\mathrm{A}$) the \qcfix{} solution is very similar to the reference Full solution, whereas the \fepn{} method achieves better accuracy as compared to the \qcadaptive{}. A different situation occurs at later stages of loading ($\tau = 0.01093\,\mu^\mathrm{A}$), where the best accuracy is achieved by the \qcadaptive{} method, while the \fepn{} method leads to the highest error. This error is quantified in Tab.~\eqref{tab:atm-res-num} by the Dislocation Position Error~($DPE$). For each reduced model, the $DPE$ value  represents how much on average the position of dislocation in this model differs (relative to~$a_0$) from the exact solution. The averaging is realized for both distinct dislocations as well as distinct time steps, i.e.,
\begin{equation}
{DPE^{\bullet}} = \frac{1}{n_s}\sum_{i_s}\frac{1}{n_d}\sum_{i_d} \frac{|x_{i_d,i_s}^{\bullet}-x_{i_d,i_s}^{\rm \full{}}|}{a_0},
\label{eq:error_dislocation_positions}
\end{equation}
where~$x_{i_d,i_s}^{\rm \full{}}$ and~$x_{i_d,i_s}^{\bullet}$ denote horizontal position of $i_d$-th dislocation at $i_s$-th time step corresponding to the \full{} and one of the effective models (\qcfix{}, \qcadaptive{}, or \fepn{}). $n_s$ denotes the number of time steps used for the error evaluation, and~$n_d$ is the total number of dislocations at the given time step. In Tab.~\eqref{tab:atm-res-num}, the $DPE$ quantity is evaluated as an average difference in position of three dislocations in the last two steps before transmission of the \fepn{} model ($\tau = 0.01093\,\mu^\mathrm{A}$). The most accurate \qcadaptive{} method exhibits an average difference of $5.83\,a_0$, while the highest average difference of $8.92\,a_0$ can be observed for the \fepn{} method.

The computational performances obtained for the individual models are summarized in Tab.~\ref{tab:atm-res-num} in terms of the number of DOFs and computing times. Both QC models achieve significant reductions in the number of DOFs, which also results in considerable computational savings. The \qcfix{} requires less computing time compared to the \qcadaptive{} method because in the latter a significant amount of time is spent on mesh refinement procedures. The \fepn{} model, on the other hand, uses relatively fine discretization along the glide plane (recall that~$h_\mathrm{max} = a_0/16$ to capture dislocation core accurately), requiring even a higher number of DOFs as compared to the full atomic model. Only a five-fold speed-up is thus obtained, which can be significantly improved by employing much coarser meshes. Additional simulations show that $4$--$8$ times coarser meshes relative to the fine~$h_\mathrm{max} = a_0/16$ discretization do not change the results by more than~$5\%$ while approximately a ten-fold speed-up is achieved, cf.~\cite{bormann2019computational, bormann2020peierls}.

Although the presented example was set to show dislocation pile up with subsequent dislocation transmission, another, rather unexpected, mechanism can be observed in the system depending on the choice of the material contrast ratio~$\rho$. In particular, when~$\rho = 4$, a dislocation pile up followed by dislocation reflection into Phase~A is observed, as shown in Figs.~\ref{fig:atm-res-position-4-1} and~\ref{fig:atm-res-position-40-2}. Since both the \qcfix{} as well as the \fepn{} model were not set up for such a situation, i.e.,~no inclined glide planes or fully-resolved regions are present (recall Figs.~\ref{fig:atm-geo-PN} and~\ref{fig:atm-mesh-QCf}), they cannot account for dislocation reflection---unlike the fully flexible adaptive QC model. 

It is worth noting that the presented speed-ups are obtained for two-dimensional atomic simulations only. In three-dimensional examples---which are more general and physically relevant---, a significantly larger number of atoms is present. The neighbor search algorithm within the considered cut-off radius will be significantly slower and the number of interatomic bonds is substantially higher. As a consequence, the QC simulations are expected to be much slower compared to the simulations of homogenized \fepn{} model.
Moreover, the presented example considers a simple Lennard--Jones pair potential. Use of a more advanced multibody potential, such as embedded atom model, will induce even further computing costs required by the QC simulations.

On the contrary, 3D extensions of the homogenized \fepn{} model are expected to be relatively efficient provided that dislocation trajectories remain a priory known, and fixed glide planes can be considered. In a fully general case with arbitrary dislocation trajectories, however, sliding of glide planes should be considered in all possible lattice directions, which will result in small elastic regions representing individual atoms with multiple glide planes among them, each requiring its associated GSFE. Such a model may become computationally very expensive compared even to a full atomic simulation. In cases without any prior information on possible trajectories of dislocations, it is thus preferable to use adaptive QC approach instead.
%
%
\section{Lattice model: crack propagation in a concrete specimen}
\label{sec:Mesoscale}
%
%
\subsection{Full lattice model}
\label{sec:Mesos-Full}
\begin{figure}
	\centering
	\subfloat[considered geometry]{\includegraphics[width=0.84\linewidth]{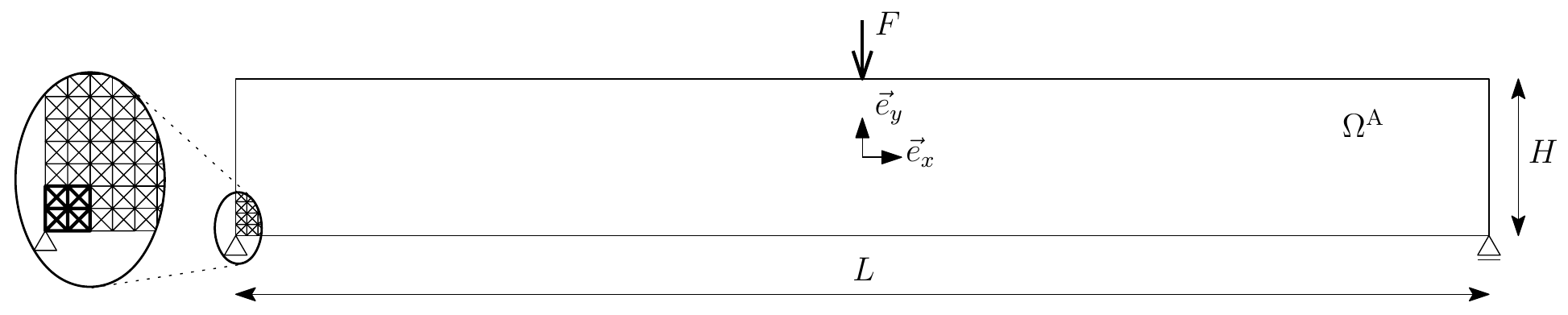}\label{fig:lat-geo-3bp}}
	\subfloat[lattice]{\includegraphics[width=0.15\linewidth]{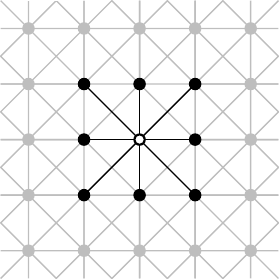}\label{fig:lat-geo-lat}}
	\caption{A quasi-brittle concrete specimen modeled as an X-braced damageable lattice of spacing~$l_0$. (a)~A sketch of the considered geometry of a three-point bending example. (b)~X-braced lattice (gray) with neighboring nodes and active interactions (black) for one particular lattice site (white circle).}
\end{figure}

The second example considers a quasi-brittle (concrete) specimen subjected to a three-point bending test, modeled at the mesoscale as a discrete lattice system shown in Fig.~\ref{fig:lat-geo-3bp}. The domain~$\Omega$ is of size~$L \times H$, inside which a homogeneous X-braced lattice with spacing~$l_0$ is considered in between~$n_\mathrm{par}$ particles~$\alpha$ positioned at~$\vec{r}^\alpha$ stored in an index set~$N$. All lattice interactions are modeled as damageable with an exponential softening law. To prevent spurious damage localization in the vicinity of prescribed displacements or applied loads, the lattice is made~$100$ times stiffer in padding regions with size~$8l_0 \times 6l_0$ under the loading force and~$22l_0 \times 4l_0$ around both supports. Upon loading, a localized crack growing along the symmetry plane from bottom to top is expected.

In analogy to the atomic system considered in the previous section, an inter-particle pair potential~$\pi^{\alpha\beta}$ is introduced. It consists of two contributions, the elastically stored energy reflected by~$\phi^{\alpha\beta}$, and a dissipation distance~$\mathcal{D}^{\alpha\beta}$,
\begin{equation}
	\pi^{\alpha\beta}_k(r^{\alpha\beta}, \omega^{\alpha\beta};\omega^{\alpha\beta}_{k-1}) = (1-\omega^{\alpha\beta})\phi^{\alpha\beta}(r^{\alpha\beta}_+) + \phi^{\alpha\beta}(r^{\alpha\beta}_-) + \mathcal{D}^{\alpha\beta}(\omega^{\alpha\beta},\omega^{\alpha\beta}_{k-1}).
	\label{eq:incrementalPair}	
\end{equation}
Because dissipative mechanisms are included, a variational formulation of rate-independent systems is considered~\cite{mielke2003energetic, mielke2005evolution}, and the interaction potential depends also on an internal variable~$\omega^{\alpha\beta}$, reflecting the level of damage in the interaction connecting particles~$\alpha$ and~$\beta$. In addition, the pair potential is considered in an incremental fashion, i.e.,~at a time instance~$t_k$, $\pi^{\alpha\beta}_k = \pi^{\alpha\beta}(t_k)$, and depends on the configuration of the system in the current as well as the previous time step. To allow damage processes to evolve only under tension and not under compression, the elastically stored energy is split into two parts, $(1-\omega^{\alpha\beta})\phi^{\alpha\beta}(r^{\alpha\beta}_+) + \phi^{\alpha\beta}(r^{\alpha\beta}_-)$, in which only the first term is affected by the damage variable. The two distance quantities, $r^{\alpha\beta}_+ = \max{(r^{\alpha\beta}, r^{\alpha\beta}_0)}$ and~$r^{\alpha\beta}_- = \min{(r^{\alpha\beta}, r^{\alpha\beta}_0)}$ (assuming~$\phi^{\alpha\beta}(r_0^{\alpha\beta}) = 0$), ensure that the damage variable weakens the interaction only under tension, i.e.,~for~$r^{\alpha\beta} > r^{\alpha\beta}_0$. A quadratic elastic potential is considered,
\begin{equation}
\phi^{\alpha\beta}(r^{\alpha\beta}) = \frac{1}{2}EA r_0^{\alpha\beta} \big( \varepsilon^{\alpha\beta}(r^{\alpha\beta}) \big)^2,
\label{Examples:Eq:1}
\end{equation}
expressed as a function of strain
\begin{equation}
\varepsilon^{\alpha\beta}(r^{\alpha\beta}) = \frac{r^{\alpha\beta} - r_0^{\alpha\beta}}{r_0^{\alpha\beta}},
\label{eq:strain}	
\end{equation}
where, in analogy to the atomic system, $\vec{r}^{\alpha\beta} = \vec{r}^\beta - \vec{r}^\alpha$ is a vector of relative positioning, $r^{\alpha\beta} = \| \vec{r}^{\alpha\beta} \|_2$ the Euclidean distance between a pair of particles~$\alpha$ and~$\beta$, and~$E A$ is the normal cross-sectional stiffness of the bond.

The dissipation distance~$\mathcal{D}^{\alpha\beta}$ measures the energy dissipated by a single interaction during the evolution of the damage variable between two consecutive states, $\omega_1^{\alpha\beta}$ and~$\omega_2^{\alpha\beta}$, i.e.,
\begin{equation}
\mathcal{D}^{\alpha\beta}(\omega_2^{\alpha\beta},\omega_1^{\alpha\beta})=
\left\{
\begin{array}{ll}
{\displaystyle D^{\alpha\beta}({\omega}_2^{\alpha\beta})-D^{\alpha\beta}({\omega}_1^{\alpha\beta})}, &\mbox{if} \quad {\omega}_2^{\alpha\beta}\geq{\omega}_1^{\alpha\beta},\\
+\infty, & \mbox{otherwise},
\end{array}
\right.\quad 
\label{SubSect:LNDamage:Eq:4}
\end{equation}
where~$D(\omega)$ is the energy dissipated during a unidirectional damage process up to the damage level~$\omega$. It is defined implicitly such that the following exponential damage law results
\begin{equation}
\omega(\varepsilon) = \left\{
\begin{aligned}
&1-\frac{\varepsilon_0}{\varepsilon}\exp{\left(-\frac{{\varepsilon}-\varepsilon_0}{\varepsilon_f}\right)}, &&\mbox{if} \quad \varepsilon \geq \varepsilon_0, \\
&0, &&\mbox{if} \quad \varepsilon < \varepsilon_0,
\end{aligned}
\right.
\label{eq:lat-damage}
\end{equation}
where~$\varepsilon_0$ is the limit elastic strain for which damage starts to evolve, and~$\varepsilon_f$ characterizes the slope of the softening branch in the associated stress-strain diagram, and where the upper index~$\alpha\beta$ has been dropped for brevity; for more details see~\cite[Section~4.1]{rokovs2017adaptive}. The adopted constitutive parameters are specified in Tab.~\ref{tab:lat-mat}.
\begin{table}
	\caption{Geometric and constitutive parameters of the X-braced lattice used for the three-point bending test. $l_0$ denotes lattice spacing, $\varepsilon_0$ limit elastic strain for an exponential damage model, $\varepsilon_f$ specifies slope of the softening branch, $EA$ the normal cross-sectional stiffness, and~$L\times H$ is size of considered domain.}
	\renewcommand*{\arraystretch}{1.3}	
	\centering
	\begin{tabular}{c|rrrrrr}
		parameters & \multicolumn{1}{c}{$l_0$} & \multicolumn{1}{c}{$EA$} & \multicolumn{1}{c}{$\varepsilon_0$} & \multicolumn{1}{c}{$\varepsilon_f/\varepsilon_0$} & \multicolumn{1}{c}{$L/l_0$} & \multicolumn{1}{c}{$H/l_0$} \\\hline
		values & $1\,{\rm [mm]}$ & $20\,{\rm [MN]}$ & $0.005$ & $10$ & $2{,}048$ & $256$
	\end{tabular}
	\label{tab:lat-mat}
\end{table}

In analogy to Eq.~\eqref{eq:ms_min}, with the introduction of a time discretization for a considered time horizon~$ t \in [0,T] $ from Eq.~\eqref{eq:timediscretization}, the evolution of the fully-resolved damageable lattice system is governed by the minimization of the total incremental energy~$\Pi_k$,
\begin{equation}
\column{q}_k \in \underset{\column{v}}{\mbox{arg min }}
\Pi_k(\column{v};\column{q}_{k-1}), \quad k = 1, \ldots, n_T,
\label{IP}
\end{equation}
minimized over all admissible configurations~$\column{v}$ with an initial condition~$\column{q}_0^\mathsf{T} = [\column{r}_0^\mathsf{T},\column{0}^\mathsf{T}]$, where $\column{q}^\mathsf{T} = [\column{r}^\mathsf{T},\column{z}^\mathsf{T}]$ is a state variable of the entire system, consisting of the kinematic variable~$\column{r}$ storing the positions of all particles, whereas~$\column{z}^\mathsf{T} = [\omega_1, \dots, \omega_{n_\mathrm{int}}]$ collects the damage variables of all~$n_\mathrm{int}$ interactions, and~$\column{q}_k = \column{q}(t_k)$. The incremental energy of the entire system is obtained by collecting contributions from all interactions, i.e.,
\begin{equation}
\Pi_k({\column{q}};\column{q}_{k-1}) 
= 
\mathcal{V}(\column{q}) + \mathcal{D}(\column{q},\column{q}_{k-1}) - \column{f}_k^\mathsf{T} {\column{r}},
\label{IE}
\end{equation}
where
\begin{equation}
\mathcal{V}(\column{q}) = \frac{1}{2}\sum_{\substack{\alpha=1 \\ \beta \in B_\alpha}}^{n_\mathrm{par}} \big[ (1-\omega^{\alpha\beta})\phi^{\alpha\beta}(r^{\alpha\beta}_+) + \phi^{\alpha\beta}(r^{\alpha\beta}_-) \big]
\label{SubSect:DissLatt:Eq:3}
\end{equation}
is the internally stored energy\footnote{The numerical implementation of Eq.~\eqref{SubSect:DissLatt:Eq:3} is again conveniently converted from a sum over all particles into a sum over all interactions, recall Eq.~\eqref{eq:energy} and the related Footnote~1.}, and
\begin{equation}
\mathcal{D}(\underline{q}_2,\underline{q}_1)=\frac{1}{2}\sum_{\substack{\alpha=1\\\beta \in B_\alpha}}^{n_\mathrm{par}}\mathcal{D}^{\alpha\beta}(\underline{z}_2,\underline{z}_1)
\label{SubSect:DissLatt:Eq:6}
\end{equation}
is the global dissipation distance (recall also Eqs.~\eqref{eq:incrementalPair} and~\eqref{SubSect:LNDamage:Eq:4}), $B_\alpha$ is the initial set of nearest-neighbors associated with a particle~$\alpha$ (cf. Fig.~\ref{fig:lat-geo-lat}), whereas~$\column{f}_k$ is a vector of external forces applied at a time instance~$t_k$. At each time step, the minimization problem~\eqref{IP} is solved to obtain a local minimum. Because multiple minima may exist, the energetic solution corresponds to the one that satisfies the following energy balance
\begin{equation}
\mathcal{V}(\column{q}_k) - \mathcal{V}(\column{q}_0) + \mathrm{Var}_{\mathcal{D}}(\column{q}, t_k) = \mathcal{W}(\column{q}, t_k), \quad k = 1, \ldots, n_T,
\label{eq:en-bal}
\end{equation}
which equates a sum of the internally stored energy and the dissipated energy, defined as
\begin{equation}
\mathrm{Var}_{\mathcal{D}}(\column{q},t_k) =  \sum_{i=1}^{k}\mathcal{D}(\column{z}_{i},\column{z}_{i-1}),
\label{Sect:VarForm:Eq:1}
\end{equation}
with the work performed by the external forces
\begin{equation}
\mathcal{W}(\column{q},t_k) = \sum_{i=1}^{k}\frac{1}{2}(\column{f}_i+\column{f}_{i-1})^\mathsf{T}(\column{r}_i-\column{r}_{i-1}).
\label{Sect:VarForm:Eq:2}
\end{equation}
To ensure that obtained minimum corresponds to a physically relevant energetic solution, the energy balance constraint of Eq.~\eqref{eq:en-bal} is closely monitored in each time increment. Upon its violation, the considered time step is restarted with a smaller loading increment. To avoid such situations from occurring, the evolution path of the system is controlled using a dissipation driven arc-length method or crack mouth opening displacement control, as elaborated in more detail below in Section~\ref{sec:Mesos-Results}.
For further details on the variational formulation of rate independent systems see, e.g.,~\cite{MieRou:2015} (Section~3.2), and for applications to lattice systems, e.g.,~\cite{Kochmann:2016,rokovs2016variational,desmoulins2017local,rokovs2017extended,rokovs2017adaptive}.
%
%
\subsection{Quasicontinuum model}
\label{sec:Mesos-QC}
In analogy to Section~\ref{ssec:Micro-QC}, the two QC steps are introduced as follows. First, the interpolation step is introduced for the kinematic variable, cf.~Eq.~\eqref{eq:qc-interp},
\begin{equation}
\column{r} \approx \mtrx{\Phi} \column{g},
\label{eq:qc2-interp}
\end{equation}
while the internal variable~$\column{z}$ is not reduced for simplicity and to avoid issues with non-uniqueness in the prolongation operation and mesh coarsening as discussed in~\cite[Section~3.1]{rokovs2016variational}. The reduced QC state variable then reads~$\column{q}_\mathrm{red}^\mathsf{T} = [\column{g}^\mathsf{T},\column{z}^\mathsf{T}]$. To allow for crack paths to be independent of the mesh topology, an extended variant of a QC methodology introduced in~\cite{rokovs2017extended} is adopted. This extension allows for an efficient mesh coarsening through Heaviside step function enrichments used in the interpolation matrix~$\mtrx{\Phi}$. The generalized kinematic variable~$\column{g}$ then stores the positions of all representative particles ($\column{r}_\mathrm{rep}$, analogous to atomic QC) as well as coefficients of linear combinations of the special function enrichments.

The summation step again samples the total incremental energy based on only a small set of selected sampling atoms stored in an index set~$N_\mathrm{S}$, i.e.,
\begin{equation}
\Pi_k ( \column{q}_k ; \column{q}_{k-1} ) \approx 
\sum_{ \alpha \in N_\mathrm{S} }
w^{\alpha}\pi^\alpha_k(\column{q}_k;\column{q}_{k-1}) - \column{f}_k^\mathsf{T} \column{r},
\label{eq:qc2-summ}
\end{equation}
where~$w^\alpha$ and~$N_\mathrm{S}$ correspond to the central summation rule of Beex~\emph{et al.}~\cite{beex2014central}, generalized for Heaviside type of enrichments according to~\cite{rokovs2017extended}. The energy associated with a lattice site~$\alpha$ is defined as
\begin{equation}
\pi^\alpha_k(\column{q}_k;\column{q}_{k-1}) = \frac{1}{2}\sum_{\beta \in B_\alpha} \pi^{\alpha\beta}_k(r^{\alpha\beta}_k, \omega^{\alpha\beta}_k;\omega^{\alpha\beta}_{k-1}),
\end{equation}
where the incremental energy of a single interaction~$\pi^{\alpha\beta}_k$ has been defined in Eq.~\eqref{eq:incrementalPair}. A QC method therefore minimizes the approximate total incremental energy of Eq.~\eqref{eq:qc2-summ} with respect to the reduced state variable~$\column{q}_\mathrm{red}$, mapped on the state of the entire system through the interpolation relation of Eq.~\eqref{eq:qc2-interp}.

The area of high interest, i.e., the fully-resolved region, is allowed to evolve adaptively throughout the simulation.
Available options for the refinement criteria have been recently discussed in~\cite{chen2021refinement}. From the presented options, only a few studies presented concepts suitable for applications to structural lattices~\cite{memarnahavandi2015goal,rokovs2017adaptive,rokovs2017extended,mikevs2018molecular, chen2021refinement}. In this work, the following mesh error indicator, employed also in~\cite{rokovs2017adaptive}, is adopted for mesh refinement as well as for mesh coarsening. Interactions that are likely to be subjected to damage are identified through the refinement criterion, which compares the energy stored in an interaction (recall that damage is only allowed in tension) against a certain threshold energy, i.e.,
\begin{equation}
\varepsilon^{\alpha\beta} > 0,   \quad \mbox{and} \quad
(1-\omega^{\alpha\beta})\phi^{\alpha\beta}(r^{\alpha\beta}) \geq \theta_\mathrm{r}\,\phi^{\alpha\beta}_\mathrm{th}, \quad \alpha\beta \in S_\mathrm{int}^K,
\label{eq:refinement}
\end{equation}
where~$\theta_\mathrm{r} \in (0,1)$ is a safety parameter and~$\phi_\mathrm{th}^{\alpha\beta} = \phi^{\alpha\beta}(r_0(1+\varepsilon_0))$ is the elastic threshold energy at which the damage starts to evolve, and~$S_\mathrm{int}^K$ denotes the set of all sampling interactions~$\alpha\beta$ located inside an element~$K$. If at least one bond in~$S_\mathrm{int}^K$ satisfies the condition of Eq.~\eqref{eq:refinement}, element~$K$ is fully refined (i.e.,~all lattice sites inside the triangle are added as repnodes, and the triangulation is updated). Alternatively, element~$K$ is coarsened if the elastic energy of all bonds is lower than a given threshold, i.e.,
\begin{equation}
(1-\omega^{\alpha\beta})\phi^{\alpha\beta}(r^{\alpha\beta}) \leq 
\theta_\mathrm{c}\,\phi^{\alpha\beta}_\mathrm{th}, \quad \alpha\beta \in S_\mathrm{int}^K,
\label{eq:coarsening}
\end{equation}
where~$\theta_\mathrm{c} \in (0,1)$ is a coarsening parameter satisfying~$\theta_\mathrm{c} < \theta_\mathrm{r}$. Note that in Eq.~\eqref{eq:coarsening} all bonds, i.e.,~including those in compression, are considered. The fully-resolved region is adaptively updated following the same three steps (i)--(iii) presented at the end of Section~\ref{ssec:Micro-QC}. No initial mesh refinement is needed, unlike in the atomic QC system, since no initiation analogous to the Volterra perturbation is employed (which needs to be resolved by the adopted QC system).
%
%
\subsection{Finite element cohesive zone model}
\label{sec:Mesos-Cohesive}
\begin{figure}
	\centering
	\includegraphics[width=0.9\linewidth]{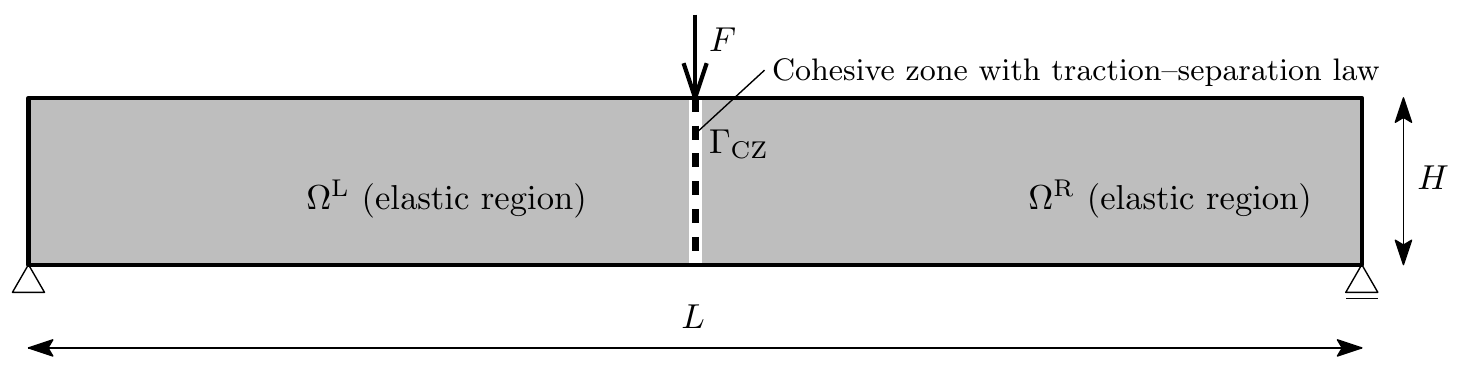}
	\caption{The \fecz{} model for a crack propagation in the three-point bending test, showing the elastic domain (gray), cohesive zone (dashed line), and applied boundary conditions.}
	\label{fig:lat-geo-cz}
\end{figure}

The initial lattice problem of Fig.~\ref{fig:lat-geo-3bp} is translated into a continuous Finite Element Cohesive Zone~(\fecz{}) model shown in Fig.~\ref{fig:lat-geo-cz}, in which it is assumed that only one potential crack path is present in an otherwise continuous specimen. In contrast to a discrete lattice, the continuous domain does not require local stiffening under the applied boundary conditions to prevent excessive deformation. A cohesive zone~$\Gamma_{\mathrm{cz}}$, located along the vertical axis of symmetry, splits the homogeneous domain~$\Omega$ into two parts~$\Omega^{\mathrm{L}}$ and~$\Omega^{\mathrm{R}}$, i.e.,
\begin{equation}
\begin{aligned}
\Gamma_{\mathrm{cz}} &= \left\{ \vec{x} \in \mathbb{R}^2: x =   x_{\mathrm{cz}}, | y | \leq H/2 \right\}, \\
\Omega^{\mathrm{L}} &= \left\{ \vec{x} \in \mathbb{R}^2: -L/2 \leq x \leq x_{\mathrm{cz}}, | y | \leq H/2 \right\}, \\
\Omega^{\mathrm{R}} &= \left\{ \vec{x} \in \mathbb{R}^2:  x_{\mathrm{cz}} \leq x \leq L/2, | y | \leq H/2 \right\}, \\
\Omega &= \Omega^{\mathrm{L}} \cup \Omega^{\mathrm{R}},
\end{aligned}
\end{equation}
where $x_{\mathrm{cz}}=0$ denotes the horizontal position of the cohesive zone.

The mechanical behavior of the \fecz{} system is governed by the minimization of the total potential energy defined as
\begin{equation}
\vec{u}_k \in \underset{{\scriptsize\vec{v}}}{\text{arg min}}\ \left(\Psi(\vec{v}) - \int_{\Gamma_t}\vec{t}\cdot\vec{v}\,\mathrm{d}\Gamma_t \right),
\label{eq:mincz}
\end{equation}
where the second term on the right-hand side represents a concentrated force~$F$ in Fig.~\ref{fig:lat-geo-cz}, distributed through tractions~$\vec{t}$ over a small portion of the specimen boundary~$\Gamma_t$. Both subdomains~$\Omega^{\mathrm{L}}$ and~$\Omega^{\mathrm{R}}$ are purely elastic isotropic and the deformation related to the crack propagation is localized within the cohesive zone~$\Gamma_{\mathrm{cz}}$. The total internal energy~$\Psi$ (cf.~also Eq.~\eqref{eq:internal-energy_atm}) reads
\begin{equation}
\label{eq:internal-energy}
\Psi(\vec{u}) = \int_{\Omega\setminus\Gamma_{\mathrm{cz}}}\psi_{\mathrm{e}}(\vec{u})\,\mathrm{d}\Omega+\int_{\Gamma_{\mathrm{cz}}}\psi_{\mathrm{cz}}(\vec{u})\,\mathrm{d}\Gamma_{\mathrm{cz}},
\end{equation}
where~$\psi_{\mathrm{e}}$ is the elastic strain energy density defined in Eq.~\eqref{eq:linearElastic}, and~$\psi_\mathrm{cz}$ is the cohesive zone potential (defined solely on~$\Gamma_{\mathrm{cz}}$). Both are obtained through the homogenization of the underlying lattice as follows.
%
%
\subsection{Calibration}
\begin{figure}
	\centering
	\subfloat[lattice unit cell]{\includegraphics[trim=0cm 0cm 0cm 0cm, clip=true, width=0.33\linewidth]	{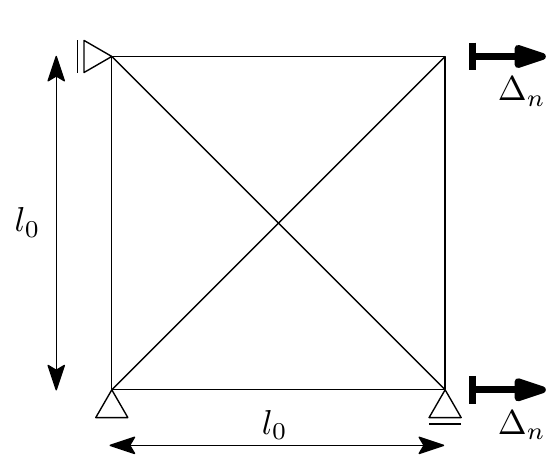}\label{fig:lat-geo-calib-a}}
	\hspace{0.5em}
	\subfloat[traction--separation law]{\includegraphics[trim=0cm 0cm 0cm 0cm, clip=true, width=0.5\linewidth]		{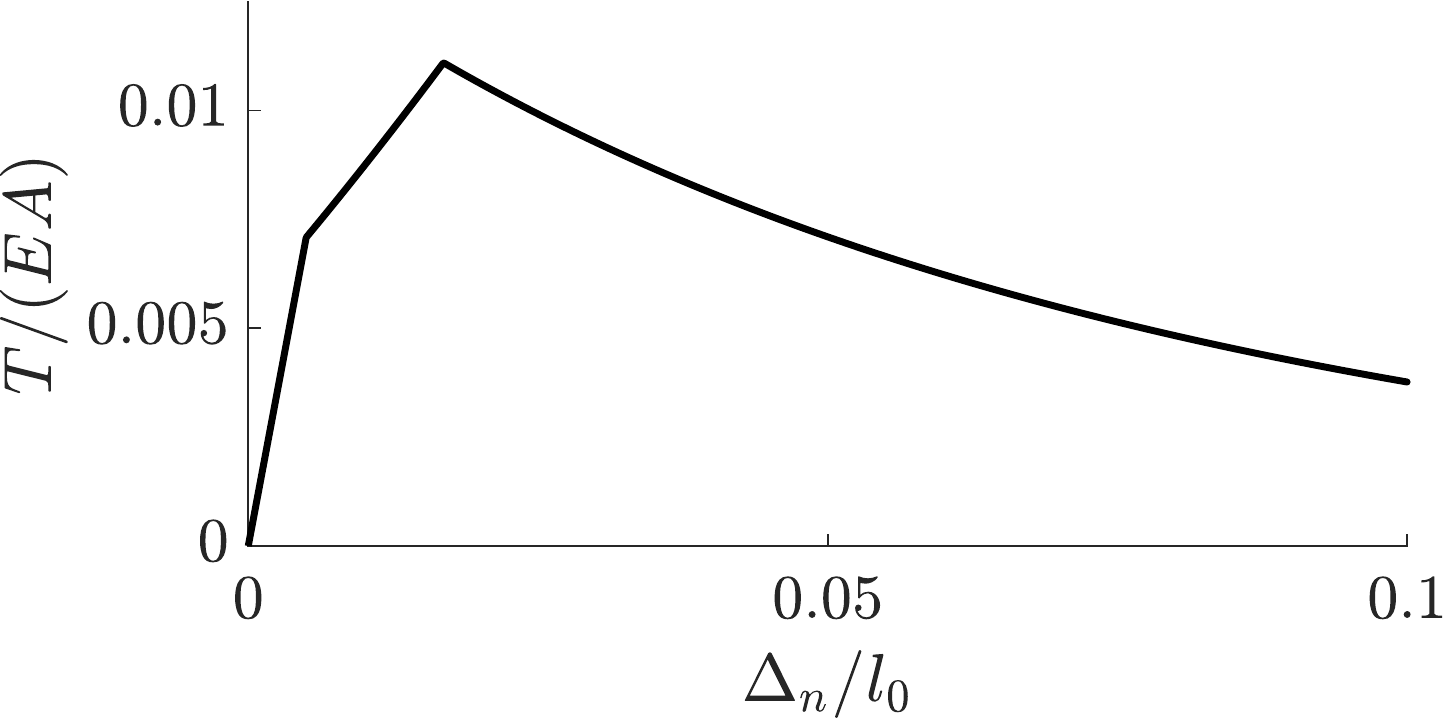}\label{fig:lat-geo-calib-b}}
	\caption{(a)~The lattice unit cell used for the computation of the traction--separation law. (b)~Normalized traction--separation law as a function of the normalized normal opening~$\Delta_n/l_0$.}
	\label{fig:lat-geo-calib}
\end{figure}

For the calibration of the \fecz{} model, a single lattice unit cell as shown in Fig.~\ref{fig:lat-geo-calib-a}, is considered. The normal cross-sectional stiffness of the horizontal and vertical bonds are reduced to half of its true value to respect the considered periodicity. The corresponding assembled effective stiffness tensor~$\tensorfour{D}^\mathrm{lat}$, listed in Tab.~\ref{tab:homogenized_lat}, is compared to an isotropic plane strain stiffness tensor~$\tensorfour{D}(E_\mathrm{iso},\nu_\mathrm{iso})$.
The resulting effective elastic parameters are identified as
\begin{equation}
E_\mathrm{iso} = \frac{5}{6}\frac{EA}{l_0 l_z} \big(1+1/\sqrt{2} \big) \quad \mbox{and} \quad \nu_\mathrm{iso} = \frac{1}{4},
\end{equation}
where $l_{z}$ is the unit out-of-plane thickness. The fixed Poisson's ratio is given by the fact that $D_{1122}^\mathrm{lat} = D_{1212}^\mathrm{lat}$ and the effective Young's modulus is obtained by fitting $D_{1111}^\mathrm{lat}=D_{1111}(E_\mathrm{iso},\nu_\mathrm{iso})$. 
Note that for the considered setting of lattice structure (identical normal stiffness of all bonds), the resulting stiffness tensor~$\tensorfour{D}^\mathrm{lat}$ is not fully isotropic, cf.,~e.g.,~\cite[Appendix~
B]{mikevs2017quasicontinuum} for more details.
\begin{table}
	\caption{Homogenized constitutive parameters corresponding to the lattice unit cell.}
	\renewcommand*{\arraystretch}{1.3}		
	\centering
	\begin{tabular}{c|cc}
		parameters & $D_{1111}^\mathrm{lat} = D_{2222}^\mathrm{lat}$ & $D_{1122}^\mathrm{lat} = D_{1212}^\mathrm{lat}$ \\\hline
		~ & $(1+1/\sqrt{2}) \, EA/(l_0 l_z)$ & $~~(1/\sqrt{2}) \, EA/(l_0 l_z)$ \\
	\end{tabular}
	\label{tab:homogenized_lat}
\end{table}
The cohesive zone potential~$\psi_{\mathrm{cz}}$ is a function of the displacement jump across the cohesive zone~$\Gamma_{\mathrm{cz}}$, 
\begin{equation}
\vec{\Delta} = \Delta_t \vec{e}_t + \Delta_n \vec{e}_n = \llbracket \vec{u} \rrbracket=\vec{u}^{\mathrm{R}}-\vec{u}^{\mathrm{L}},
\label{eq:Disregistry}
\end{equation} 
see also Eq.~\eqref{eq:Disregistry_atm}. Due to the symmetry of the considered problem, the tangential component~$\Delta_t$ vanishes, i.e.,
\begin{equation}
\vec{u}^{\mathrm{L}} \cdot \vec{e}_y = \vec{u}^{\mathrm{R}} \cdot \vec{e}_y, \quad\mathrm{on}\quad\Gamma_{\mathrm{cz}},
\label{eq:pb-traction}
\end{equation}
and hence the cohesive zone potential~$\psi_\mathrm{cz}$ becomes a function of the crack normal opening~$\Delta_n$ only. The cohesive zone potential is obtained numerically from the response of the lattice unit cell subjected to a horizontal stretch of~$\Delta_n/l_0$ with the boundary conditions shown in Fig.~\ref{fig:lat-geo-calib-a}. The corresponding traction acting on the cohesive zone as a function of the normal crack opening is defined as
\begin{equation}
T(\Delta_n) = \frac{\mathrm{d}\, \psi_{\mathrm{cz}}(\Delta_n)}{\mathrm{d}\, \Delta_n},
\end{equation}
and is plotted for the lattice constitutive parameters of Tab.~\ref{tab:lat-mat} in Fig.~\ref{fig:lat-geo-calib-b}.
%
%
\subsection{Results and comparison}
\label{sec:Mesos-Results}
Due to the present damage processes, arbitrarily small increments in externally applied boundary conditions might lead to an uncontrolled evolution of the crack, and hence also of the internal variable~$\column{z}$ in both the fully-resolved and QC system (i.e.,~a snap-back may occur), which in turn might result in the violation of the energy balance constraint of Eq.~\eqref{eq:en-bal}. To avoid such situations, the evolution path is closely controlled using a dissipation driven arc-length method, see, e.g.,~\cite{verhoosel2009dissipation,Gutierrez:2004,May:2016} for more details. For the \fecz{} approach, the standard Crack Mouth Opening Displacement~(CMOD) control method~\cite{Jirasek:2002} is sufficient to reliably compute a mechanical response of the system.

\begin{figure}
	\centering
	\subfloat[\qccoarse{}]{\includegraphics[width=0.9\linewidth]{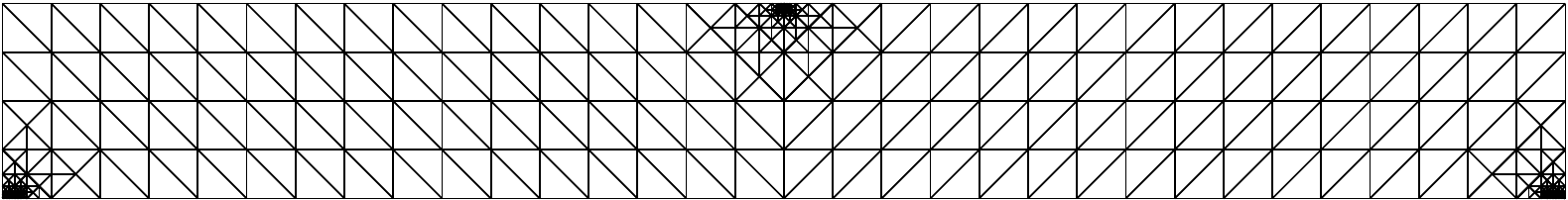}\label{fig:lat-mesh-QCc}}\\
	\vspace{0em}
	\subfloat[\qcfine{}]{\includegraphics[width=0.9\linewidth]{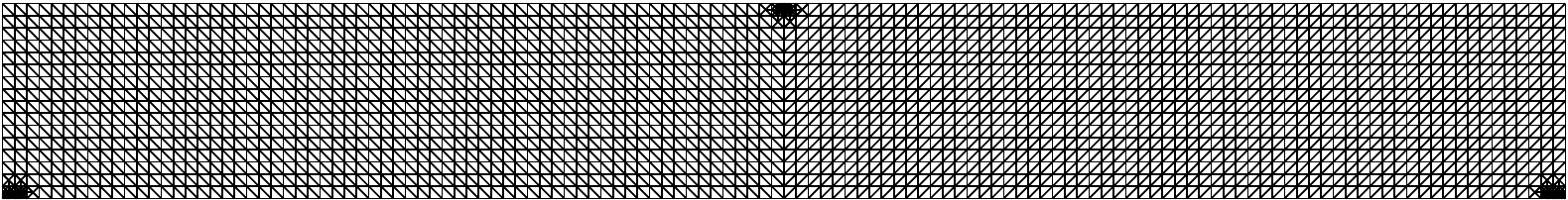}\label{fig:lat-mesh-QCf}}\\
	\vspace{0em}
    \subfloat[\fecz{}]{\includegraphics[width=0.9\linewidth]{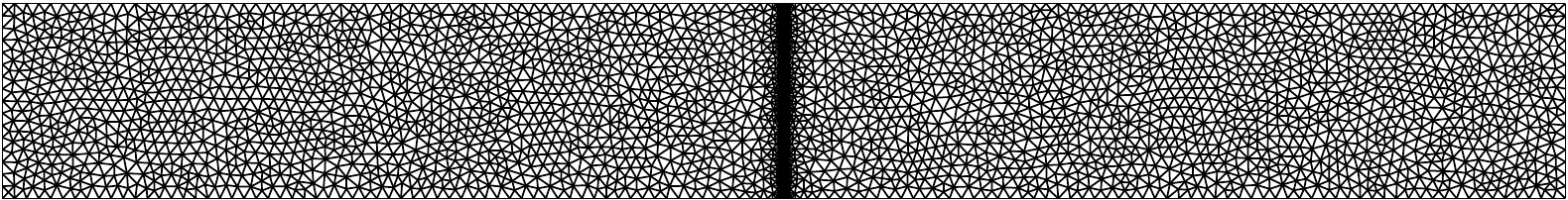}\label{fig:lat-mesh-CZ}}\\
	\caption{Employed initial triangulations for crack propagation in the three-point bending test of concrete specimen, corresponding to~(a) \qccoarse{} model, (b)~\qcfine{} model, and~(c) \fecz{} model.}
	\label{fig:lat-mesh}
\end{figure}

The mechanical behavior of the fully-resolved mesoscopic system of Section~\ref{sec:Mesos-Full} (referred to as \full{}) is adopted as the underlying reference solution. Two QC models, as described in Section~\ref{sec:Mesos-QC}, are then considered: (i)~one with a coarse triangulation of maximum element size~$H/4$ (referred to as \qccoarse{}), and~(ii) one with a fine triangulation of maximum element size~$H/16$ (referred to as \qcfine{}), see Figs.~\ref{fig:lat-mesh-QCc} and~\ref{fig:lat-mesh-QCf}. The \fecz{} model, detailed in Section~\ref{sec:Mesos-Cohesive}, is considered for only one triangulation with a fixed mesh. A maximum element size~$H/16$ adopted far from the cohesive zone, which is gradually refined down to~$H/256=l_0$ towards the assumed crack path~$\Gamma_{\mathrm{cz}}$, cf. Fig.~\ref{fig:lat-mesh-CZ}. A mesh convergence study has been performed to verify that the adopted element size provides converged results. The initial triangulations of all reduced models are shown in Fig.~\ref{fig:lat-mesh}, whereas the corresponding number of DOFs are listed in Tab.~\ref{tab:lat-res-num}. All three effective models are compared against the fully-resolved simulation in terms of force--displacement curves, peak forces, crack lengths, and crack opening profiles.
\begin{figure}
	\flushleft
	\subfloat[\processZone{} before crack localization]{\includegraphics[width=0.9\linewidth]{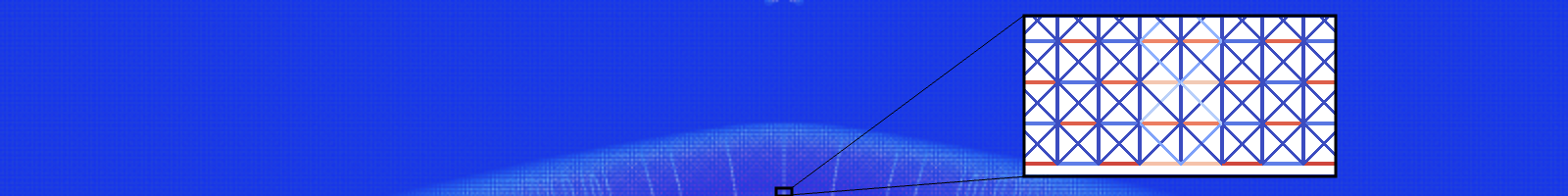}\label{fig:lat-res-zonea}}\\
	\vspace{-6em}
	\subfloat[\processZone{} with a fully developed and localized crack]{\includegraphics[width=0.9\linewidth]{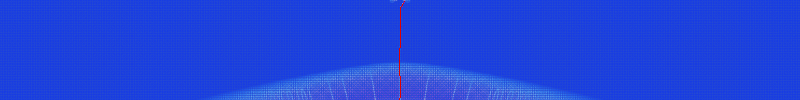}\label{fig:lat-res-zoneb}}
	\put(2,-10){\stackunder{\includegraphics[scale=0.37]{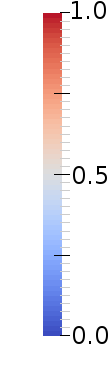}{\hspace{-3em}\scriptsize Damage}}}
    \caption{The damage level~$\omega$ in all interactions computed with the full lattice model for the three-point bending test. The \processZone{}, consisting of alternatingly unloading interactions under damage and elastically, is shown with a close-up of the lattice one step before crack initialization in~(a), and in the last step of the simulation for the final \processZone{} and fully developed crack path in~(b).}
	\label{fig:lat-res-zone}
\end{figure}

\begin{figure}
	\centering
	\subfloat[force--displacement diagrams]{\includegraphics[trim=0cm 0cm 0cm 0cm, clip=true,scale=0.75]{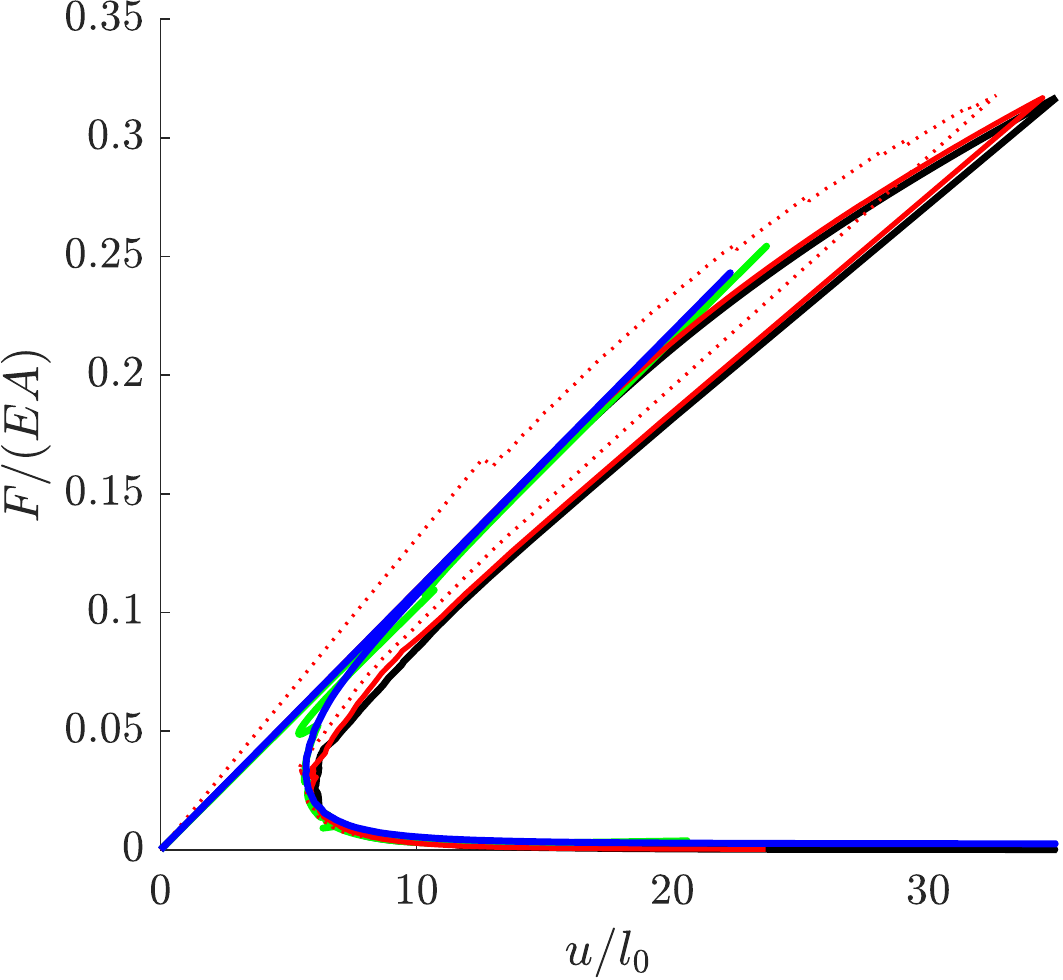}\label{fig:lat-res-fu-elast}\label{fig:lat-res-fu-elast-inelasta}}
	\hspace{2em}
	\subfloat[inelastic part of force--displacement diagrams]{\includegraphics[trim=0cm 0cm 0cm 0cm, clip=true,scale=0.75]{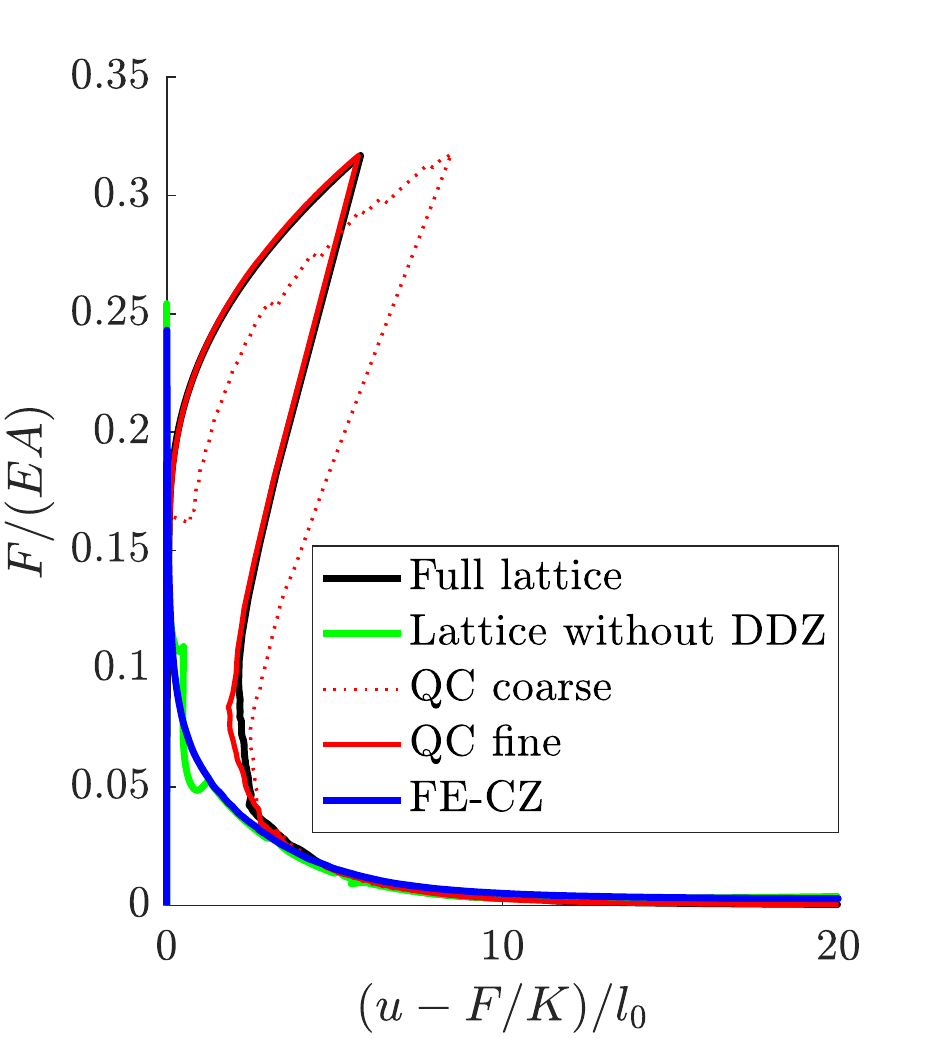}\label{fig:lat-res-fu-inelast}\label{fig:lat-res-fu-elast-inelastb}}
	\caption{Normalized force--displacement diagrams for the three-point bending test. (a)~Complete diagrams. (b) Inelastic parts of the force--displacement diagrams, in which the elastic part of the displacement, $F/K$, is subtracted from the total macroscopic displacement $u$ and only the inelastic part is plotted.}
	\label{fig:lat-res-fu-elast-inelast}
\end{figure}

With the increasing external load, individual bonds near the bottom center region start to damage until localization occurs, forming a full central crack propagating almost vertically across the specimen's height, cf. Fig.~\ref{fig:lat-res-zoneb}. The corresponding evolutions of the normalized magnitude of the externally applied force $F/(EA)$ are shown in Fig.~\ref{fig:lat-res-fu-elast-inelast} as a function of the normalized vertical displacement measured under the force, $u/l_0$. The specimen exhibits, after the initial linear response, a significant softening followed by a severe snap-back once the central crack localizes. While both QC approaches manage to capture the initial ductile behavior (the \qcfine{} being almost indistinguishable from the \full{} solution), the \fecz{} completely omits this mechanism. Upon crack localization, however, all effective methods provide satisfactory qualitative description of the force--displacement curve. The \qccoarse{} method suffers from an initially too stiff response as a consequence of a rather coarse mesh, although with adaptive refinement this discrepancy drops rapidly. To compare all methods quantitatively, the relative errors in the initial elastic stiffnesses and peak forces are summarized in Tab.~\ref{tab:lat-res-num}. Here we conclude that the \fecz{} model underestimates the peak force by more than~$23\%$, while being very accurate in the initial elastic stiffness. Although the \qccoarse{} method overestimates the initial elastic stiffness by approximately~$21\%$, it delivers an accurate peak force. The \qcfine{} method provides very accurate results in both considered quantities.

The rather unexpected pronounced softening before crack localization is caused by an extensive damage region in which individual horizontal bonds are damaging or unload elastically in an alternating way, forming a checker-board pattern, cf. Fig.~\ref{fig:lat-res-zonea}. This region is hereafter referred to as the Distributed Damage Zone~(DDZ). This partially damaged region expands progressively along the bottom edge of the specimen and upwards, featuring multiple possible crack paths (emerging in Fig.~\ref{fig:lat-res-zonea}) until localization occurs and the central crack is formed. A similar behavior is observed for both \full{} and QC models, whereas the \fecz{} model is unable to capture the initial DDZ as a consequence of the linear-elastic constitutive law used in the bulk material, which results in significant inaccuracies before the crack localizes (recall Fig.~\ref{fig:lat-res-fu-elast-inelast}). Additional simulations (not shown) indicate that the observed errors further increase with increasing ductility of the considered exponential softening law of the individual bonds.

To verify that the \processZone{} is at the root of the inaccuracy observed in the \fecz{} model, an additional simulation has been performed, in which damage was confined to evolve only in bonds located within~$2\,l_0$ distance from the vertical axis of the symmetry, whereas the remainder of the specimen was considered as purely elastic. This version of the full model provided results (referred to as Lattice without DDZ) closely resembling those of the \fecz{} model. In particular, the relative error in the peak force of the \fecz{} model drops to~$4.41\%$ compared to this reduced version of the full model, thus revealing that the error in the peak force indeed results from the neglected \processZone{}. To increase the accuracy of the \fecz{} model, multiple cohesive cracks should be realized in the \processZone{} or a more involved constitutive model with damage should be employed for the bulk material.

\begin{figure}
	\centering
	\includegraphics[width=0.8\linewidth]{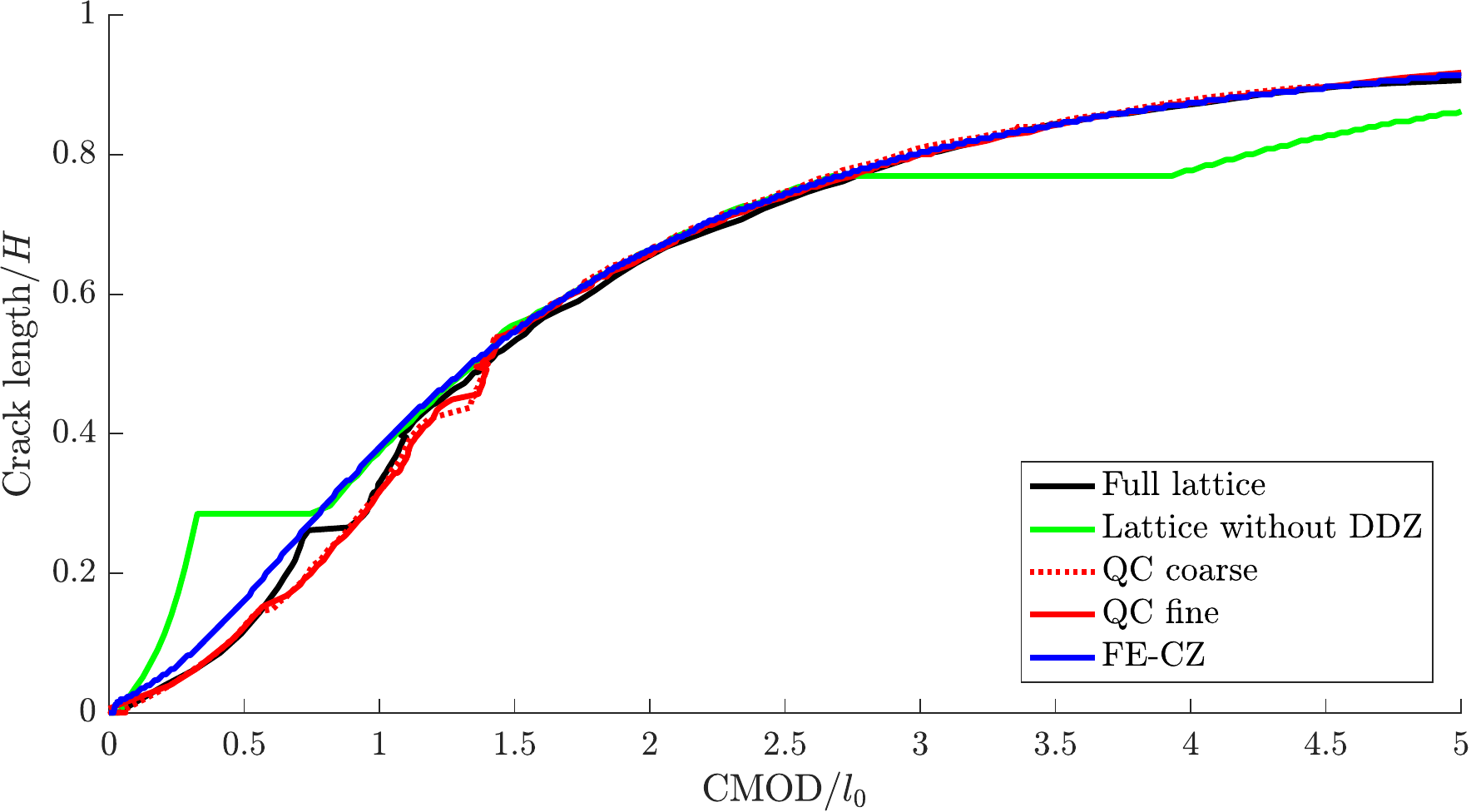}
	\caption{The normalized crack length as a function of the Crack Mouth Opening~(CMOD).}
	\label{fig:lat-res-crack-length}
\end{figure}

\begin{figure}
	\centering
	\subfloat[crack opening, pre-peak load]{\includegraphics[trim=0cm 0cm 0cm 0cm,clip=true,scale=0.75]{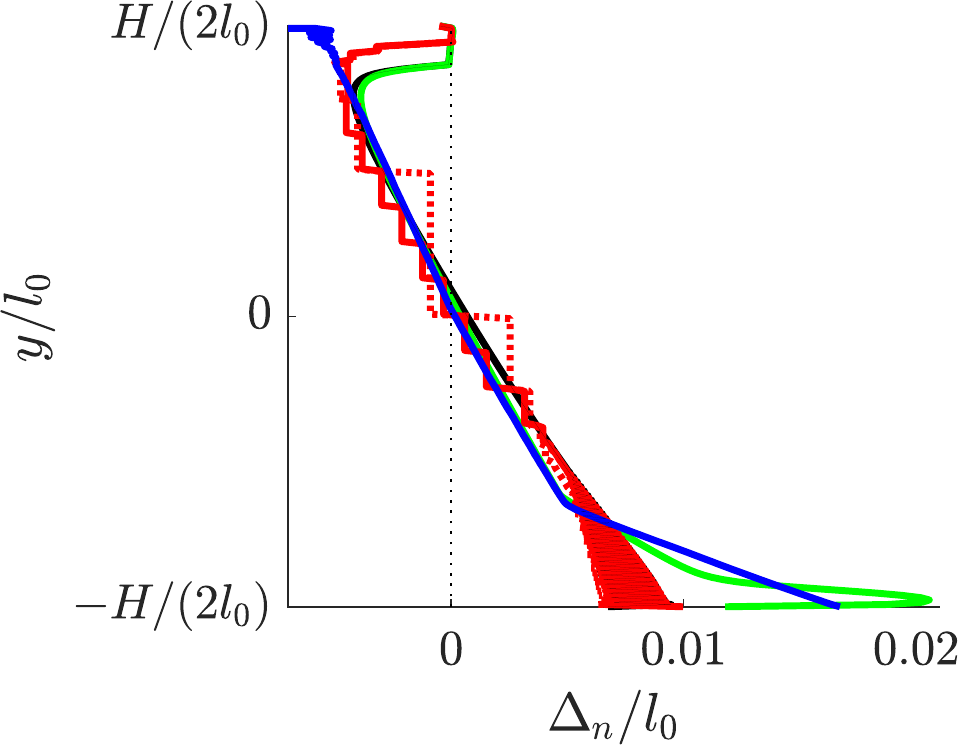}\label{fig:lat-res-opening_pre}}
	\hspace{2em}
	\subfloat[crack opening, post-peak load]{\includegraphics[trim=0cm 0cm 0cm 0cm,clip=true,scale=0.75]{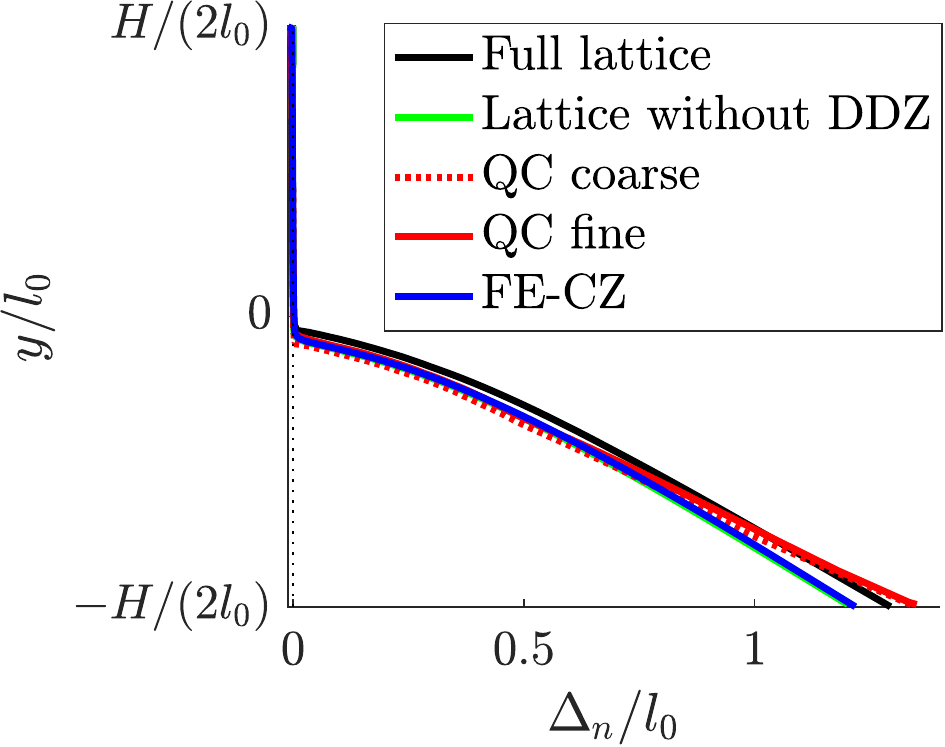}\label{fig:lat-res-opening_post}}
	\caption{Normalized crack opening profiles for the three-point bending test. (a)~Pre-peak opening corresponding to loading force~$ F/EA = 0.24 $, and~(b) post-peak opening corresponding to loading force~$ F/EA = 0.035 $. Due to lack of distributed damage, the \fecz{} model initially significantly deviates~(a), whereas upon crack localization it becomes accurate~(b).}
	\label{fig:lat-res-opening}
\end{figure}

To compare the crack lengths (Fig.~\ref{fig:lat-res-crack-length}) and the crack openings (Fig.~\ref{fig:lat-res-opening}), the central crack needs to be identified and its opening measured in discrete lattice systems. To this end, a fixed crack width of~$l_0$ is considered, consisting of the most damaged bonds in each horizontal layer of bonds; if no damage occurs in a horizontal layer, the central bond (located at the specimen's vertical axis of symmetry) is used. The normal crack opening~$\Delta_n$ is then defined as the difference between the horizontal displacements of the two end nodes of all crack bonds. Due to elasticity, negative crack openings are observed in compressive regions. The crack length, expressed as a function of the CMOD, is shown in Fig.~\ref{fig:lat-res-crack-length}. All approximate methods show comparable results, although both QC approaches are slightly more accurate, capturing also some of the initial irregularities of the \full{} model response. The \fecz{} method in general shows a smooth crack evolution as compared to the discrete models, initially overestimating the crack length, but then converging rapidly to the reference.

The corresponding crack opening diagrams, represented as the normalized normal opening~$\Delta_n/l_0$ and plotted against the normalized vertical coordinate~$y/l_0$, are shown in Fig.~\ref{fig:lat-res-opening}. 
The presented results correspond to a pre-peak configuration of~$ F/EA = 0.24 $ (Fig.~\ref{fig:lat-res-opening_pre}), and to a post-peak configuration of~$ F/EA = 0.035 $ (Fig.~\ref{fig:lat-res-opening_post}). Here we clearly see that the \fecz{} and \full{} model without \processZone{} (Lattice without \processZone{}) tend to show significantly larger crack mouth openings before the crack localizes, whereas both QC models provide accurate predictions. Upon crack localization, in accordance with previously discussed results, all effective methods achieve a good accuracy.

\begin{figure}
	\centering
	\includegraphics[scale=0.55]{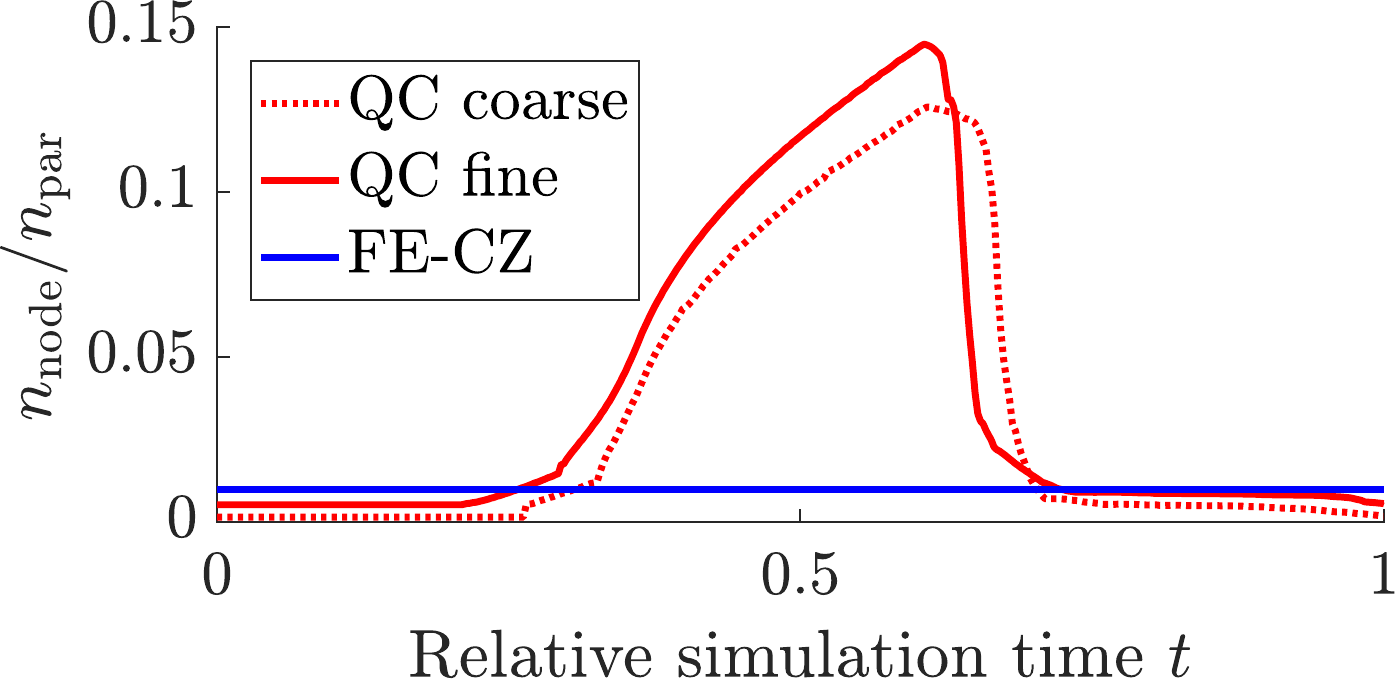}
	\caption{Evolution of the number of DOFs relative to that of the fully-resolved model.}
	\label{fig:lat-res-dof}
\end{figure}

The numerical performances of all methods are shown in Fig.~\ref{fig:lat-res-dof} in terms of the number of DOFs relative to the number of DOFs of the fully-resolved system. Here we see that although both QC models start with noticeably different meshes (Figs.~\ref{fig:lat-mesh-QCc} and~\ref{fig:lat-mesh-QCf}), once the fully-resolved DDZ develops, the relative number of DOFs increases from approximately~$0.5\%$ to roughly~$15\%$ for both QC approaches. Upon crack localization, the mesh coarsens and the relative number of DOFs drops to~$0.5\%$ again. The \fecz{} model has no adaptivity, having a constant~$1\%$ relative number of DOFs. The achieved performance in terms of computing times is summarized in Tab~\ref{tab:lat-res-num}. Here we see that in spite of relatively large savings in the number of DOFs, both QC methods attain a speed-up of only a factor of~$2.5$. This is caused mainly by the initial full refinement of the entire \processZone{} (covering almost~$1/7$ of the domain in both QC models) and its subsequent coarsening, which require a substantial number of mesh iterations. This behavior can be avoided by using more aggressive mesh refinement/coarsening strategies, in which the mesh is updated only in selected steps or only a limited number of mesh iterations is allowed in each time step. Such modifications can significantly reduce the number of mesh iterations and thus speed up the simulation. However, if too aggressive, they may compromise the accuracy of the QC simulations. In cases with more localized phenomena, a higher computational gain by the QC methods is expected. The \fecz{} model, on the contrary, provides a substantial speed-up of the order of~$50$, but at the cost of a lower accuracy as compared to the \full{} solution.

Unlike the atomistic problem of Section~\ref{sec:Microscale}, extensions to more relevant 3D structural lattices can be realized without significant changes to both the general QC as well as the homogenized model with fixed cohesive zones. In situations in which a crack trajectory is not known a priori and fixed cohesive zones cannot be used, alternative options for the homogenized model may be used. For instance, either multiple cohesive zones can be considered in the region in which crack is expected, or cohesive zones can be built in between all elements~\cite{xu1994numerical}. However, the latter option significantly increases the induced computing efforts and the resulting crack trajectory may moreover be mesh dependent. The extended finite element method, with cohesive formulation, is another interesting alternative~\cite{wang2015numerical,benvenuti2008regularized,xu1994numerical}, suffering, nevertheless, from technical difficulties for situations involving crack curving and crack branching.
\begin{table}
	\caption{The numerical performance corresponding to the individual computational models used for the three-point bending test example.}
	\renewcommand*{\arraystretch}{1.3}
	\centering
	\begin{tabular}{l|rrrr}
		~ & \shortstack{elastic stiffness \\ error} & \shortstack{peak force \\ error} & \shortstack{minimal/maximal \\ DOFs} & \shortstack{computational \\ demand} \\\hline
		Full solution	& $0$ & $0$ & $1,053,183$	& $1.0$ \\
		QC coarse		& $+21.32\%$ & $+0.28\%$ & $1,464/132,380$	& $\approx 1/2.5$ \\
		QC fine			& $+1.61\%$	& $-0.03\%$ & $5,408/152,456$	& $\approx 1/2.5$ \\
		FE-CZ			&  $+0.78\%$	& $-23.30\%$ & $10,175$			& $\approx 1/50$ \\ 
	\end{tabular}
	\label{tab:lat-res-num}
\end{table}
%
%
\section{Summary and conclusion}
\label{sec:Conclusion}
This paper has provided a detailed description and thorough comparison of two classes of homogenization techniques towards effective representation of two-dimensional discrete systems. The first class, referred to as QuasiContinuum~(QC) based methods, considers the fully-resolved underlying discrete system, which is subsequently reduced through suitable mathematical tools such as projection and reduced integration. The second class, referred to as homogenization based methods, consists in homogenizing the underlying discrete system first, into which localized discrete mechanisms are subsequently embedded. The necessary theoretical basis for both classes of methods has been reviewed, and their performance demonstrated on two representative examples considered at the nano- and meso-scale, revealing their strengths and weaknesses.

The first class, i.e.,~QC based models, is capable of capturing all important phenomena of the underlying full model. The obtained results are thus sufficiently accurate for the examples considered at both scales, reproducing also unexpected mechanical behavior such as dislocation reflection and distributed cracking. At the same time, QC based models provide a significant reduction of the original problem in terms of the number of Degrees Of Freedom~(DOFs), although the achieved savings in computing times might be less significant due to the computational cost involved with mesh adaptivity. Typical speed-ups achieved in this work are of the order of~$2.5$--$15$.

The second class, i.e.,~homogenization based approaches, represented by the Peierls--Nabarro model at the nanoscale and by the cohesive zone model at the mesoscale, is often more efficient as compared to the QC based methods in terms of computational speed-up, although it is usually less accurate because the effects that have their origin in the discreteness of a system (e.g., Peierls stress) cannot be captured by the continuum approach. Typical savings in computing times corresponded to~$5$--$50$ in comparison with the full model.
Although primal kinematic quantities such as overall displacement, dislocation positions, or crack opening and crack length are usually captured with adequate accuracy, the associated conjugate quantities such as transmission stress or maximum peak force suffer from more significant errors.
The most significant weakness of the homogenization-based models is that they are constructed based on prior assumptions on the response of the system, thus being unable to capture general behavior and unexpected phenomena such as dislocation reflection, distributed cracking, or crack branching. 
This may be acceptable if one is certain that no such phenomena will occur, or even desired if one particular phenomena is to be studied that would only be clouded by the unexpected phenomena. If this is not the case, model limitations can be lifted by adopting, e.g.,~multiple glide planes of different orientations, assuming all inter-element interfaces as cohesive zones, or using an extended formulation combined with cohesive zone models. Such extensions might, nevertheless, significantly complicate the entire procedure and increase the associated computing time.

The resulting computational savings achieved by all reduced models are substantially affected by the level of scale separation, which is not very large in the presented representative examples. With increasing scale separation, e.g.,~increasing the size of specimen domain while fixing the lattice spacing, a significant improvement in observed speed-ups may be reached.

Both presented classes are sufficiently general tools for predicting localized phenomena in discrete systems. The homogenization based models are usually more efficient, although they might lack a sufficient level of detail and accuracy. The QC based methods, on the other hand, exhibit greater flexibility compensated by higher computing costs.
Which method to adopt thus strongly depends on the desired accuracy and the acceptable computational costs. However, one should keep in mind the potential occurrence of unexpected phenomena. The QC based methods can thus be considered a safe option that may bring, nevertheless, only limited gain. Larger gain may be achieved by dedicated homogenized models, but safe option would be to check their results by a flexible QC method to verify occurrence of any unexpected phenomena.
%
%
\section*{Acknowledgements}
KM and OR would like to acknowledge financial support from the Czech Science Foundation (GA\v{C}R), projects No.~17-04150J~(KM) and No.~19-26143X~(OR). All authors would furthermore like to thank Milan Jir\'{a}sek and Jan Zeman from the Czech Technical University in Prague for useful discussions and critical comments on the manuscript.
%
%
%
\bibliography{mybibfile}
\end{document}